%% file: main.tex
\documentclass[journal=ancac3,manuscript=article,layout=twocolumn,10pt]{achemso}

\usepackage[utf8]{inputenc}
\usepackage[T1]{fontenc}
\usepackage{lmodern}

\usepackage{cancel}
\usepackage{amsmath}
\usepackage{amssymb}
\usepackage{textcomp}
\usepackage{gensymb}
\usepackage{siunitx}
\DeclareSIUnit\angstrom{\text {Å}}
\DeclareSIUnit\amu{\text {amu}}
\DeclareSIUnit\atm{\text {atm}}
\DeclareSIUnit\atom{\text {atom}}

\usepackage{nicefrac}
\usepackage{graphicx}
\usepackage{float}
\usepackage{wrapfig}

\usepackage{titletoc}
\contentsmargin{2.55em}
\dottedcontents{section}[3.8em]{\vspace{\baselineskip}\bfseries}{2.3em}{1pc}
\dottedcontents{subsection}[6.1em]{\vspace{0.5\baselineskip}}{3.2em}{1pc}

\usepackage[colorlinks,allcolors=black,citecolor=blue,urlcolor=blue]{hyperref}
\usepackage{cleveref}
\usepackage{placeins}
\usepackage{booktabs}
\usepackage{makecell}
\usepackage{multirow}
\usepackage{tabularx}
\usepackage{xcolor}

\DeclareUnicodeCharacter{2212}{-}
\graphicspath{{figures/}}

\let\oldmaketitle\maketitle
\let\maketitle\relax

\mciteErrorOnUnknownfalse

\newcommand*\dd{\mathop{}\!\mathrm{d}}   %

\def\mytitle{
Strain-Dependent Wetting of Graphene
}
\title{\mytitle}

\author{Darren Wayne Lim}%
\affiliation{Cavendish Laboratory, Department of Physics, University of Cambridge, Cambridge, CB3 0HE, UK}
\alsoaffiliation{Lennard-Jones Centre, University of Cambridge, Trinity Ln, Cambridge, CB2 1TN, UK}

\author{Xavier R.~Advincula}%
\affiliation{Yusuf Hamied Department of Chemistry, University of Cambridge, Lensfield Road, Cambridge CB2 1EW, UK}
\alsoaffiliation{Cavendish Laboratory, Department of Physics, University of Cambridge, Cambridge, CB3 0HE, UK}
\alsoaffiliation{Lennard-Jones Centre, University of Cambridge, Trinity Ln, Cambridge, CB2 1TN, UK}

\author{William C.~Witt}%
\affiliation{Harvard John A.~Paulson School of Engineering and Applied Sciences, Harvard University, Cambridge, MA, USA}

\author{Angelos Michaelides}%
\affiliation{Yusuf Hamied Department of Chemistry, University of Cambridge, Lensfield Road, Cambridge CB2 1EW, UK}
\alsoaffiliation{Lennard-Jones Centre, University of Cambridge, Trinity Ln, Cambridge, CB2 1TN, UK}

\author{Fabian L.~Thiemann}%
\affiliation{Microsoft Research AI for Science,  Cambridge, CB1 2FB, UK}
\alsoaffiliation{Cavendish Laboratory, Department of Physics, University of Cambridge, Cambridge, CB3 0HE, UK}

\author{Christoph Schran}%
\email{cs2121@cam.ac.uk}
\affiliation{Cavendish Laboratory, Department of Physics, University of Cambridge, Cambridge, CB3 0HE, UK}
\alsoaffiliation{Lennard-Jones Centre, University of Cambridge, Trinity Ln, Cambridge, CB2 1TN, UK}

\date{\today}

\begin{document}

\twocolumn[
\begin{@twocolumnfalse}
\oldmaketitle
\begin{abstract}
Understanding how water wets graphene is critical for predicting and controlling its behaviour in nanofluidic, sensing, and energy applications.
A key measure of wetting is the \textit{contact angle} made by a liquid droplet against the surface, yet experimental measurements for graphene span a wide range, with no consensus for free-standing graphene.
Here, we use a machine learning potential with \textit{ab initio} accuracy to provide an atomistic first-principles prediction for this unsolved problem, finding a weakly hydrophilic contact angle of $72.1 \pm 1.5 \degree$.
More importantly, we unveil that graphene's wetting properties are highly sensitive to mechanical strain: tensile strain makes graphene significantly less hydrophilic, while compressive strain induces coherent ripples around the droplet, resulting in pronounced anisotropic wetting and contact angle hysteresis.
We show that there is a strong coupling between the three-phase contact line and the intrinsic thermal ripples of free-standing graphene, which contributes to this strain sensitivity.
Our results demonstrate that the wettability of 2D membranes are governed not only by their chemistry but also by their dynamic morphology, introducing a new class of wetting behaviour unique to atomically thin materials that offers an additional explanation for variability in experimental measurements.
These findings reveal that mechanical strain may be a practical route to controlling wetting in 2D nanomaterials-based technologies, with promising consequences for nanofluidic and nano-filtration applications.
\end{abstract}
\end{@twocolumnfalse}
]

\clearpage

\section*{Introduction}

The water-graphene interface is one of, if not the most, scientifically fascinating and technologically important water-solid interfaces.
It, for example, exhibits strongly altered dielectric properties\cite{Fumagalli2018/10.1126/science.aat4191,Wang2025/10.1038/s41586-025-09558-y}, anomalous friction\cite{Radha2016/10.1038/nature19363,Secchi2016/10.1038/nature19315}, a potential superionic state\cite{Kapil2022/10.1038/s41586-022-05036-x,Jiang2024/10.1038/s41567-023-02341-8}, and may or may not be ``wetting transparent''\cite{Hou2026/10.1038/s41467-026-71053-3,Shih2012/10.1103/PhysRevLett.109.176101,Wang2026/10.1016/j.chempr.2026.103023}.
Coupled with this, it has potential applications in water filtration and desalination\cite{OHern2014/10.1021/nl404118f,Surwade2015/10.1038/nnano.2015.37}, blue energy\cite{Siria2017/10.1038/s41570-017-0091,Zhang2021/10.1038/s41578-021-00300-4}, and biomolecular translocation\cite{Garaj2010/10.1038/nature09379,Luan2018/10.1021/acs.jpclett.8b01340}.
The single most important and most basic property of the interface is its wettability.
Yet remarkably, after so much work, it is still not clear if graphene is hydrophobic or hydrophilic\cite{Cicero200810.1021/ja074418+,Belyaeva2020/10.1016/j.surfrep.2020.100482}, nor is it known how the intrinsically rippled nature of graphene\cite{Fasolino2007/10.1038/nmat2011,Meyer2007/10.1038/nature05545,Wang2012/10.1021/nl300071y} impacts its wettability.

The wettability is most directly quantified by the water droplet contact angle.
Unfortunately, experimental measurements of the graphene-water contact angle vary drastically\cite{Carlson2024/10.1021/acs.jpclett.4c01143}, from values as low as $10\degree$ \cite{Belyaeva2018/10.1002/adma.201703274} to as high as $143\degree$ \cite{Hsieh2013/10.1016/j.tsf.2012.05.056}.
Some of this variation can be attributed to the choice of substrate used to support the graphene sheet.
Wetting transparency to certain materials\cite{Belyaeva2018/10.1002/adma.201703274, Rafiee2012/10.1038/nmat3228} might contribute, although there is disagreement as to the exact nature of this transparency\cite{Hou2026/10.1038/s41467-026-71053-3,Shih2012/10.1103/PhysRevLett.109.176101,Wang2026/10.1016/j.chempr.2026.103023,Li2013/10.1038/nmat3709,Raj2013/10.1021/nl304647t,Zhao2016/10.1039/C5RA13916C}.
Some experimental studies have attempted to measure the contact angle on free-standing graphene, in order to isolate the above-mentioned effects\cite{Ondarcuhu2016/10.1038/srep24237,Prydatko2018/10.1038/s41467-018-06608-0}; however, reported results disagree.
This may be due to airborne contaminants, defects, and other imperfections during sample preparation, which have been found to affect the contact angle\cite{Aria2016/10.1021/acs.jpcc.5b10492,Kozbial2016/10.1021/acs.accounts.6b00447,Li2013/10.1038/nmat3709,Shin2010/10.1021/la100231u}, providing another source of noise complicating experimental measurements.

Computational estimates of the graphene-water contact angle are likewise limited, either in system size or in accuracy.
While contributing tremendously to the qualitative understanding of water-carbon behaviour, simulations based on empirical force fields yield inconsistent results for the contact angle, depending on the parametrization.
For example, Taherian et al.\cite{Taherian2013/10.1021/la304645w}~report a contact angle of $101\degree$, whereas Ma et al.\cite{Ma2016/10.1038/nmat4449}~find a contact angle of $89\degree$, and Carlson et al.\cite{Carlson2024/10.1021/acs.jpclett.4c01143}~predict $80\degree$.
The contact angle is also sensitive to the choice of water model\cite{Liao2022/10.1016/j.apsusc.2022.154477}, further compounding these discrepancies.
These issues highlight the need for simulations in which all interatomic interactions are treated from first principles.\cite{Li2023/10.1103/PhysRevResearch.5.023018}
A pioneering \textit{ab initio} molecular dynamics (AIMD) study by Li and Zeng\cite{Li2012/10.1021/nn204661d} provided early qualitative insights into this system, reporting a contact angle of $87\degree$.
While a \textit{tour de force} effort for its time, true quantitative predictions were hindered by the steep computational cost of AIMD, greatly limiting the system size and thereby making finite-size effects unavoidable\cite{Amirfazli2004/10.1016/j.cis.2004.05.001}.
More generally, the computational studies discussed above impose `frozen' graphene sheets with fixed carbon atoms; yet free-standing graphene is fully flexible, and ripples thermally in order to lower its vibrational free energy\cite{Fasolino2007/10.1038/nmat2011,Meyer2007/10.1038/nature05545}.
This flexibility is already known to impact water diffusion and transport\cite{Marbach2018/10.1038/s41567-018-0239-0}, which is inherently linked to wetting\cite{Agosta2026/10.1038/s41524-026-02079-w}.
As such, capturing this membrane-like behaviour is essential for a realistic description of graphene's intrinsic wettability.
Therefore, despite extensive experimental and computational efforts, the intrinsic graphene–water contact angle remains unresolved.

Here we provide a best-effort atomistic first-principles prediction for the contact angle of dynamical water droplets on free-standing pristine graphene using molecular dynamics (MD) simulations.
By focusing on free-standing graphene, we isolate its intrinsic wetting behaviour without the influence of substrates or contamination.
To this end, we develop a novel methodology to define the contact angle for a spherical nanodroplet on a non-flat surface in a spatially-resolved manner, allowing us to properly account for surface rippling in free-standing graphene.
Through the use of a machine learning potential (MLP)\cite{Bartok2010/10.1103/PhysRevLett.104.136403,Behler2007/10.1103/PhysRevLett.98.146401,Deringer2019/10.1002/adma.201902765,Kang2020/10.1021/acs.accounts.0c00472,Thiemann2024/10.1088/1361-648X/ad9657,Unke2021/10.1021/acs.chemrev.0c01111} trained to replicate density functional theory (DFT) calculations, we are able to simulate droplets orders of magnitude larger than in previous \textit{ab initio} studies, which are essential for resolving and correcting the finite-size effects known to influence nanoscale contact angles.\cite{Amirfazli2004/10.1016/j.cis.2004.05.001, Kanduc2017/10.1063/1.4990741, Kanduc2018/10.1103/PhysRevE.98.032804}

Furthermore, we study the interplay between surface ripples in free-standing graphene and the wetting of nanoscale water droplets.
This interplay is probed by applying mechanical strain to the graphene sheet, thereby modifying its rippling dynamics.
We discover that the application of strain has an outsized effect on the graphene-water contact angle, significantly larger than expected from energy scaling arguments.
We find a two-way coupling of the dynamics of surface rippling to the three-phase contact line at the droplet edge, establishing that surface rippling plays a major role in the wetting of graphene by water.
These findings imply that strain is a control lever for wettability, and contributes to the large variability in experimental measurements.

\section*{Results and Discussion}

\subsubsection*{Water contact angle on free-standing graphene is weakly hydrophilic}

\begin{figure}
    \centering{}
    \includegraphics[width=1.0\linewidth]{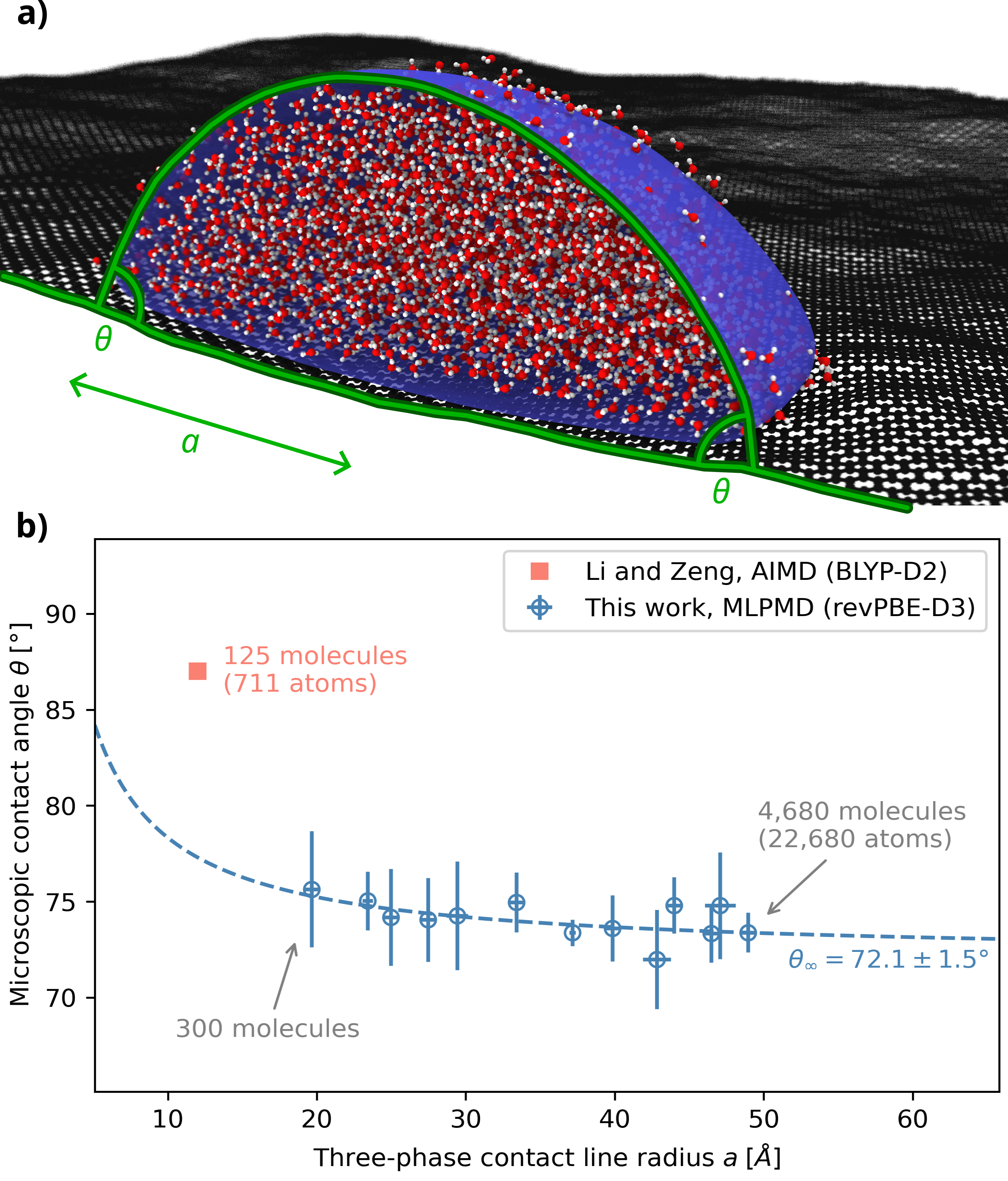}
    \caption{
    \textbf{The macroscopic contact angle of water on free-standing graphene is $72.1 \degree$.}
    \textbf{(a)} Cross-section of a snapshot of a spherical droplet of 4,680 water molecules on a free-standing, fully-dynamical graphene sheet, with the droplet's time-averaged interface overlaid in blue.
    The contact angle is determined as described in the \hyperref[sec:methods]{Methods} section and schematically illustrated here in green.
	\textbf{(b)} The microscopic contact angle for droplets of varying size, plotted against the radius of the three-phase contact line $a$.
    The result obtained by Li and Zeng (Ref.~\citenum{Li2012/10.1021/nn204661d}) using \textit{ab initio} MD is also plotted for comparison.
    The dashed line indicates the best-fit corrections following \cref{eq:finite-size-correction}, whose extrapolation yields the finite-size corrected macroscopic contact angle $\theta_{\infty}$.
    }
    \label{fig:fig1}
\end{figure}

To determine the intrinsic wetting behaviour of water on free-standing graphene, we simulate water droplets of varying sizes and extract their contact angles (\Cref{fig:fig1}a).
Using a MLP with first-principles accuracy at the revPBE-D3\cite{Perdew1996/10.1103/PhysRevLett.77.3865,Grimme2010/10.1063/1.3382344} level, we are able to perform MD for systems containing up to 23k atoms and nanosecond timescales, enabling a systematic assessment of finite-size effects.
These system sizes and timescales were enabled by \texttt{symmetrix}, a CPU-based accelerator for MLPs\cite{symmetrix,Kovacs2025/10.1021/jacs.4c07099,Batatia2025/10.1063/5.0297006}.
The contact angles were determined using a novel methodology based on the geometrical intersection of the droplet's \textit{time-averaged interface} with the \textit{time-averaged graphene heightmap}; see \hyperref[sec:methods]{Methods} for details.

For any droplet of finite size, the observed (or ``microscopic'') contact angle $\theta$ deviates from the true macroscopic limit $\theta_{\infty}$ due to a generalized line tension effect\cite{Kanduc2018/10.1103/PhysRevE.98.032804}, which causes a linear dependence between $\cos\theta$ and $1/a$ of the form
\begin{equation}
    \cos\theta \;\;=\;\; \underbrace{\rule[-1.5em]{0pt}{1.5em}\frac{\gamma_{\text{sv}}-\gamma_{\text{sl}}}{\gamma_{\text{lv}}}}_{\cos\theta_{\infty}} \;-\; \frac{\kappa}{\gamma_{\text{lv}}a} \; , \label{eq:finite-size-correction}
\end{equation}
where $\gamma_{\text{sv}}$, $\gamma_{\text{sl}}$, and $\gamma_{\text{lv}}$~are the surface energies of the solid-vapour, solid-liquid, and liquid-vapour interfaces respectively; $\kappa$ is the apparent line tension\cite{Kanduc2018/10.1103/PhysRevE.98.032804,Kanduc2017/10.1063/1.4990741}; and $a$ is the radius of the three-phase contact line.

To obtain the finite-size corrected contact angle $\theta_{\infty}$, we therefore plot the microscopic contact angles $\theta$ of all simulated droplets against their corresponding radii $a$ and extrapolate the resulting trend to $a\to\infty$ in \Cref{fig:fig1}b.
The apparent line tension $\kappa$ required to make this best-fit is $(7.8\pm3.3)\times 10^{-12}\,\unit{\newton}$, which is within the spread of previously reported line tensions in the literature.\cite{Amirfazli2004/10.1016/j.cis.2004.05.001,Dobbs1993/10.1016/0378-4371(93)90120-S,Getta1998/10.1103/PhysRevE.57.655,Heim2013/10.1021/la402932y,Sergi2012/10.1016/j.fluid.2012.07.010,Werder2003/10.1021/jp0268112,Zhang2018/10.1063/1.5040574}
This procedure yields a macroscopic contact angle of $\theta_{\infty} = 72.1 \pm 1.5 \degree$.
We argue that this is a good estimate for the contact angle of water on pristine, free-standing graphene, as the simulation accounts for thermal and dynamical effects, including surface ripples of the graphene sheet, at first-principles level accuracy with sufficient timescale and finite-size correction.

The contact angle directly yields the surface energy difference $\Delta\gamma=\gamma_{\text{sv}}-\gamma_{\text{sl}}=\gamma_{\text{lv}}\cos\theta$ via the Young equation\cite{Young1805/10.1098/rstl.1805.0005}. 
Based on the MLP's liquid-vapour surface tension (see Supporting Information), our result for the contact angle translates to a surface energy difference of $22.9\pm\SI{1.9}{\milli\newton\meter^{-1}}$, where the positive value indicates graphene's weak hydrophilicity.
Our estimate is consistent with the experimental measurement of van Engers et al.\cite{Van_Engers2017/10.1021/acs.nanolett.7b01181}~at $32\pm\SI{8}{\milli\newton\meter^{-1}}$, which was obtained using a `direct' pull-off force method on glass-supported epoxy-bonded graphene samples.
Remaining discrepancies might be explained by the substrate effect.

We briefly compare this result with simulations using empirical force fields.
For example, Ma et al.\cite{Ma2016/10.1038/nmat4449}~model the droplet using the TIP4P potential for water, plus a Lennard--Jones 6-12 potential between oxygen and carbon, fitted to a diffusion quantum Monte Carlo benchmark\cite{Ma2011/10.1103/PhysRevB.84.033402}; they report a graphene-water contact angle of $89\degree$ based on these calculations.
On the other hand, Carlson et al.\cite{Carlson2024/10.1021/acs.jpclett.4c01143}~propose a model using the SPC/E potential for water, plus a Lennard--Jones 6-12 potential also between oxygen and carbon but with different parameters, fitted to reproduce a finite-size corrected graphite-water contact angle of $60\degree$.
This predicts a $80\degree$ graphene-water contact angle, close to our estimate.
As another point of comparison, Taherian et al.\cite{Taherian2013/10.1021/la304645w}~report a contact angle of $101\degree$ based on phantom-wall thermodynamic integration calculations from the Werder et al.\cite{Werder2003/10.1021/jp0268112}~potential.
The disagreement between these empirical force field predictions indicates that graphene–water wetting is sensitive to the choice of parametrization, even when nominally fitted on first-principles data or experimental graphite-water contact angles.
As such, our MLP-based approach provides a necessary benchmark where all degrees of freedom are described at first-principles level.

We also compare our result against the AIMD study by Li and Zeng (Ref.~\citenum{Li2012/10.1021/nn204661d}), which reports a contact angle of $87\degree$ at the BLYP-D2 level of theory.
However, it should be noted that their results were affected by their small system sizes, restricted to a 125-molecule droplet and a $\SI{20}{\ps}$ sampling window, owing to the high cost of AIMD.
Assuming a similar value of line tension for the BLYP-D2 prediction to revPBE-D3, this implies that their finite-size corrected result should be around ${\sim}82\degree$.
Some numerical disagreement should be expected, as the underlying electronic structure of the water-graphene interaction differs, but the overall result of graphene being weakly hydrophilic is in agreement.
It should also be noted that they used a spatially fixed graphene sheet, which does not account for the dynamical ripples of free-standing graphene.

In summary, we propose our macroscopic estimate of $\theta_{\infty} = 72.1 \pm 1.5\degree$ as a reference for the contact angle of water on pristine, free-standing graphene.
Crucially, our methodology maintains a fully flexible representation of the entire system, explicitly capturing the natural thermal rippling and dynamic response of the graphene substrate.
Given the known impact of surface vibrations on water diffusion and transport~\cite{Marbach2018/10.1038/s41567-018-0239-0}, this naturally raises the question of how rippling might impact the wetting behaviour.

\subsubsection*{Strained graphene is less hydrophilic}

\begin{figure*}
    \centering{}
    \includegraphics[width=1.0\textwidth]{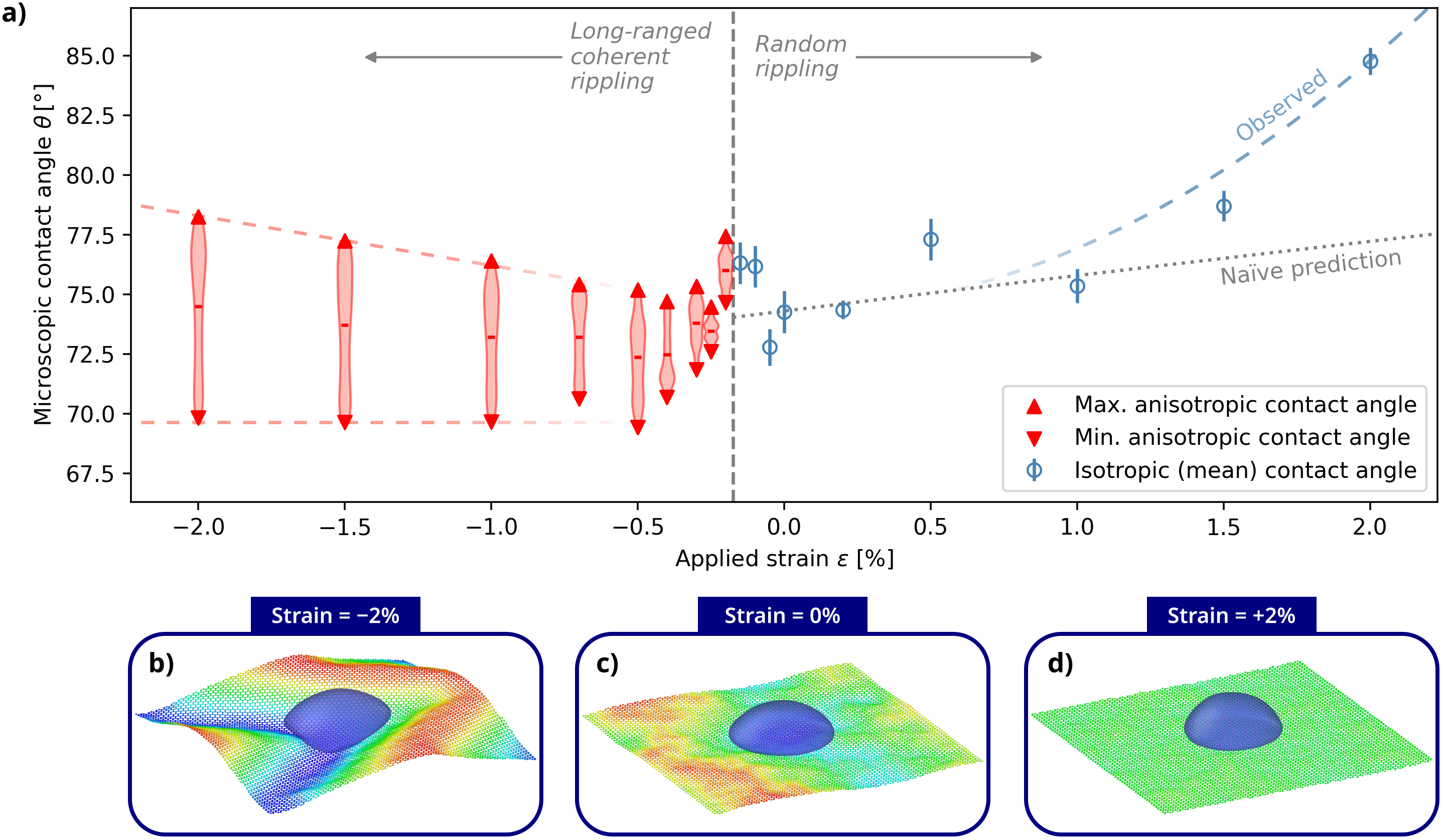}
    \caption{
    \textbf{Strong strain-dependence, from compressive (negative) to tensile (positive), on the graphene-water contact angle.}
    \textbf{(a)} The microscopic contact angle for a droplet of 1,000 water molecules is affected by the biaxial strain applied upon the graphene sheet.
    When unstrained, or placed under tensile strain (right side of vertical divider), the graphene sheet ripples randomly due to thermal fluctuations, and increasing tensile strain suppresses these ripples and also increases the contact angle (dashed blue line is a guide to the eye).
    This increase in contact angle is significantly larger than the ``na{\"i}ve'' prediction (dotted grey line) derived from the expected scaling of static intermolecular interaction energies with carbon density.
    On the other hand, compressive strains greater than or equal to approximately $-0.2\%$ (left side of vertical divider) cause the graphene sheet to form a long-ranged coherent ripple wave which the droplet ``surfs''; as such the droplet exhibits continuous contact line motion, with an anisotropic distribution of contact angles in the co-moving frame, shown here as a violin plot with maximum and minimum values marked.
    These maximum and minimum anisotropic contact angles vary with the applied strain (dashed red lines are guides to the eye).
    \textbf{(b--d)} Representative snapshots of the graphene sheet and time-averaged droplet interface illustrated in blue under (b) $-2.0\%$ compressive strain, (c) free-standing conditions, and (d) $+2.0\%$ tensile strain, with the carbon atoms colour-coded according to their instantaneous $z$-coordinate.
    Snapshots (c) and (d) share the same colour scale, while snapshot (b) is shown with a different colour scale as the displacements are an order of magnitude larger.
    The coherent rippling of the graphene sheet, and ``surfing'' position of the anisotropic droplet, under compressive strain is seen in snapshot (b).
    }
    \label{fig:fig2}
\end{figure*}

Mechanical strain provides a natural parameter to tune the dynamical properties of free-standing graphene.
For example, it is known that the application of tensile strain modifies the phonon spectrum\cite{Bonini2012/10.1021/nl202694m,Huang2009/10.1073/pnas.0811754106,Xu2025/10.1103/q8xg-qn4g}, leading to a suppression of thermal ripples\cite{Singh2013/10.1103/PhysRevB.87.094112}.
Under compression, on the other hand, graphene is known to undergo a mechanical phase transition, spontaneously ``buckling'' and forming a long-ranged coherent ripple wave with amplitude much larger than thermal random ripples\cite{Ma2016/10.1038/nmat4449, Meng2013/10.1063/1.4857115, Thiemann2020/10.1021/acs.jpcc.0c05831, Wang2017/10.1088/2053-1591/aa7324}.
The question, therefore, is whether these dynamical effects change the water contact angle.

We performed a series of MD simulations in which mechanical strain was applied biaxially to tune the rippling dynamics.
The simulated systems explore a range of compressive and tensile strains, reaching up to $2.0\%$ in each direction.
We find that tensile strain increases the microscopic contact angle from $74.3\pm 2.8\degree$ when relaxed up to $84.8\pm 1.2\degree$ at $+2.0\%$ tensile strain (right half of \Cref{fig:fig2}a).
The observed increase in contact angle is noteworthy, as it represents a \textit{decrease} in the magnitude of the surface energy difference $\Delta\gamma$ by nearly 67\% in the most extreme case.
In other words, tensile strain makes graphene substantially less hydrophilic.

This effect cannot be explained merely by the increased interatomic spacing of the graphene sheet: na{\"i}vely, the strength of water-carbon interactions is expected to scale with the area density of carbon atoms, giving a $\Delta\gamma\propto(1+\varepsilon)^{-2}$ proportionality for strain $\varepsilon$ if static effects are the only significant factor.
This implies that a $+2.0\%$ strain should only drive a ${\sim}4\%$ decrease in $\Delta\gamma$.
Indeed, DFT calculations (see Supporting Information) show that the interaction energy of a water slab upon a fully wetted graphene sheet follows the ``na{\"i}ve'' scaling law.
Such a small change in static intermolecular interactions is therefore insufficient to explain the much larger weakening of hydrophilicity.

To further confirm whether the observed increase of contact angle under tension is related to the dynamical rippling motion of the graphene membrane, we simulated the same droplet on spatially fixed, flat graphene sheets of the same strain condition (see Supporting Information).
When no strain is applied, the microscopic contact angle on the spatially fixed unstrained graphene ($72.4 \pm 1.5\degree$) is within the error range of the result for dynamical free-standing graphene ($74.3 \pm 2.8\degree$), which agrees with earlier findings in literature that spatially fixed and free-standing graphene have similar contact angles\cite{Werder2003/10.1021/jp0268112, Liao2022/10.1016/j.apsusc.2022.154477}.
However, this changes under tensile strain: while the contact angle on spatially fixed graphene does increase with strain, the increase is less dramatic (only to $80.8 \pm 1.4\degree$) than on dynamical graphene ($84.8 \pm 1.2\degree$).
This is not a fair one-to-one comparison for the dynamical component of the interfacial free energy, as spatially fixing the graphene sheet removes only the carbon degrees of freedom non-adiabatically, but this nonetheless highlights that some of the strain-driven weakening of hydrophilicity must stem from the unique dynamical properties of graphene's surface ripples and how they are modified under strain.

As for the simulations with biaxial compressive strains, we find that the graphene sheet undergoes the previously-reported mechanical phase transition for strains of approximately $-0.2\%$ or larger, where it forms a large-amplitude long-ranged coherent ripple wave (see \Cref{fig:fig2}b).
Our observations of this phenomenon are mostly in line with previous literature\cite{Ma2016/10.1038/nmat4449, Meng2013/10.1063/1.4857115, Thiemann2020/10.1021/acs.jpcc.0c05831, Wang2017/10.1088/2053-1591/aa7324}, although in our case, the critical strain of this phase transition is slightly affected by the presence of the droplet.
Furthermore, in this coherent rippling state, the wave travels at a roughly constant velocity plus a diffusive component, and the droplet is consequently carried along in the valley of the wave in the same ``surfing'' motion as first reported for force field-based simulations by Ma et al.\cite{Ma2016/10.1038/nmat4449}.
This random-to-coherent rippling phase transition, and subsequent ``surfing'' motion, is illustrated in more detail in the Supporting Information.

Owing to this long-ranged coherent rippling of the graphene sheet, and the constant ``surfing'' motion of the droplet, the droplet does not relax into a static equilibrium configuration, but instead settles into a strongly anisotropic steady-state even in the long-time average.
In our case, the droplet fits entirely within a single ripple trough, and its contact line motion continuously follows the surrounding crests and valleys under the constant propagation of the ripple wave.
This gives rise to distinct advancing and receding sides of the three-phase contact line in the frame co-moving with the droplet's centre-of-mass, producing contact-angle hysteresis and breaking the rotational symmetry of the droplet about the $z$-axis.
Consequently, the contact angle is no longer uniform, but spans a range of anisotropic values along the steady-state co-moving contact line.
The distribution of these time-averaged spatially-resolved contact angles as a function of compressive strain is plotted in the left half of \Cref{fig:fig2}a.
For sufficiently large compression, the minimum anisotropic contact angle (representing the \textit{receding} boundary) is roughly constant with strain, whereas the maximum anisotropic contact angle (representing the \textit{advancing} boundary) increases with strain.
The mean contact angle across the three-phase contact line increases slightly with compression, as a result of this asymmetric shift.

The overall picture of \Cref{fig:fig2}a is that mechanical strain has a non-monotonic yet qualitatively consistent effect: both tensile and compressive strain increase the contact angle, meaning that graphene becomes effectively less hydrophilic under either type of strain.
It should be noted that, for compressive strains, the dynamic angles at the receding and advancing boundaries cannot be directly compared to the static equilibrium angles under tensile strains, as contact angle hysteresis is driven by the kinetics of contact line motion rather than the equilibrium interfacial thermodynamics.\cite{Gao2006/10.1021/la060254j}
Nonetheless, the two-sided trend of \Cref{fig:fig2}a shows that mechanical strain has an outsized effect on graphene's wettability.
Given the strong modulation of surface dynamics by strain, we turn next to a detailed analysis of the coupling of surface rippling and water droplet formation.

\subsubsection*{Two-way coupling of surface ripples to the droplet edge}

\begin{figure*}
    \centering{}
    \includegraphics[width=1.0\textwidth]{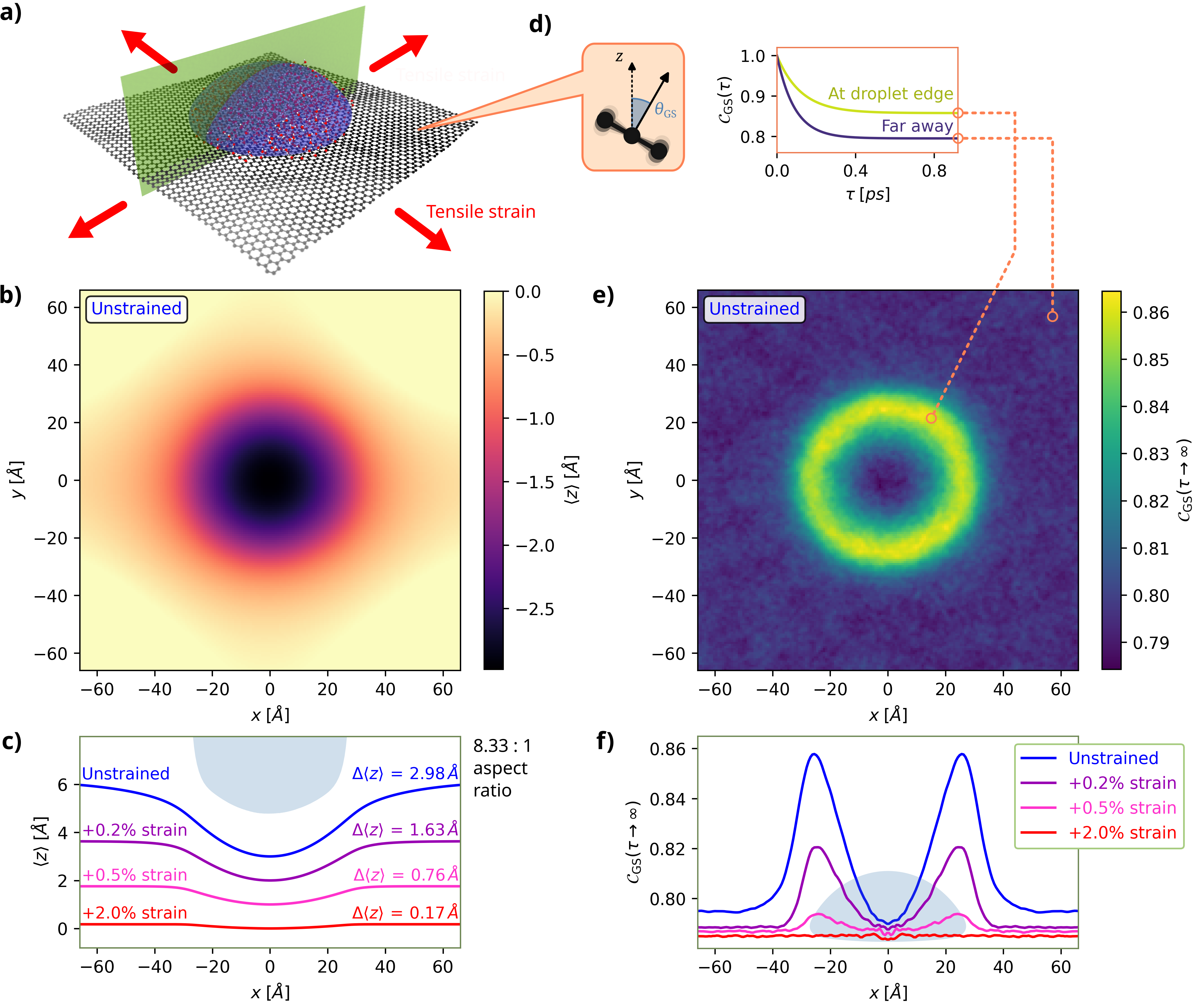}
    \caption{
    \textbf{The ``footprint'' of the water droplet on graphene and its effects on graphene's rippling dynamics under strain.}
    \textbf{(a)} A representative snapshot of a droplet of 1,000 molecules, with tensile strain applied to the graphene sheet.
    The translucent green plane indicates the cross-section that panels (c) and (f) pertain to.
    \textbf{(b)} The time-averaged graphene heightmap (TAGH), representing the average $z$-coordinate of the carbon atoms in droplet-centred coordinates, for the unstrained case.
    \textbf{(c)} The TAGH along the cross-section for each strain condition, with the time-averaged droplet interface for the unstrained case shown at the top.
    The graph is plotted with a 8.33:1 aspect ratio to highlight the distortion of the sheet.
    \textbf{(d)} The dynamics of surface rippling were studied via the graphene sheet's local inclination angle $\theta_{\text{GS}}$ relative to the $z$-axis, and its normalized temporal autocorrelation function $\mathcal{C}_{\text{GS}}(\tau)$ as defined in \cref{eq:long_time_incl_corr}; this is further elaborated in \hyperref[sec:methods]{Methods}.
    The graph shows two examples of $\mathcal{C}_{\text{GS}}(\tau)$ plotted against $\tau$ for two locations on the free-standing graphene sheet, which decay exponentially from unity to some long-time limit.
    \textbf{(e)} The long-time limit $\mathcal{C}_{\text{GS}}(\tau\to\infty)$ of the graph from panel (d), illustrated for all locations on the graphene sheet for the unstrained case.
    This ``long-time inclination correlation'' peaks at the droplet edge and returns to a uniform background value everywhere else, including at the droplet centre.
    \textbf{(f)} The long-time inclination correlation $\mathcal{C}_{\text{GS}}(\tau\to\infty)$ along the cross-section for each strain condition.
    The time-averaged droplet interface for the free-standing case is overlaid in the background, to illustrate the location of the droplet edge.
    }
    \label{fig:fig3}
\end{figure*}

In all simulations for either unstrained or tensile strained graphene, the presence of the water droplet induces a persistent distortion on the graphene sheet, resembling a ``footprint'' (see \Cref{fig:fig3}b).
The distortion is reduced, but never eliminated, by applying tensile strain (see \Cref{fig:fig3}c).
This latent curvature of the graphene surface towards the droplet can be attributed to weakly hydrophilic attractions, which favour an increased graphene-water contact area compared to a perfectly flat surface.
Such curvature must be localized in the vicinity of the droplet edge in the limit of large droplet sizes, due to translational symmetry along the $xy$ plane in both the deep droplet interior and the faraway unwetted region forbidding any curvature.

The graphene-water-vacuum interface along the droplet perimeter, i.e.~the three-phase contact line, must also perturb the dynamics of graphene's thermal ripples.
This arises from the breaking of translational symmetry at the contact line, disrupting Bloch symmetry for all phonon dispersions\cite{Zheng2016/10.1016/j.commatsci.2016.08.036}.
To study this, we examine the \textit{long-time inclination correlation} of the graphene sheet, denoted $\mathcal{C}_{\text{GS}}(\tau\to\infty)$, which was first introduced by Thiemann et al.\cite{Thiemann2025/10.1073/pnas.2416932122}~as a spatially-localized indicator of phase transitions in rippling dynamics; see \hyperref[sec:methods]{Methods} for details.
The long-time inclination correlation quantifies how strongly the sheet's ripples are correlated over time, with values close to 1 suggesting a constrained local inclination and static rippling, and lower values suggesting greater flexibility and dynamic rippling.
For reference, free-standing unwetted graphene at $\SI{300}{\kelvin}$ has a uniform value of $\mathcal{C}_{\text{GS}}(\tau\to\infty)=0.804\pm 0.005$, and a graphene sheet fully covered by water has $\mathcal{C}_{\text{GS}}(\tau\to\infty)=0.791\pm 0.008$.
In the presence of a droplet, however, $\mathcal{C}_{\text{GS}}(\tau\to\infty)$ exhibits a pronounced, highly localized peak of approximately $\sim$0.86 at the droplet perimeter (\Cref{fig:fig3}e).
This enhancement suggests that the wetting-induced curvature leads to a local stiffening of the membrane, `freezing out' orientational fluctuations of ripples in the direction tangential to the contact line.

Under the application of tensile strain, we observe a decrease in the magnitude of thermal random ripples, as consistent with literature\cite{Singh2013/10.1103/PhysRevB.87.094112}.
Moreover, strain leads to a flattening of the droplet-induced curvature, seen in \Cref{fig:fig3}c.
Finally and most strikingly, strain suppresses the localized perturbation of $\mathcal{C}_{\text{GS}}(\tau\to\infty)$, seen in \Cref{fig:fig3}f.
In the most extreme case of the $+2.0\%$ tensile strain, the time-averaged $z$-coordinate of the graphene sheet only has a displacement of $\SI{0.17}{\angstrom}$ compared to $\SI{2.98}{\angstrom}$ in the unstrained case, and there is no perturbation seen in $\mathcal{C}_{\text{GS}}(\tau\to\infty)$ at all.
A similar trend is seen under compressive strains, where the localized perturbation of $\mathcal{C}_{\text{GS}}(\tau\to\infty)$ at the droplet perimeter vanishes with increasing compression, though in this regime the long-ranged coherent ripple wave presents a non-uniform ``background'' for the long-time inclination correlation (see Supporting Information).

Although some disturbance of rippling dynamics at the droplet boundary is expected from symmetry arguments alone, the magnitude and sharpness of the observed jump in $\mathcal{C}_{\text{GS}}(\tau\to\infty)$ for the unstrained case is striking.
Rather than transitioning smoothly from the fully wetted value of ${\sim}0.79$ to the unwetted value ${\sim}0.80$, the correlation increases dramatically only at the droplet edge.
Furthermore, this disturbance is sensitive to modifications of the graphene sheet's vibrational spectrum via mechanical strain.
This behaviour points toward a two-way coupling in which wetting modifies graphene's ripples, and the ripples, in turn, contribute to the free-energy balance at the contact line.
We argue that this mechanism explains the anomalous increase of contact angle under tensile strain, as shown in \Cref{fig:fig2}a on dynamical graphene, beyond any effect attributable to spatially fixed graphene.

The coupling of water wetting to surface ripples is reminiscent of the ``dripplons'' reported by Yoshida et al.\cite{Yoshida2018/10.1038/s41467-018-03829-1}.
A dripplon is a localized water droplet, formed within graphene-nanoconfined water with an accompanying deformation of the graphene sheets; they are created by water's layering preference being stronger than the graphene bending energy.
In particular, dripplons represent a coupling between confined water's mixed hydration states and graphene's flexibility, which is a mechanism unique to flexible confining materials, as opposed to rigid materials which prefer interstratification of the mixed hydration states.
In our case, the unconfined droplets result in coupling between the three-phase contact line and graphene's surface ripples instead, being uniquely enabled by the flexibility of graphene.

\section*{Conclusion}

In this work, we used a MLP with first-principles accuracy to perform nanosecond-scale MD simulations of water droplets on graphene to deliver a best-effort prediction of the contact angle of $72.1\pm 1.5\degree$.
These results show that free-standing graphene is weakly hydrophilic, at the revPBE-D3 level of theory.
We expect this contact angle estimate to be a reference for future experiments, in particular for settling the ``wetting transparency'' controversy by measuring the contact angle on free-standing graphene unsupported by any substrate.
While the exact value of the contact angle may potentially depend on the description of the underlying electronic structure, we expect the general conclusions to be robust.

Furthermore, we show that the application of mechanical strain has a strong effect on the wetting of dynamical graphene.
Tensile strain causes the water contact angle for a nanoscopic droplet of fixed size to increase drastically, implying that tension makes graphene less hydrophilic.
This increase of contact angle is only partially accounted for by the weakening of static water-graphene interactions, and is also related to dynamical rippling of the graphene sheet.
We demonstrate that the presence of a water droplet not only induces an average curvature in the membrane, but also causes a localized disturbance of the vibrational dynamics of the graphene sheet in the vicinity of the droplet boundary, thus revealing a two-way coupling between wetting at the three-phase contact line and graphene surface ripples.
Tensile strain then results in a suppression of this dynamical perturbation, which contributes to the strong increase of contact angle.

The application of compressive strain induces a phase transition from thermally random ripples to long-ranged coherent rippling.
For droplets of lengthscale comparable to the wavelength, the coherent wave `straddles' the droplet, causing the droplet to ``surf'' the valley of the wave.
This wave-driven motion gives rise to an anisotropic droplet geometry with a range of non-uniform contact angles, and in particular the advancing contact angle increases with compression due to corrugation stiffening.
For wetting at lengthscales much larger than the wavelength of the coherent wave, we imagine that the nanoscopic roughness of the compressed sheet might also lead macroscopically to either the Wenzel or Cassie--Baxter wetting states\cite{Murakami2014/10.1021/la4049067,Onda2022/10.1021/acs.langmuir.2c01847,Park2023/10.1002/admi.202202439}, although this only occurs far beyond the nano scale.

Our results show that mechanical strain affects the contact angle non-monotonically, and is an important parameter in studying the wetting of graphene.
Qualitatively similar results, where solid-solid interfacial energies can be affected by strain, have been previously demonstrated in the literature\cite{Adinehloo2024/10.1039/d3na01079a, Schweika2004/10.1103/PhysRevB.70.041401, Xu2017/10.1038/s41467-017-00636-y}; these reported results, however, are driven by mechanical effects limited to solid-solid interfaces, e.g.~lattice mismatch and surface stress.
Such mechanisms do not apply to the graphene-water system, which is laterally isotropic and atomically thin.
Instead, we propose that mechanical strain has a significant effect on wetting via its coupling to graphene surface ripples (ZA phonons), even though such modes are typically neglected in continuum wetting theories.
This may be a further explanation for the large range of reported experimental results, where the contact angle seems to be greatly affected by the substrate hosting the graphene.
While strain \textit{can} be controlled experimentally to fairly good precision\cite{Bao2009/10.1038/nnano.2009.191}, some degree of strain is always unavoidable in real experiments.
As such, even if graphene is fully chemically opaque to wetting, the presence of the substrate (and any strain induced by lattice mismatch or thermal expansion) fundamentally alters the vibrational spectrum on the graphene and thus changes its wetting properties.

These observations suggest that, in general, wetting on a 2D membrane is fundamentally different compared to rigid solids: wetting is governed not only by surface chemistry --- where e.g.~the presence of chemically polar interactions is already known to greatly affect hydrophobicity and hydrophilicity\cite{Giovambattista2007/10.1021/jp071957s} --- but also by vibrational entropy.
This introduces a new class of wetting behaviour unique to atomically thin materials.
On the flip side, our results also suggest that it might be possible to modulate and control the wetting of 2D nanomaterials, particularly graphene, via mechanical strain engineering for potential technological applications.
We propose that such ``strain-dependent hydrophilicity'' could be used to develop nanofluidic pumps for transporting water through graphene-confined or carbon nanotube-confined systems\cite{Bocquet2010/10.1039/B909366B,Secchi2016/10.1038/nature19315}.
This could be confirmed experimentally using e.g.~high-resolution atomic force microscopy (AFM) to detect the wetting-induced curvature of a nanodroplet\cite{Bonn2009/10.1103/RevModPhys.81.739,Seveno2013/10.1103/PhysRevLett.111.096101}, or electron spectroscopy to detect ripple suppression at the droplet edge\cite{Li2023/10.1038/s41467-023-38053-z,Qi2021/10.1038/s41586-021-03971-9}.

\section*{Methods}
\label{sec:methods}

\subsubsection*{Machine learning potential}

Throughout this work, we use the same MLP model as Ref.~\citenum{Advincula2026/10.1126/sciadv.aeb5772}.
This MLP was developed using the MACE architecture\cite{Batatia2022/mace, Batatia2025/10.1038/s42256-024-00956-x, Kovacs2023/10.1063/5.0155322} with 128 invariant channels ($L=0$), two layers, and a $\SI{6}{\angstrom}$ cutoff distance per layer.
The model thus captures semi-local interactions through an effective receptive field of $\SI{12}{\angstrom}$.

To accurately represent the potential energy surface, the MLP was trained using energies and atomic forces obtained from DFT calculations using the CP2K/Quickstep code.\cite{Kuhne2020/10.1063/5.0007045} We specifically used the revPBE functional\cite{Perdew1996/10.1103/PhysRevLett.77.3865} with D3 dispersion correction with zero damping\cite{Grimme2010/10.1063/1.3382344}, due to its robust performance in reproducing the structure and dynamics of liquid water\cite{Gillan2016/10.1063/1.4944633,Marsalek2017/10.1021/acs.jpclett.7b00391,Morawietz2016/10.1073/pnas.1602375113}, while also effectively capturing protonic defects and the interaction energies between water and graphene\cite{Brandenburg2019/10.1063/1.5121370}.

Training data included a wide variety of configurations representing different carbon-water environments, such as: bulk and implicitly confined water; free-standing graphene sheets, and AA and AB stacked graphene sheets of varying distance; water on graphene; and water confined between graphene sheets or within carbon nanotubes. In total, the training set comprised 5,845 structures, yielding root-mean-square errors of $\SI{0.9}{\milli\electronvolt/\atom}$ for energies and $\SI{26.3}{\milli\electronvolt/\angstrom}$ for forces.

The MLP was extensively benchmarked against unseen test data, and shown to reproduce an accurate description of water across a wide range of temperatures, demonstrating its transferability and accuracy for the thermodynamic landscape of bulk liquid water. It was also shown to be able to predict the bending rigidity of graphene, which is critical for studying graphene rippling dynamics.

Further details of this MLP development and benchmarking are provided in the Supporting Information.

\subsubsection*{Molecular dynamics simulations}

All MD simulations reported herein with the MLP were performed using the LAMMPS\cite{Thompson2022/10.1016/j.cpc.2021.108171} software patched with \texttt{symmetrix}\cite{symmetrix,Kovacs2025/10.1021/jacs.4c07099,Batatia2025/10.1063/5.0297006} --- an optimized C++ and Kokkos implementation that accelerates machine learning potentials for efficient large-scaled inference --- for fast CPU evaluation on the ARCHER2\cite{Beckett2024/10.5281/zenodo.14507040} cluster.
Unless explicitly stated otherwise, all simulations were performed in the NVT ensemble at a temperature of $\SI{300}{\kelvin}$.
A time step of $\SI{1}{\fs}$ was employed, and simulations utilized a Nos{\'e}--Hoover thermostat with damping time $\SI{100}{\fs}$.
The total simulation time, across all configurations, was $\SI{48.88}{\ns}$.

The simulations reported for the finite-size corrected water contact angle on free-standing graphene consist of 13 different spherical water droplets, of sizes ranging from 300 to 4,680 water molecules, placed on a free-standing graphene sheet of dimensions $\SI{154.0}{\angstrom}\times\SI{148.2}{\angstrom}$; the total number of atoms thus range from 9,540 to 22,680.
Each configuration was simulated for $\SI{1.30}{\ns}$, with the first $\SI{0.10}{\ns}$ used for equilibration and the remaining $\SI{1.20}{\ns}$ broken up into 5 statistically-independent blocks for the contact angle analysis and uncertainty estimate.

The simulations reported for the effects of mechanical strain on the contact angle consist of the droplet of 1,000 water molecules placed on the same graphene sheet, but with mechanical strain applied biaxially by horizontally scaling the $x$ and $y$ directions of the fixed periodic boundaries.
Each configuration thus contains 11,640 atoms.
Each configuration was simulated for $\SI{1.06}{\ns}$, with the first $\SI{0.10}{\ns}$ used for equilibration and the remaining $\SI{0.96}{\ns}$ broken up into 4 statistically-independent blocks for the contact angle analysis and uncertainty estimate.

Further details of the MD simulations are provided in the Supporting Information.

\subsubsection*{Obtaining the contact angle on non-flat surfaces}

In previous works, the contact angle is typically defined atomistically using the angle of intersection between the Gibbs equimolar dividing surface of the liquid-vapour interface, and the solid surface \cite{Carlson2024/10.1021/acs.jpclett.4c01143}.
Various methods exist for calculating this Gibbs dividing surface, with the most common approaches relying on slicing and binning molecular positions relative to the (fixed, flat) solid surface, and then fitting sigmoidal distributions to the binned densities, before extracting the angle of intersection from a best-fit sphere of the Gibbs dividing surface\cite{Werder2003/10.1021/jp0268112}.

These approaches, however, require a planar solid surface and thus cannot be applied to free-standing graphene.
Instead, we introduce a novel method inspired by the Willard--Chandler instantaneous interface\cite{Willard2010/10.1021/jp909219k} to compute an effective dividing surface in a symmetry-agnostic manner.

We consider only the oxygen atoms to describe the positions of the liquid water molecules.
In this section, we work in droplet-centred coordinates, i.e.~the coordinate system is translated such that, at all times $t$, the centre-of-mass of the droplet is located along the $x=y=0$ axis.

Given the positions $\mathbf{R}_{i}(t)$ of the $i$\textsuperscript{th} oxygen atom at time $t$, the coarse-grained density field at a space-time point $\mathbf{r}$, $t$ is defined as
\begin{equation}
    \bar{\rho}(\textbf{r},t) \,=\, \sum_{i=1}^{N_{\mathrm{oxy}}} \left(2\pi\xi^2\right)^{\nicefrac{-3}{2}} \exp\!\left[\,-\,\frac{\left|\mathbf{r}-\mathbf{R}_{i}(t)\right|^2}{2\xi^2}\right],
    \label{eq:inst_density}
\end{equation}
\noindent where $\xi$ is the coarse-graining length.
The Willard--Chandler instantaneous interface is then defined as the 2-dimensional isosurface $\mathbf{r}=\mathbf{s}$ where the coarse-grained density field reaches a cut-off value $\bar{\rho}(\mathbf{s},t)=c$; typically $c$ is chosen to be half of the bulk liquid density. 
The values of the parameters are $\xi=\SI{2.4}{\angstrom}$ and $c=\SI{0.016}{\angstrom^{-3}}$ for water \cite{Willard2010/10.1021/jp909219k}.

Here we introduce the notion of the \textit{time-averaged interface} to similarly be the isosurface of the time average of the density field, that is:
\begin{equation}
    \frac{1}{t_2-t_1}\,\int_{t_1}^{t_2}\bar{\rho}(\textbf{s},t)\dd{t} \;=\; c  \mbox{ . }
    \label{eq:t_ave_interface}
\end{equation}

For sufficiently long times, the time-averaged interface converges almost to the Gibbs dividing surface except ``rounded'' by lengthscale $\xi$.
To calculate the interface, instead of interpolating eq.~(\ref{eq:inst_density}) over a spatial grid as conventionally done for the Willard--Chandler method \cite{Willard2010/10.1021/jp909219k}, we use a ray-tracing binary search algorithm (detailed in the Supporting Information) to sample points on the interface in logarithmic runtime.
This approach is therefore advantageous over the Gibbs dividing surface, in that no initial assumptions are made on the geometry of the droplet other than being simply connected.

Next, we define the \textit{instantaneous graphene heightmap} to be a piecewise cubic function $h_{\mathrm{GS}}(x,y;t)$ satisfying the following conditions:

\begin{itemize}
    \item $z_i(t)=h_{\mathrm{GS}}(x_i(t),y_i(t);t)$ for all carbon atoms at position $\mathbf{R}_i(t)=(x_i(t),y_i(t),z_i(t))$ at time $t$;
    \item $h_{\mathrm{GS}}(x,y;t)$ is $C^{1}$ smooth over all $x$ and $y$ for all $t$;
    \item Amongst all functions which satisfy the above two, $h_{\mathrm{GS}}$ is the one which minimizes total curvature $\iint\left|\nabla_{x,y}^{2}h_{\mathrm{GS}}\right|^2 \dd{x}\dd{y}$;
\end{itemize}
This is achieved implicitly using a Clough--Tocher interpolation scheme \cite{Alfeld1984/10.1016/0167-8396(84)90029-3}.
The time average of this instantaneous heightmap is then taken, yielding $\langle h_{\mathrm{GS}}(x,y)\rangle_t$ as the \textit{time-averaged graphene heightmap} or TAGH for short.

With these definitions, the contact angle is obtained by scanning for the time-averaged interface eq.~(\ref{eq:t_ave_interface}) along randomly-selected rays which do not intercept the TAGH, fitting a least-squares best-fit sphere to these sampled points, and calculating the intersection angle between this best-fit sphere and the TAGH.
This procedure can either be performed isotropically, in which case the TAGH is averaged over azimuthal rotations to produce a radially-symmetric function $\langle h_{\mathrm{GS}}(r)\rangle_t$ with which the intersection is computed; or anisotropically, to produce a range of contact angles along the entire span of the intersection line.

\subsubsection*{Graphene sheet long-time inclination correlation}

The \textit{graphene sheet long-time inclination correlation}, first introduced by Thiemann et al.\cite{Thiemann2025/10.1073/pnas.2416932122}, is a spatially-localized measure of the degree to which graphene surface wave fluctuations are constrained.
Specifically, it is the long time limit of the temporal autocorrelation function of the local inclination angle $\theta_{\mathrm{GS}}(x,y;t)$, i.e.~the angle between the normal vector of a \textit{regularized} instantaneous graphene heightmap and the $z$-axis
\begin{equation}
    \theta_{\mathrm{GS}}(x,y;t) \;=\; \arctan\!\left(\left|\nabla_{x,y}\bar{h}_{\mathrm{GS}}(x,y;t)\right|\right),
    \label{eq:local_incl_angle}
\end{equation}
where the regularized heightmap $\bar{h}_{\mathrm{GS}}=h_{\mathrm{GS}}\ast\Pi_{\sigma}$ is the convolution of the instantaneous graphene heightmap with a uniform disk function $\Pi_{\sigma}$ of radius $\sigma$, which has the effect of smoothing out sharp gradients.
We choose $\sigma=\SI{4.5}{\angstrom}$ to match the procedure employed by Thiemann et al.\cite{Thiemann2025/10.1073/pnas.2416932122}.

The normalized temporal autocorrelation is thus
\begin{equation}
    \mathcal{C}_{\mathrm{GS}}(x,y;\tau) \,=\, \frac{\langle\theta_{\mathrm{GS}}(x,y;t+\tau)\theta_{\mathrm{GS}}(x,y;t)\rangle_{t}}{\langle\left(\theta_{\mathrm{GS}}(x,y;t)\right)^2\rangle_{t}}
    \label{eq:long_time_incl_corr}
\end{equation}
and the long-time inclination correlation is the limit $\tau\to\infty$ as obtained by fitting exponential decay curves to the temporal autocorrelation.

\section*{Acknowledgments}

D.W.L.\ acknowledges financial support from Trinity College, Cambridge.
X.R.A.\ and A.M.\ acknowledge support from the European Union under the ``n-AQUA'' European Research Council project (grant no.~101071937).
W.C.W.\ acknowledges support from the EPSRC (grant EP/V062654/1).
C.S.\ acknowledges financial support from the Royal Society (grant no.~RGS/R2/242614).
This work used the ARCHER2 UK National Supercomputing Service via a mixture of the UK Car--Parrinello Consortium (funded by EPSRC grant reference EP/X035891/1), as well as through the APP59749: ML4HetCat project.

This work was inspired by prior research from the ICE Group, and benefited from stimulating discussions within n-AQUA.

\section*{Competing interests}
The authors declare no competing interests.

\section*{Data availability}
All data required to reproduce the findings of this work will be made openly available on GitHub upon acceptance of this manuscript.

\section*{Author contribution statement}
D.W.L., A.M., F.L.T., and C.S.\ conceived of the presented ideas.
D.W.L.\ carried out all simulations, created the contact angle methodology, and performed the data analysis.
X.R.A.\ performed the reference DFT calculations for the MLP training data, and also developed and benchmarked the MLP.
X.R.A.\ also assisted D.W.L.\ with the DFT calculations for single molecule interaction energies (in Supporting Information).
W.C.W.\ created the \texttt{symmetrix} implementation.
A.M.\ and F.L.T.\ contributed to the interpretation of the results.
C.S.\ supervised the project.
D.W.L.\ took the lead in writing the manuscript.
All authors provided critical feedback, and helped shape the research, analysis, and manuscript.

\onecolumn

\input{main.bbl}

\end{document}


\maketitle

\clearpage

\tableofcontents
\clearpage

\section*{Details of molecular dynamics simulations}
\label{sec:SI_MD_details}

Except where otherwise noted, all MD simulations performed in this work were based on the MLP, and performed using the LAMMPS\cite{Thompson2022/10.1016/j.cpc.2021.108171} software patched with the \texttt{symmetrix}\cite{symmetrix} library for fast, scalable, and highly parallelisable CPU evaluation of the MACE MLP. This allows us to perform MLP-based MD at a very large scale, thus accessing the approaching \textit{ab initio} accuracy of the MLP's potential energy surface without needing to expend unreasonable computational effort.

All simulations used orthorhombic cells with periodic boundary conditions in all three dimensions. A timestep of $\SI{1}{\fs}$ was employed throughout. Most simulations were performed in the NVT ensemble at a temperature of $\SI{300}{\kelvin}$, which was maintained by a Nos{\'e}--Hoover thermostat with damping time $\SI{100}{\fs}$ as implemented in LAMMPS; one simulation was performed in the NpT ensemble at a pressure of $\SI{1}{\atm}$, which was maintained by a Nos{\'e}--Hoover barostat with damping time $\SI{1,000}{\fs}$.

\subsection*{Droplets of varying sizes, no strain applied}

The graphene sheets were constructed by repeating the unit cell dimensions $a=3d_c$ and $b=\sqrt{3}d_c$ along the $x$ and $y$ directions respectively, where $d_c=\SI{1.426}{\angstrom}$ is the carbon-carbon bond length obtained by energy minimization of the MLP at $\SI{0}{\kelvin}$ (rather than the experimentally measured value of $d_c=\SI{1.42}{\angstrom}$\cite{Castro_Neto2009/10.1103/RevModPhys.81.109}). As such, the $\SI{154.0}{\angstrom}\times\SI{148.2}{\angstrom}$ graphene sheets were obtained by repeating the unit cell 36 times in the $x$ direction, and 60 times in the $y$ direction. This yields a graphene sheet containing 8,640 carbon atoms. Sufficient vacuum space was added in the $z$ direction to ensure that periodic images across the $z$ direction do not interact.

The spherical water droplets of varying sizes were simulated in a sequential approach, where the equilibration of each droplet was used as the starting point for the next smaller droplet. The first system to be created was the 4,680 water molecules droplet, which was set up by generating a $\SI{77.0}{\angstrom}\times\SI{74.1}{\angstrom}\times\SI{24.6}{\angstrom}$ box of liquid water on top of the graphene sheet, and then equilibrating for $\SI{200}{\ps}$. This equilibrated droplet was thus the starting point for the production run of $\SI{1.20}{\ns}$ for the measurement of the contact angle. At the same time, the 4,300 water molecules droplet was generated by deleting 380 water molecules at random from the 4,680 water molecules droplet, and then further equilibrated for $\SI{100}{\ps}$ before starting the $\SI{1.20}{\ns}$ production run.

Smaller droplets were similarly generated by deleting the relevant number of water molecules from the equilibrated configuration of the next bigger droplet, and then equilibrating for $\SI{100}{\ps}$ more. A total of 13 different droplet sizes were investigated. All of the simulations described above were performed using the MLP with fully flexible, dynamical free-standing graphene. This set of simulations are relevant to obtaining the finite-size corrected contact angle for water on pristine, free-standing graphene.

Further to this, in order to investigate if the contact angle is significantly affected on free-standing graphene as opposed to spatially fixed, flat graphene, three more droplets (of sizes 4,680; 2,000; and 1,000 water molecules) were prepared on spatially fixed graphene sheets. These droplets were prepared by taking the corresponding starting state of the free-standing graphene simulation, resetting the positions of the carbon atoms back to the $z=0$ plane, and translating the water molecules along the $z$-axis so as to put the closest water molecule $\SI{1}{\angstrom}$ away from the graphene sheet. The droplets were then equilibrated for $\SI{100}{\ps}$ before starting the $\SI{1.20}{\ns}$ production run, with the graphene held fixed always. This set of simulations are relevant for comparing the contact angles between free-standing and spatially fixed graphene sheets, and also for comparing the contact angle methodology introduced in this work to an established literature method.

\begin{figure*}[p!]
    \centering{}
    \includegraphics[width=0.95\textwidth]{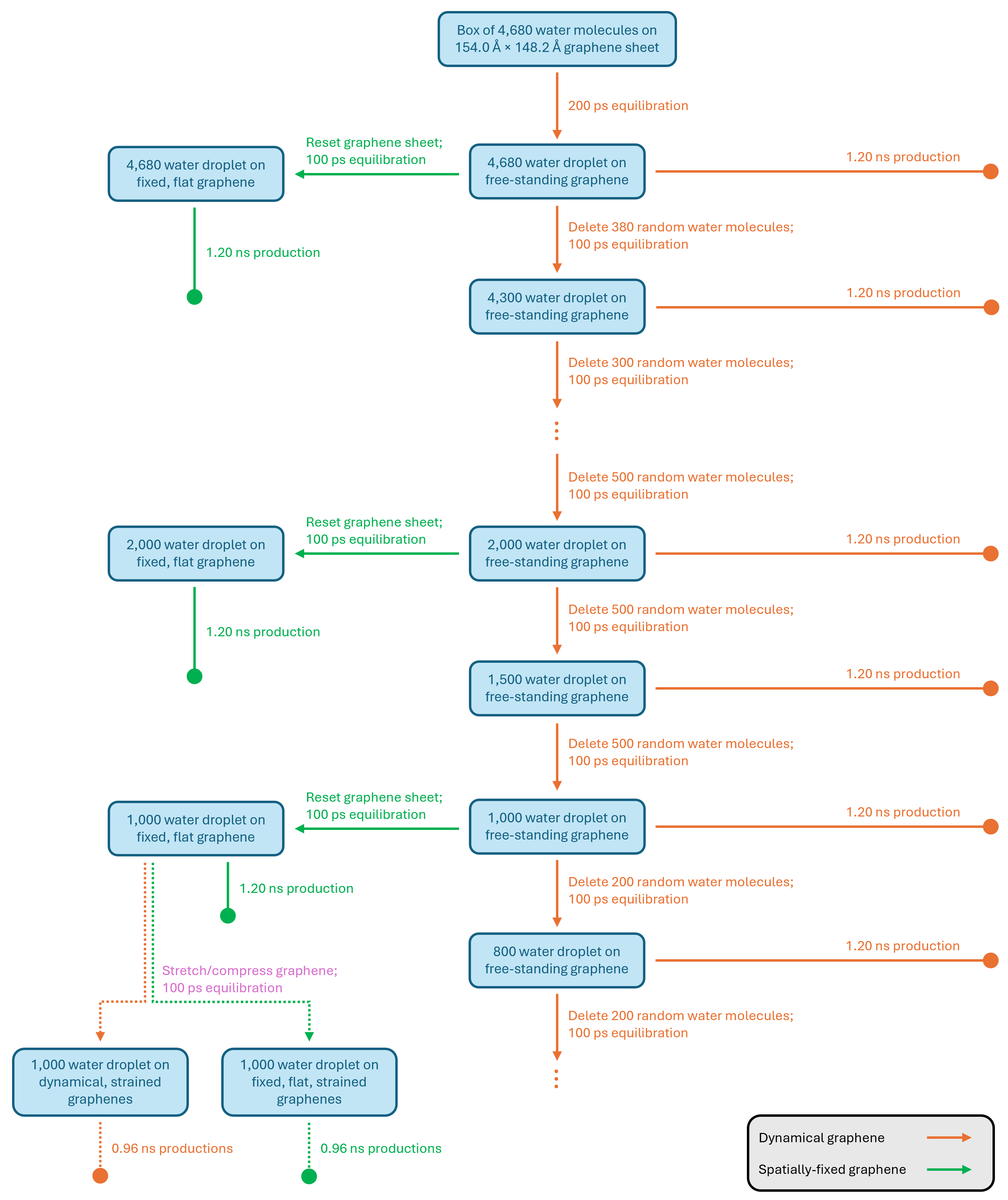}
    \caption{
    \textbf{Flowchart for the sequence of simulations performed in this work involving water droplets of varying size on strained and unstrained graphene sheets.} Arrows are colour-coded based on whether the graphene sheet was simulated dynamically or kept fixed in place.
    }
    \label{fig:SI-MD-workflow}
\end{figure*}

A flowchart summarizing this sequence of simulations is presented in \Cref{fig:SI-MD-workflow}; the configurations are listed in Table~\ref{table:MD-configurations}.

\subsection*{Droplets of 1,000 water molecules, varying strain applied}

To study how biaxial mechanical strain on the graphene sheet affects the wetting of graphene, we simulated an applied strain by re-scaling both the simulation periodic unit cell, and also the initial positions of the carbon atoms, in the $x$ and $y$ directions but not in the $z$ direction. The fixed periodic boundaries thus maintain a constant-strain condition.

These strained simulations were prepared by taking the equilibrated droplet of 1,000 water molecules on spatially fixed, flat, unstrained graphene (equilibrated using the MLP), performing the relevant re-scaling in the $x$ and $y$ directions, and then propagating the MD simulation with fully dynamical graphene. Values of tensile strains applied are: $\varepsilon = +0.20\%$, $+0.50\%$, $+1.00\%$, $+1.50\%$, $+2.00\%$. Values of compressive strains applied are: $\varepsilon = -0.10\%$, $-0.15\%$, $-0.20\%$, $-0.25\%$, $-0.30\%$, $-0.40\%$, $-0.50\%$, $-0.70\%$, $-1.00\%$, $-1.50\%$, $-2.00\%$. The strained systems were equilibrated for a further $\SI{100}{\ps}$, before starting the $\SI{0.96}{\ns}$ production run. This set of simulations are relevant for elucidating the coupling between graphene surface rippling and the three-phase contact line, and the effect of mechanical strain on the droplet contact angle.

An additional ``strained + spatially fixed'' simulation was also prepared, at the tensile strain of $\varepsilon=+2.0\%$, using the same procedure above but propagating the MD simulation with the graphene sheet still spatially fixed. This simulation is relevant for comparing the effect of strain on the droplet contact angle with or without the dynamical motion of the graphene sheet.

\subsection*{Graphene, with and without water}

To measure the long-time inclination correlation $\mathcal{C}_{\text{GS}}(\tau\to\infty)$ associated with pristine, free-standing graphene both unwetted and wetted uniformly, we also ran short simulations of (a) the graphene sheet on its own, and (b) covered uniformly on one side with a uniform water slab. In case (a), the graphene sheet was constructed the same way as in the water droplet simulations, while in case (b), a smaller $\SI{38.5}{\angstrom}\times\SI{37.1}{\angstrom}$ graphene sheet of 540 carbon atoms was used in order to save on computational cost. The graphene sheet was then covered uniformly by a box of 1,170 water molecules. In both cases, the system was equilibrated for $\SI{100}{\ps}$, before starting the $\SI{300}{\ps}$ production run.

For the dry free-standing graphene sheet, the mean value of the long-time inclination correlation $\mathcal{C}_{\text{GS}}(\tau\to\infty)$ across the sheet was $0.804\pm 0.005$. For the uniformly-wetted graphene sheet, the water slab was seen to have an average thickness of ${\sim}\SI{25}{\angstrom}$, and the mean value of $\mathcal{C}_{\text{GS}}(\tau\to\infty)$ was $0.791\pm 0.008$.

\subsection*{Bulk water, and water slabs}

To measure the properties of bulk liquid water as predicted by the MLP, we:

\begin{itemize}
    \item simulated a box of 64 water molecules with fully periodic boundaries, to find the radial distribution functions of bulk liquid water under NVT conditions;
    \item simulated a box of 520 water molecules with fully periodic boundaries, to find the radial distribution functions and equilibrium density of bulk liquid water under NpT conditions;
    \item simulated five slabs of liquid water, of varying sizes, to find the surface tension.
\end{itemize}

\clearpage
\setcellgapes{1em}
\makegapedcells
\begin{longtable}[c]{ c l c }
    \caption{
    \textbf{Detailed overview of the simulations presented in this work.} For each system, we report, where relevant, the total number of atoms $N_{\text{atoms}}$; the corresponding number of water molecules $N_{\text{H}_2\text{O}}$; the equilibration time $t_{\text{eq}}$; the simulation production time $t_{\text{sim}}$; and the pressure $p$. All systems used the same timestep $\delta t=\SI{1}{\fs}$ and same temperature $T=\SI{300}{\kelvin}$.\\}
    \label{table:MD-configurations}
    \\ \hline\hline
    \makecell[c]{\textbf{System}\\\textbf{(Dimensions)}} & \makecell[c]{\textbf{Simulation details}} & \makecell[c]{\textbf{Illustration of largest example}} \endfirsthead\hline

    \multicolumn{3}{c}{\tablename\ \thetable\ (continued)}\\
    \\ \hline\hline
    \makecell[c]{\textbf{System}\\\textbf{(Dimensions)}} & \makecell[c]{\textbf{Simulation details}} & \makecell[c]{\textbf{Illustration of largest example}} \endhead \hline

    \makecell{Droplets on free-\\standing graphene\\($\SI{154.0}{\angstrom}\times\SI{148.2}{\angstrom}$\\$\times\SI{151.1}{\angstrom}$)} & \makecell[l]{$\begin{aligned}N_{\text{atoms}}\in & \left\{ 22680; 21540; 20640; \right. \\ & \; 19140; 17640; 16140; \\ & \; 14640; 13140; 11640; \\ & \; 11040; 10440; 10140; \\ & \;\left. 9540 \right\} \end{aligned}$ \\ $\begin{aligned}N_{\text{H}_2\text{O}}\in & \left\{ 4680; 4300; 4000; \right. \\ & \; 3500; 3000; 2500; \\ & \; 2000; 1500; 1000; \\ & \;\left. 800; 600; 500; 300 \right\} \end{aligned}$ \\ $t_{\text{eq}}=\SI{100}{\ps}$ ($+\SI{100}{\ps}$ for \\ largest droplet) \\ $t_{\text{sim}}=\SI{1.20}{\ns}$} & \parbox[c]{0.35\textwidth}{\centering \includegraphics[width=0.35\textwidth]{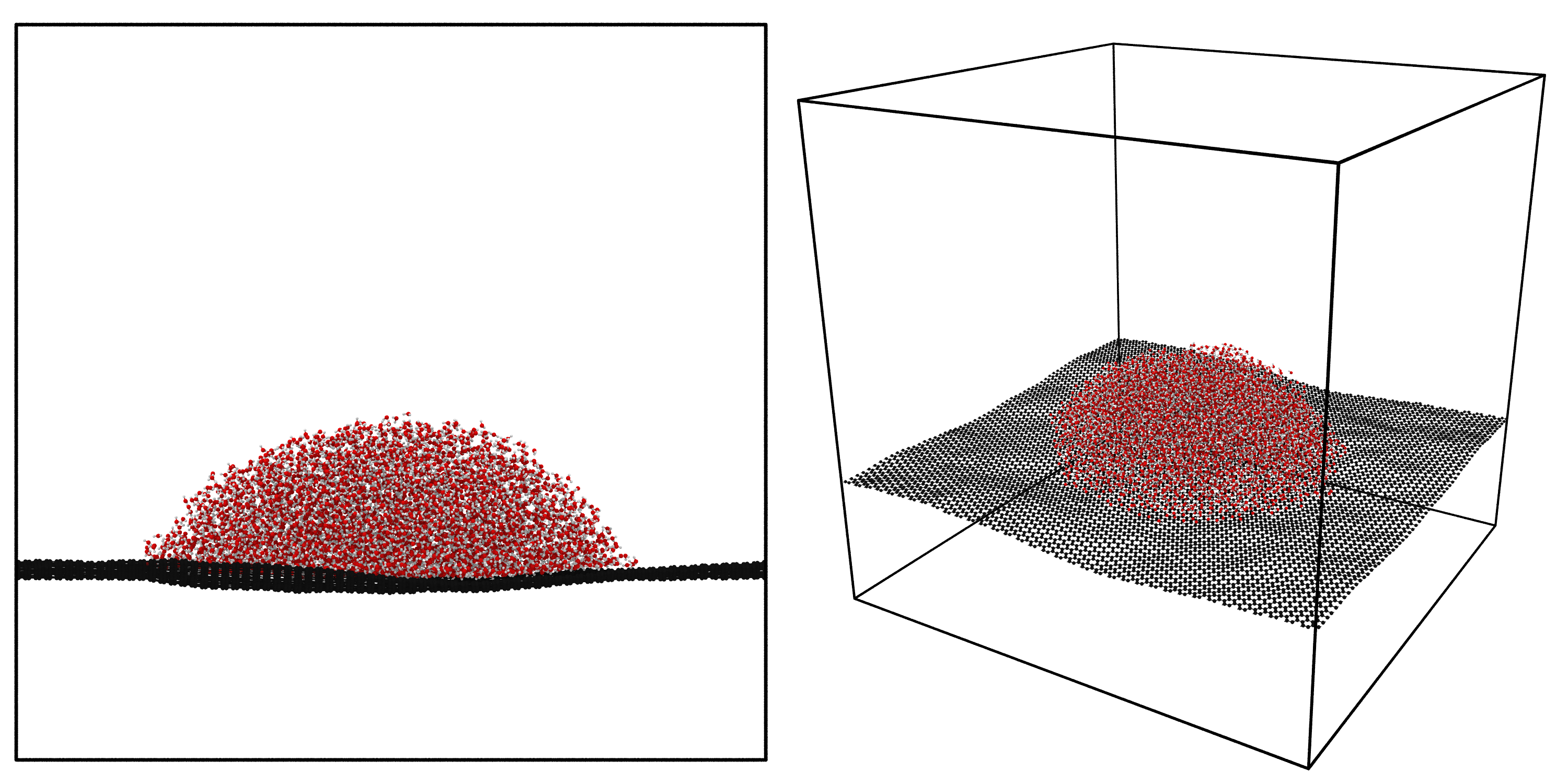}} \\ \hline
    \makecell{Droplets on fixed\\graphene\\($\SI{154.0}{\angstrom}\times\SI{148.2}{\angstrom}$\\$\times\SI{151.1}{\angstrom}$)} & \makecell[l]{$N_{\text{atoms}}\in\left\{ 22680; 14640; 11640 \right\}$ \\ $N_{\text{H}_2\text{O}}\in\left\{ 4680; 2000; 1000 \right\}$ \\ $t_{\text{eq}}=\SI{100}{\ps}$ \\ $t_{\text{sim}}=\SI{1.20}{\ns}$} & \parbox[c]{0.35\textwidth}{\centering \includegraphics[width=0.35\textwidth]{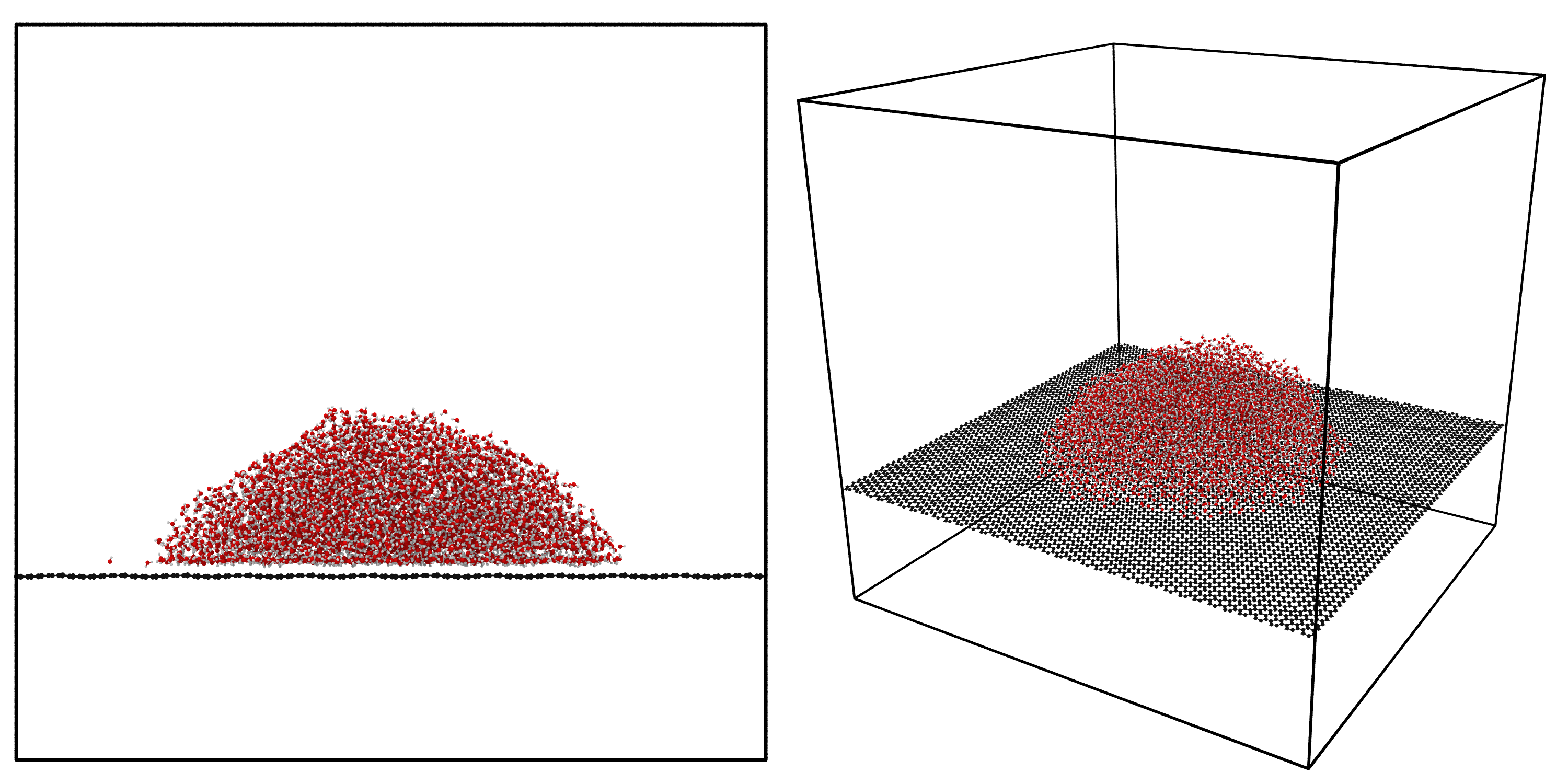}} \\ \hline
    \makecell{Droplets on tensile-\\strained graphene\\($\SI{154.3}{\angstrom}\times\SI{148.5}{\angstrom}$ to\\$\SI{157.1}{\angstrom}\times\SI{151.2}{\angstrom}$\\$\times\SI{151.1}{\angstrom}$)} & \makecell[l]{$N_{\text{atoms}}=11640$ \\ $N_{\text{H}_2\text{O}}=1000$ \\ $t_{\text{eq}}=\SI{100}{\ps}$ \\ $t_{\text{sim}}=\SI{960}{\ps}$} & \parbox[c]{0.35\textwidth}{\centering \includegraphics[width=0.35\textwidth]{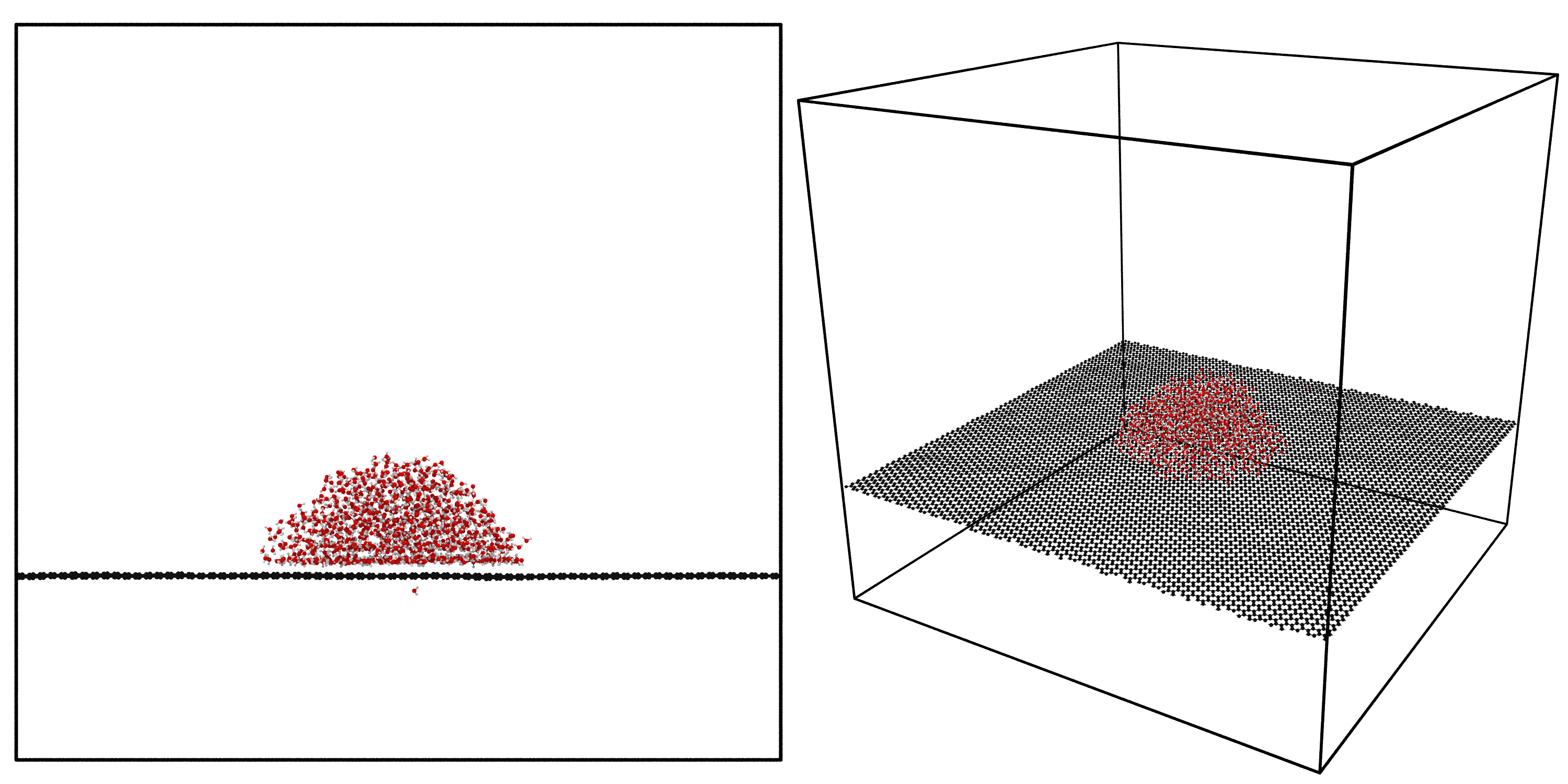}} \\ \hline
    \makecell{Droplets on fixed,\\tensile-strained\\graphene\\($\SI{157.1}{\angstrom}\times\SI{151.2}{\angstrom}$\\$\times\SI{151.1}{\angstrom}$)} & \makecell[l]{$N_{\text{atoms}}=11640$ \\ $N_{\text{H}_2\text{O}}=1000$ \\ $t_{\text{eq}}=\SI{100}{\ps}$ \\ $t_{\text{sim}}=\SI{960}{\ps}$} & \parbox[c]{0.35\textwidth}{\centering \includegraphics[width=0.35\textwidth]{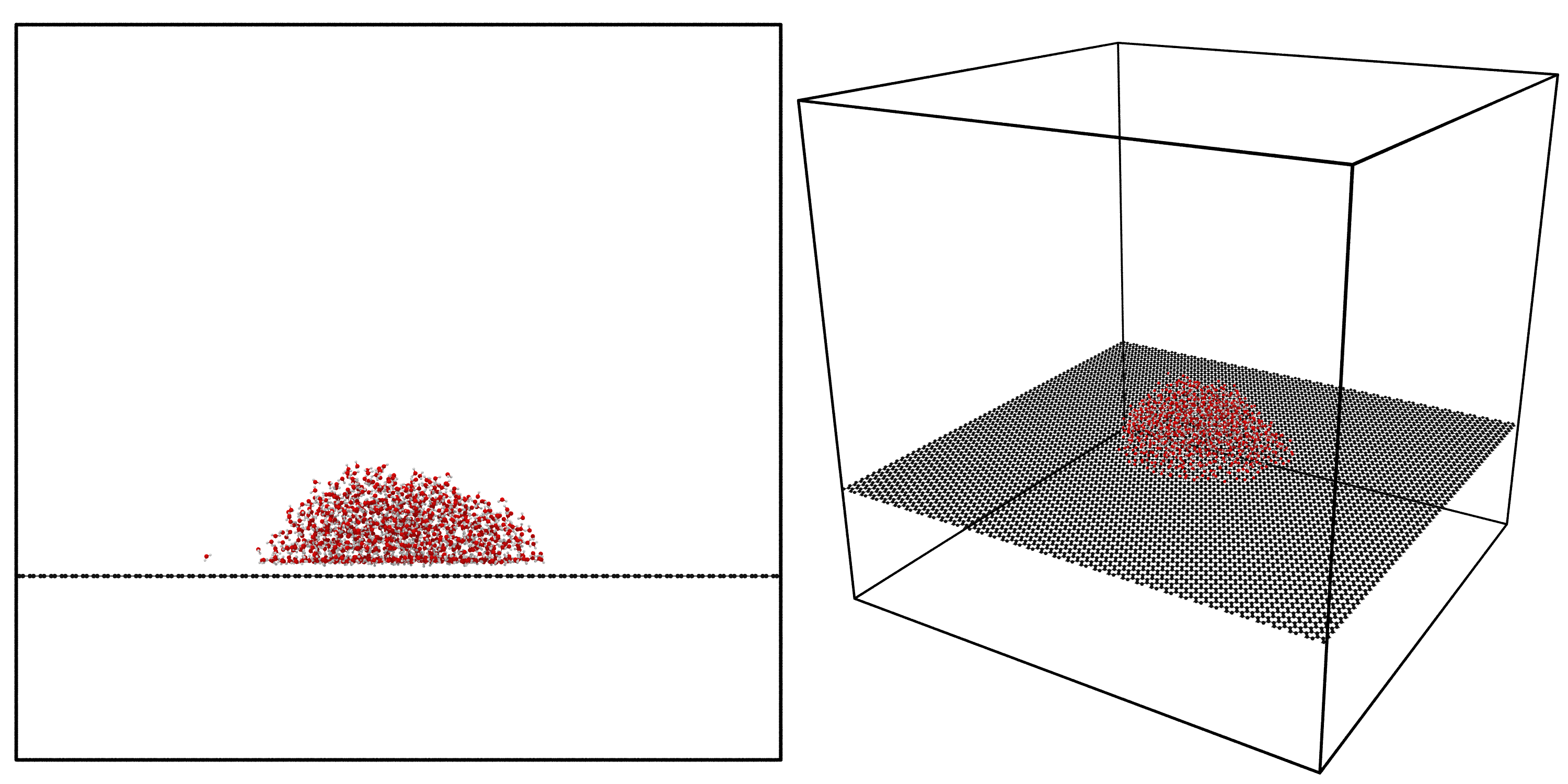}} \\ \hline
    \makecell{Droplets on\\compression-\\strained graphene\\($\SI{150.9}{\angstrom}\times\SI{145.2}{\angstrom}$ to\\$\SI{153.9}{\angstrom}\times\SI{148.1}{\angstrom}$\\$\times\SI{151.1}{\angstrom}$)} & \makecell[l]{$N_{\text{atoms}}=11640$ \\ $N_{\text{H}_2\text{O}}=1000$ \\ $t_{\text{eq}}=\SI{100}{\ps}$ \\ $t_{\text{sim}}=\SI{960}{\ps}$} & \parbox[c]{0.35\textwidth}{\centering \includegraphics[width=0.35\textwidth]{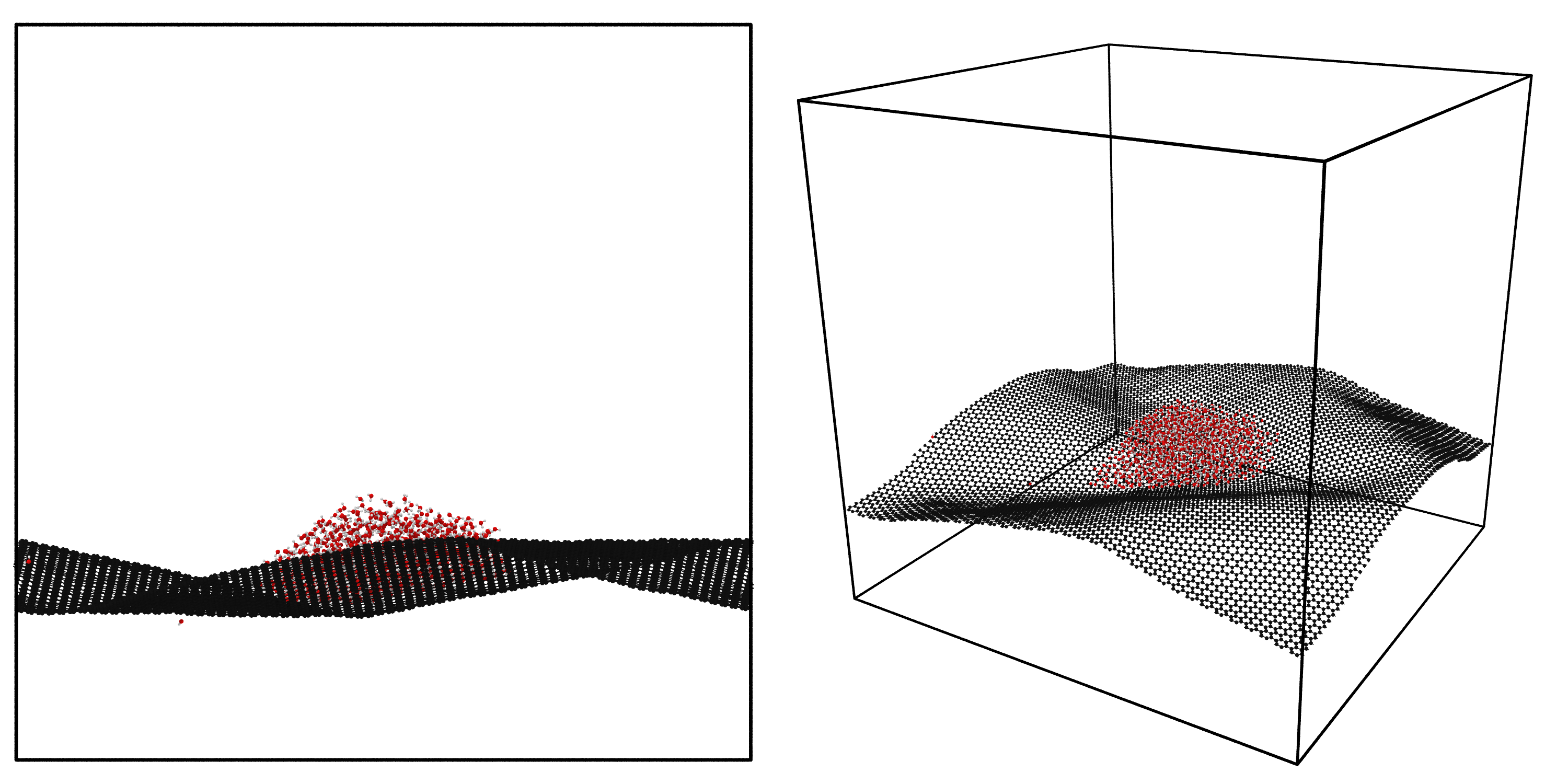}} \\ \hline
    \makecell{Free-standing\\graphene\\($\SI{154.0}{\angstrom}\times\SI{148.2}{\angstrom}$\\$\times\SI{151.1}{\angstrom}$)} & \makecell[l]{$N_{\text{atoms}}=8640$ \\ $t_{\text{eq}}=\SI{100}{\ps}$ \\ $t_{\text{sim}}=\SI{300}{\ps}$} & \parbox[c]{0.35\textwidth}{\centering \includegraphics[width=0.35\textwidth]{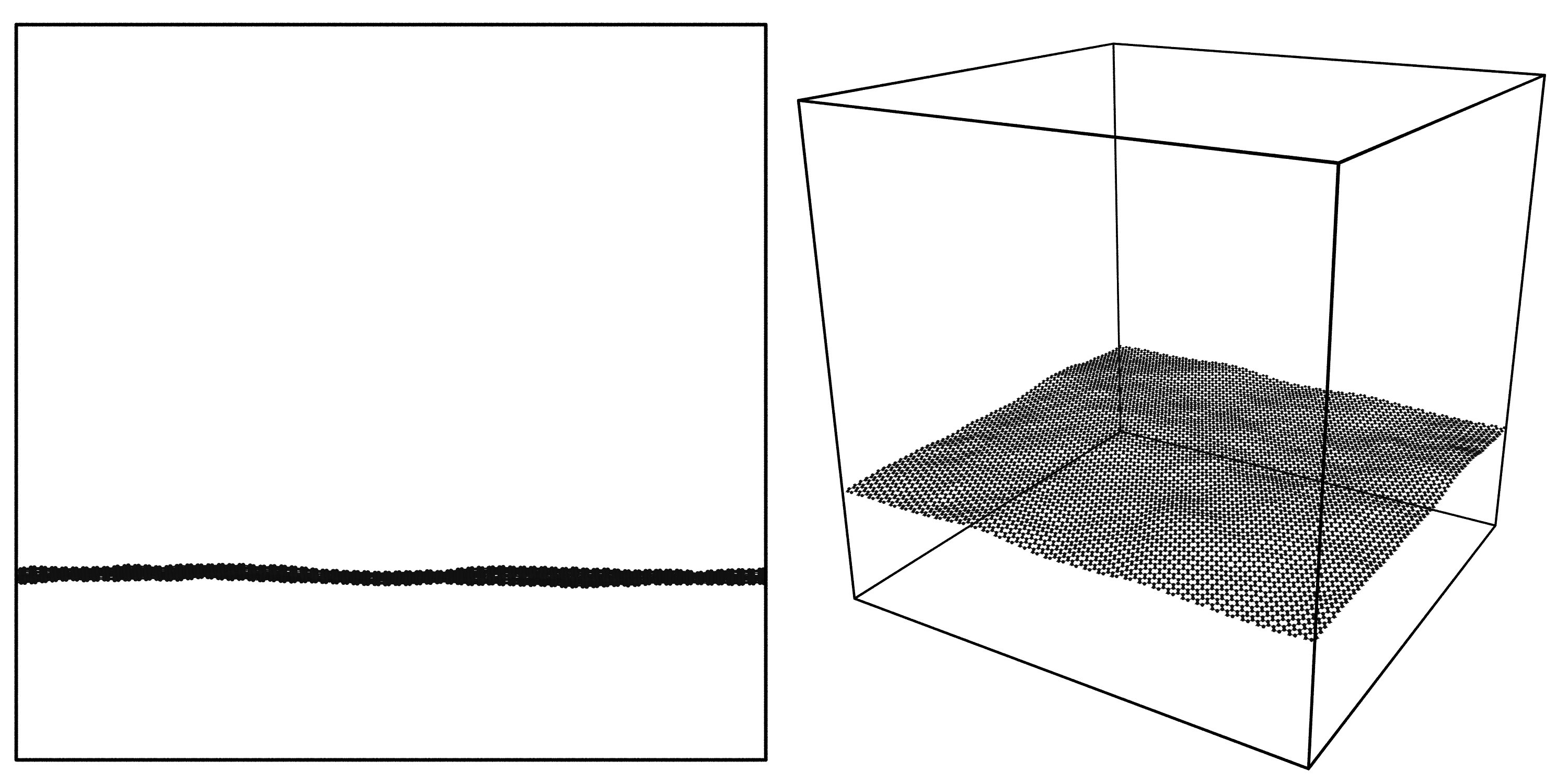}} \\ \hline
    \makecell{Graphene covered\\by water\\($\SI{38.5}{\angstrom}\times\SI{37.1}{\angstrom}$\\$\times\SI{100.0}{\angstrom}$)} & \makecell[l]{$N_{\text{atoms}}=4050$ \\  $N_{\text{H}_2\text{O}}=1170$ \\ $t_{\text{eq}}=\SI{100}{\ps}$ \\ $t_{\text{sim}}=\SI{300}{\ps}$} & \parbox[c]{0.35\textwidth}{\centering \includegraphics[width=0.35\textwidth]{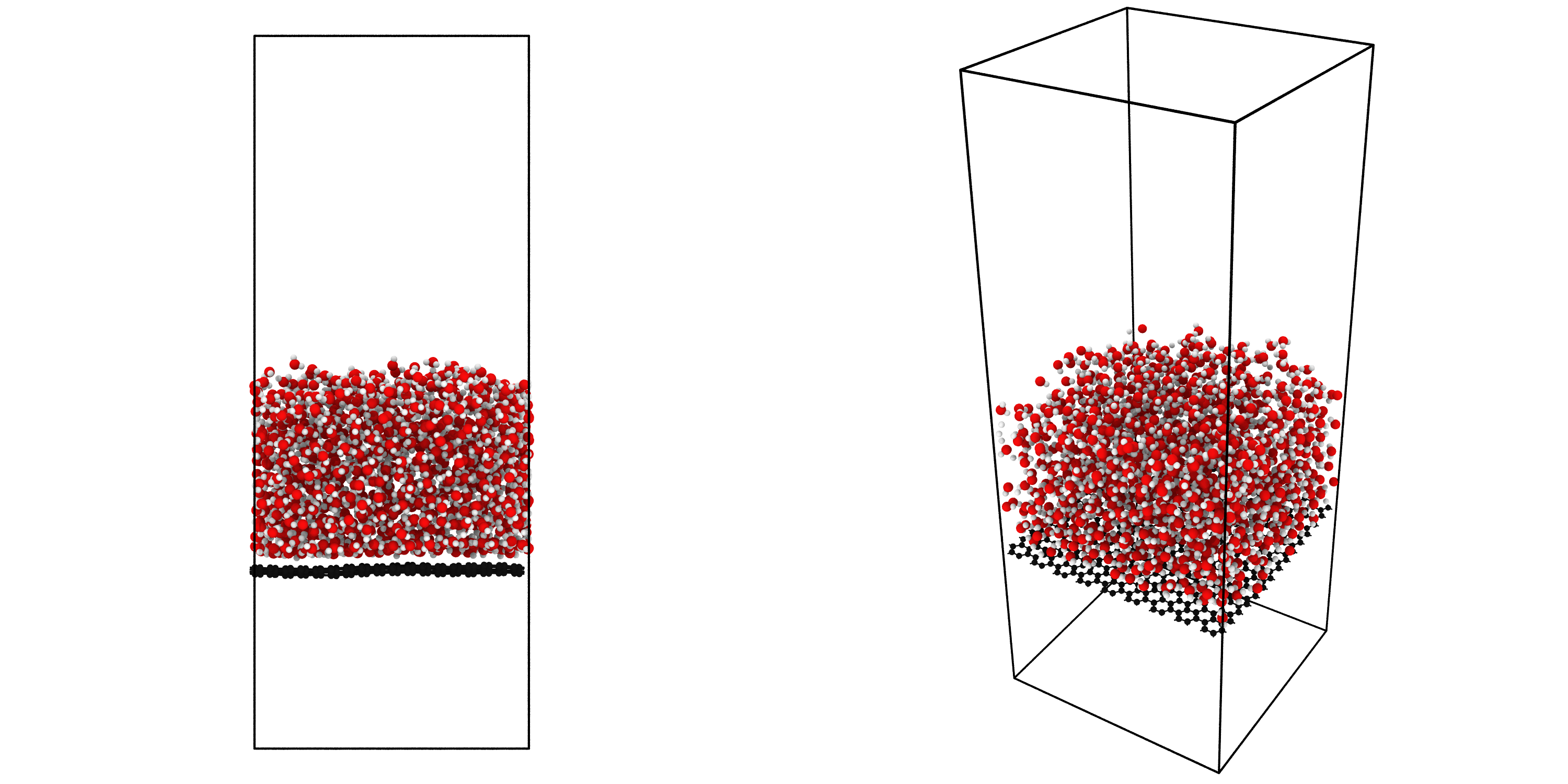}} \\ \hline
    \makecell{Bulk water, NVT\\($\SI{12.42}{\angstrom}$ cubic)} & \makecell[l]{$N_{\text{atoms}}=192$ \\ $N_{\text{H}_2\text{O}}=64$ \\ $t_{\text{eq}}=\SI{300}{\ps}$ \\ $t_{\text{sim}}=\SI{1}{\ns}$} & \parbox[c]{0.35\textwidth}{\centering \includegraphics[width=0.35\textwidth]{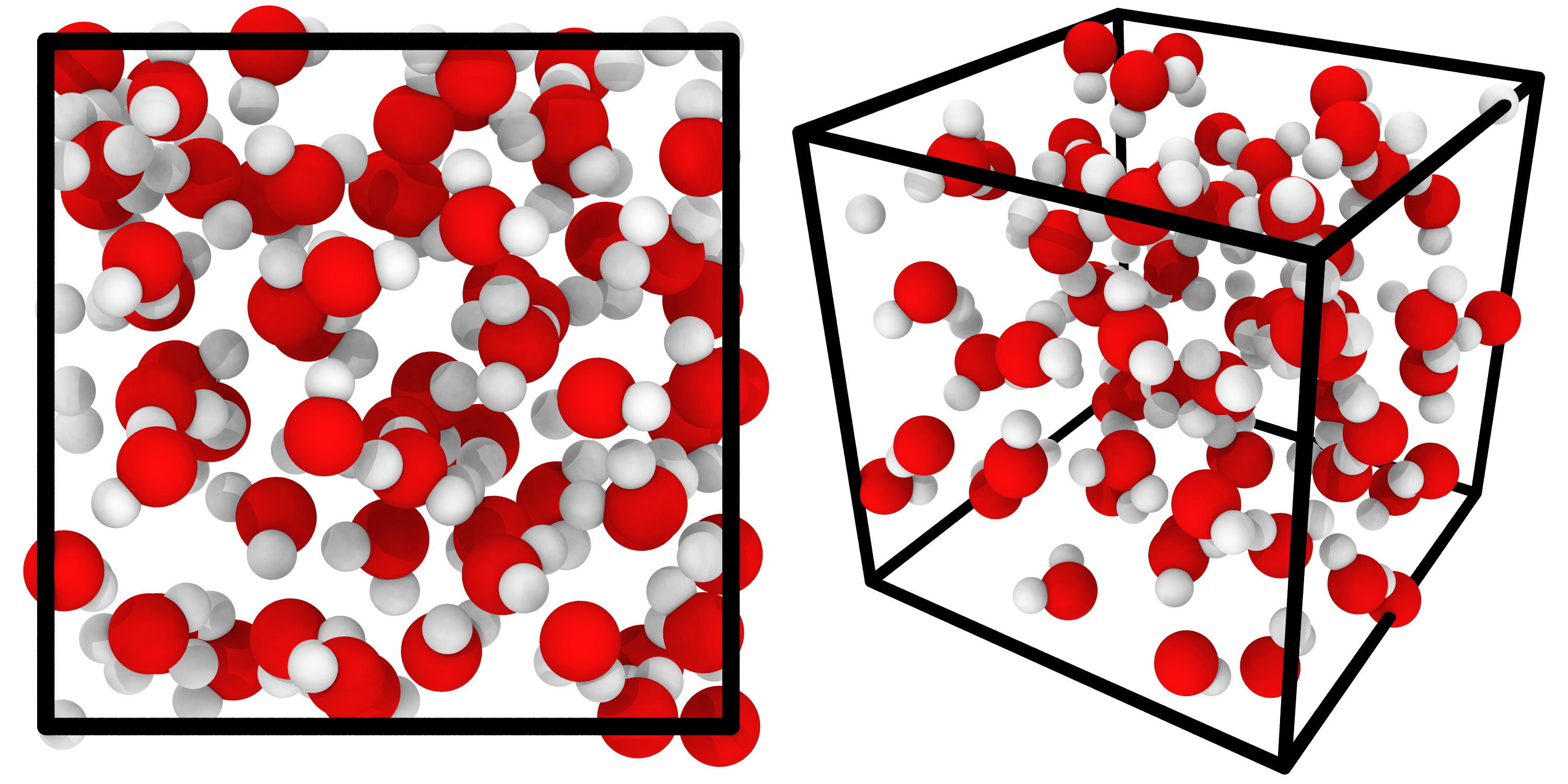}} \\ \hline
    \makecell{Bulk water, NpT\\(Average: $\SI{25.1}{\angstrom}$ cubic)} & \makecell[l]{$N_{\text{atoms}}=1560$ \\ $N_{\text{H}_2\text{O}}=520$ \\ $t_{\text{eq}}=\SI{300}{\ps}$ \\ $t_{\text{sim}}=\SI{1}{\ns}$ \\ $p=\SI{1}{\atm}$} & \parbox[c]{0.35\textwidth}{\centering \includegraphics[width=0.35\textwidth]{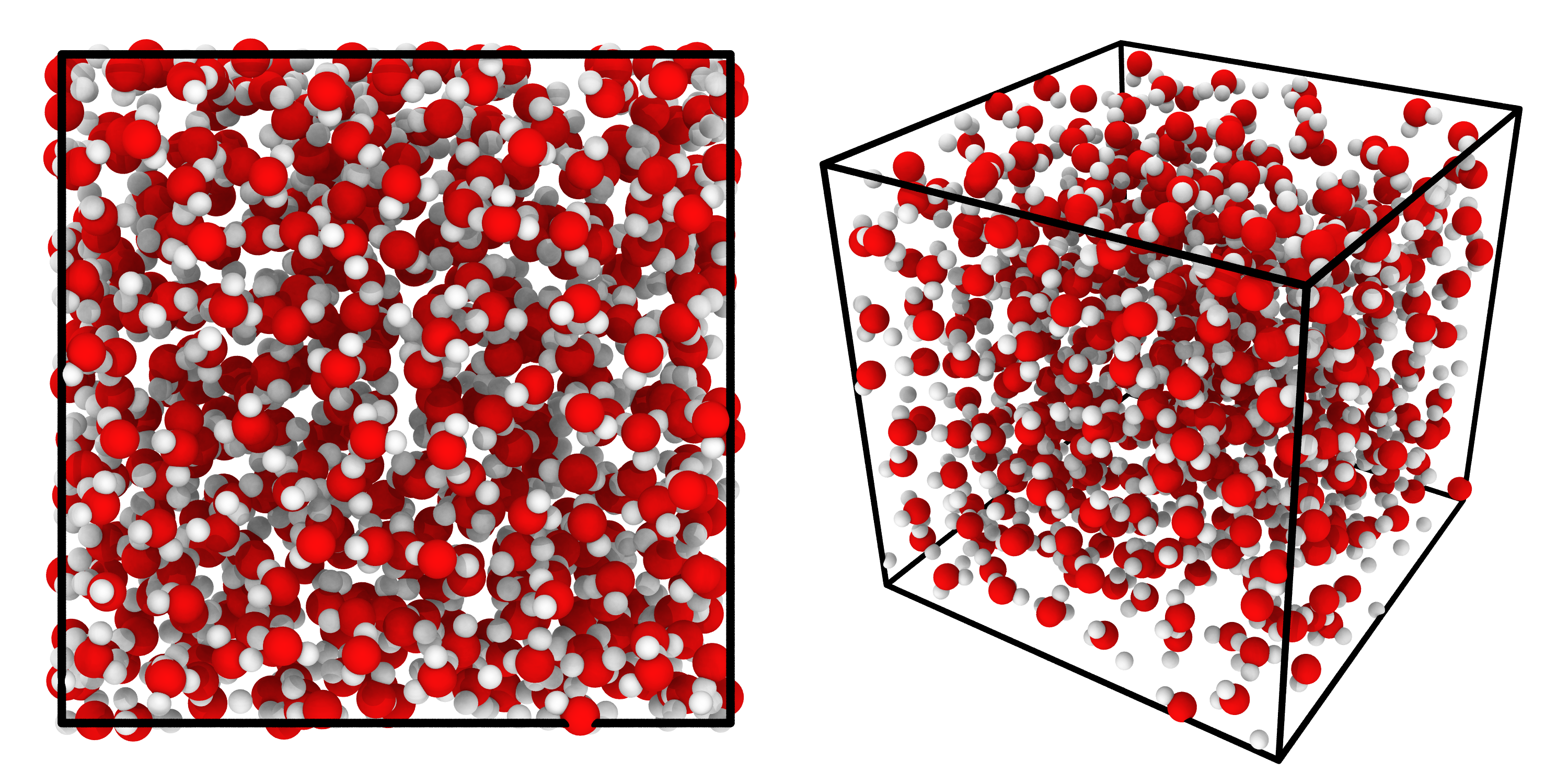}} \\ \hline
    \makecell{Water slab\\($\SI{25.2}{\angstrom}\times\SI{25.2}{\angstrom}$ to\\$\SI{75.5}{\angstrom}\times\SI{75.5}{\angstrom}$\\$\times\SI{75.0}{\angstrom}$)} & \makecell[l]{$\begin{aligned}N_{\text{atoms}}\in & \left\{14040; 9750; 6240; \right. \\ & \;\left. 3510; 1560 \right\}\end{aligned}$ \\ $\begin{aligned}N_{\text{H}_2\text{O}}\in & \left\{4680; 3250; 2080; \right. \\ & \;\left. 1170; 520 \right\}\end{aligned}$ \\ $t_{\text{eq}}=\SI{100}{\ps}$ \\ $t_{\text{sim}}=\SI{1}{\ns}$} & \parbox[c]{0.35\textwidth}{\centering \includegraphics[width=0.35\textwidth]{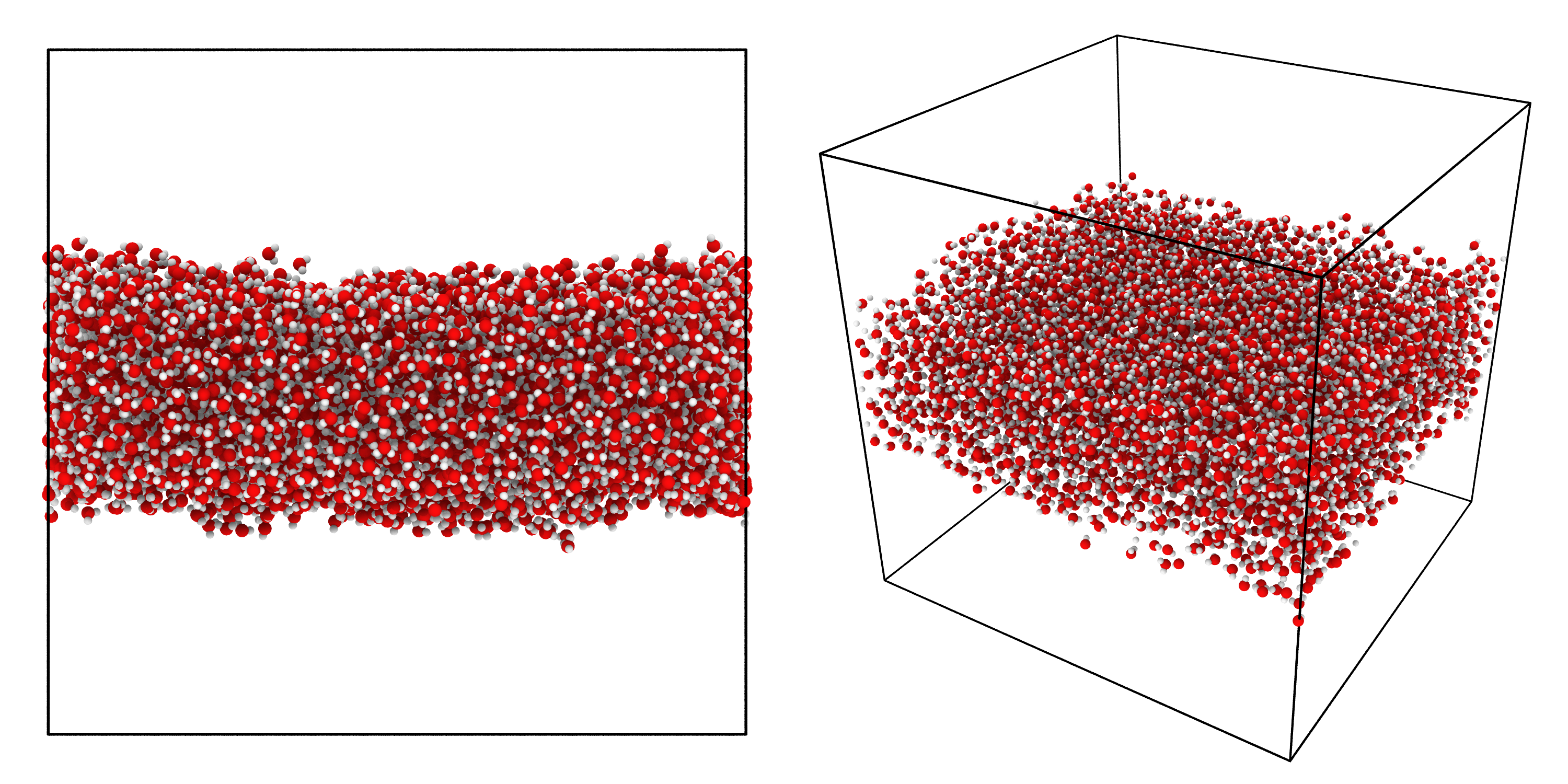}} \\ \hline\hline
\end{longtable}
\nomakegapedcells

\clearpage

\section*{Machine learning potential}

The MLP used in this work is the same as that of Ref.~\citenum{Advincula2026/10.1126/sciadv.aeb5772}, where it has already been extensively benchmarked. Here, we briefly recap some of the details regarding the development and validation of this model as reported in Ref.~\citenum{Advincula2026/10.1126/sciadv.aeb5772}, as well as further tests performed for this work. Beyond the validations presented in this section, we also performed additional benchmarks for the density, radial distribution function, and surface tension of bulk liquid water (see next section).

\subsection*{Model development}

The MLP was developed iteratively over several generations, beginning with training data from Refs.~\citenum{Advincula2025/10.1021/acsnano.5c02053} and \citenum{Thiemann2022/10.1021/acsnano.2c02784}, selected to provide broad coverage of the relevant physical regimes. These datasets provide (i) self-dissociated water configurations across a wide range of densities under graphene confinement, including the ultra-confined limit, and (ii) water-graphene interfaces spanning from flat sheets to highly curved environments within carbon nanotubes of varying radii. Together, they ensure that the model captures both the physics of graphene-water interactions and the bending rigidity of graphene, which is an essential feature for accurately modelling graphene surface rippling.

Additional structures, involving graphene layers stacked in both the AA and AB configurations at varying interlayer distances, were also introduced in order to refine the description of graphene-graphene interactions, although such interactions are not significant for this work. The training data was also extended to include graphene sheets of large dimensions, to ensure the model's applicability to the large scales demonstrated in this work.

In total, the training set comprised 5,845 structures, which were all labelled with DFT energies and forces computed by the CP2K/Quickstep code using the revPBE-D3 functional.

\subsection*{Model validation}

To evaluate the validity of the MLP for the systems studied in this work, we quantified the root-mean-square errors (RMSEs) in energies and forces predicted by the model. Specifically, 300 snapshots were randomly selected from $\SI{100}{\ps}$ MLP-based MD simulations of either bulk liquid water, or monolayer water confined between two $\SI{12.350}{\angstrom}\times\SI{12.834}{\angstrom}$ graphene sheets. For each configuration, we performed single-point DFT calculations at the same level of theory used to train the MLP (i.e.~revPBE-D3). This provides a direct and robust measure of the MLP's accuracy, as it compares predictions for structures outside of the training set sampled from its potential energy surface against the reference ab initio values.

In the case of bulk liquid water configurations, the RMSEs for energies were found to be $\SI{0.5}{\milli\electronvolt/\atom}$, and the RMSEs for forces were found to be $\SI{15.4}{\milli\electronvolt/\angstrom}$; decomposing these force RMSEs by atomic species gives $\SI{21.8}{\milli\electronvolt/\angstrom}$ for O atoms, and $\SI{11.0}{\milli\electronvolt/\angstrom}$ for H atoms. For the graphene nanoconfined water configurations, the RMSEs for energies were found to be $\SI{0.5}{\milli\electronvolt/\atom}$, and the RMSEs for forces were found to be $\SI{26.9}{\milli\electronvolt/\angstrom}$; decomposing these force RMSEs by atomic species gives $\SI{26.5}{\milli\electronvolt/\angstrom}$ for C atoms, $\SI{35.8}{\milli\electronvolt/\angstrom}$ for O atoms, and $\SI{21.8}{\milli\electronvolt/\angstrom}$ for H atoms.

\begin{figure*}[hbtp!]
    \centering{}
    \includegraphics[width=\textwidth]{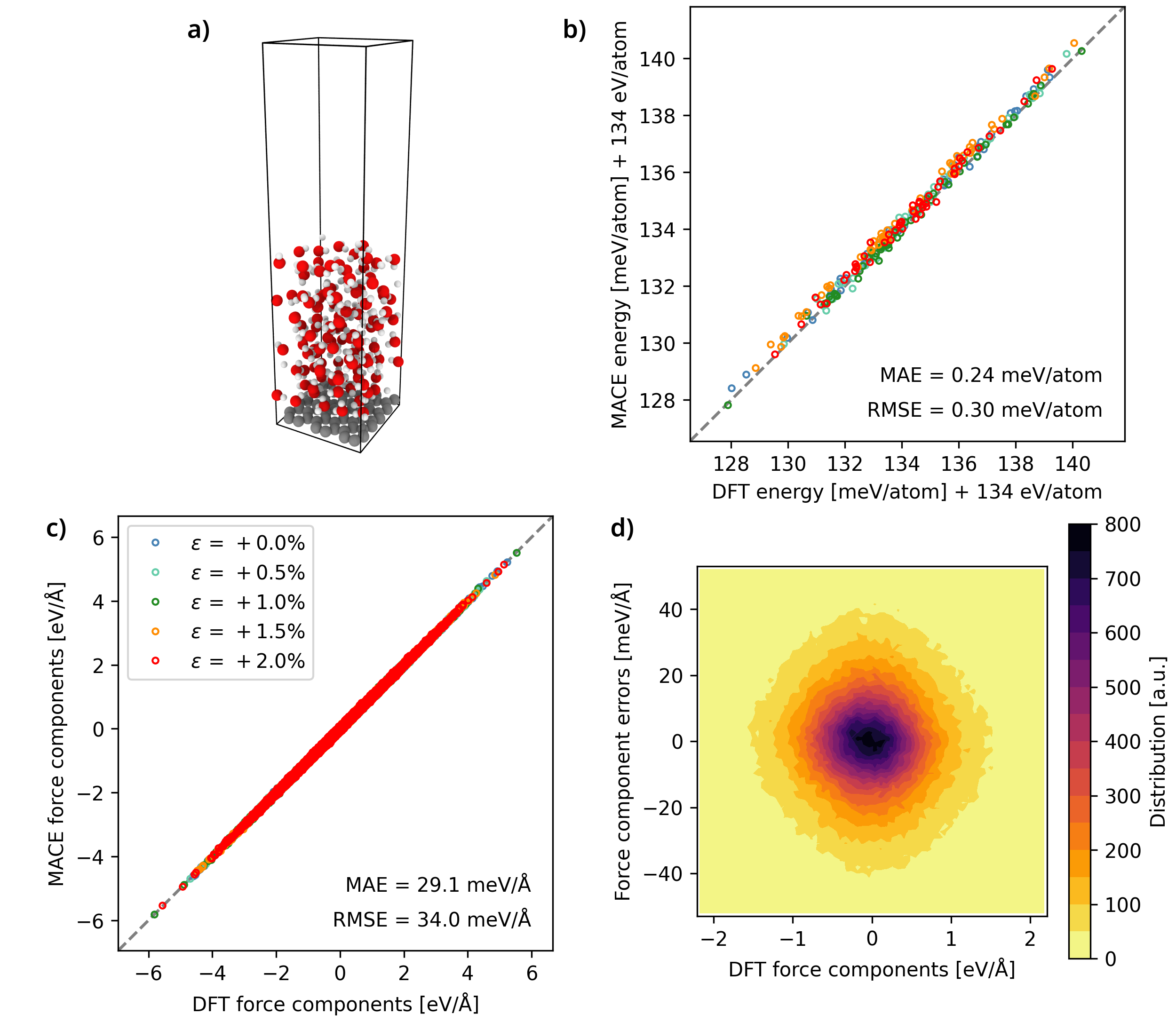}
    \caption{
    \textbf{MLP validation on fully-wetted graphene.} (a) Representative snapshot of one of the previously-unseen test configurations, which were generated from short MLP-based MD simulations, containing both water-graphene and water-vapour interfaces. 50 test configurations each were extracted from 5 strain conditions, ranging from free-standing to $\varepsilon = +2.0\%$. (b) Parity plot of MLP-predicted energies per atom ($y$-axis) against DFT-calculated reference energies per atom ($x$-axis); note that these reported energies have been shifted by a flat graphene baseline at the same strain condition, to condense the large spread of energies associated with graphene's elastic modulus. (c) Parity plot of MLP-predicted force components against DFT-calculated force components. (d) Joint distribution of MLP-predicted force component errors against DFT-calculated force components, which shows no significant correlations.
    }
    \label{fig:SI-MLP-validation-slab}
\end{figure*}

Additionally, new previously-unseen test configurations were generated from MD simulations of a liquid water slab on tensile-strained graphene, as shown in \Cref{fig:SI-MLP-validation-slab}a. These configurations therefore represent both the water-graphene and water-vapour interfaces. The RMSEs for energies were found to be $\SI{0.3}{\milli\electronvolt/\atom}$, and for forces to be $\SI{34.0}{\milli\electronvolt/\angstrom}$, with no significant correlations between force errors and DFT-calculated force magnitudes.

\begin{figure*}[hbtp!]
    \centering{}
    \includegraphics[width=0.45\textwidth]{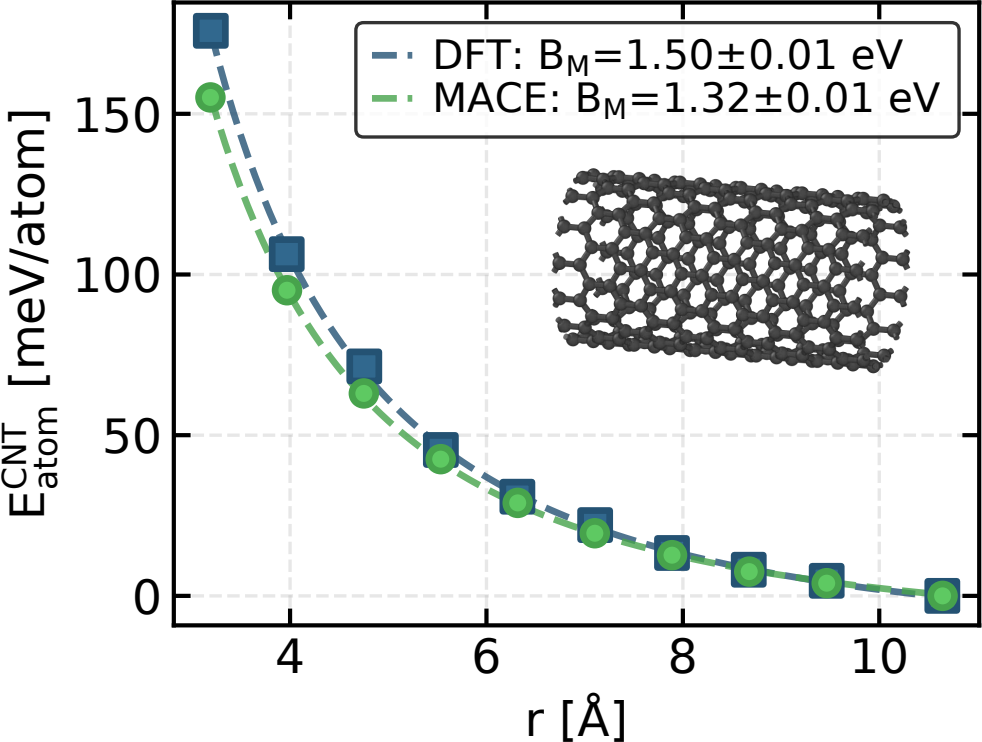}
    \caption{
    Energy per atom as a function of nanotube radius for SWCNTs rolled along zigzag directions. The dashed line indicates the fit used to extract the bending rigidity $B_M$ of graphene. A representative SWCNT structure is shown as an inset.
    }
    \label{fig:SI-MLP-validation-S5}
\end{figure*}

Because a central focus of this work is the surface rippling dynamics of free-standing and mechanically strained graphene sheets, it is essential that the model accurately captures the bending rigidity of graphene, which determines how the material deforms. For graphene, the bending rigidity $B_M$ can be obtained by fitting the energy per atom in single-wall carbon nanotubes (SWCNTs) of varying radii $r$, using the following expression:\cite{Wei2013/10.1021/nl303168w}

\begin{equation}
E_{\text{atom}}^{\text{CNT}} \;\;=\;\; E_0 \;+\; \frac{S_{0}B_{M}}{2r^2}
\end{equation}

\noindent where $E_{\text{atom}}^{\text{CNT}}$ is the energy per atom in a SWCNT, $E_0$ is the energy per atom in flat unstrained graphene, and $S_0 = \SI{2.63}{\angstrom^2}$ is the planar footprint of a carbon atom in graphene. By computing $E_{\text{atom}}^{\text{CNT}}$ for nanotubes with different radii, $B_M$ can be obtained from the curvature dependence. As shown in \Cref{fig:SI-MLP-validation-S5}, our MLP accurately reproduces this curvature dependence, in excellent agreement with the reference \textit{ab initio} values.

\begin{figure*}[hbtp!]
    \centering{}
    \includegraphics[width=\textwidth]{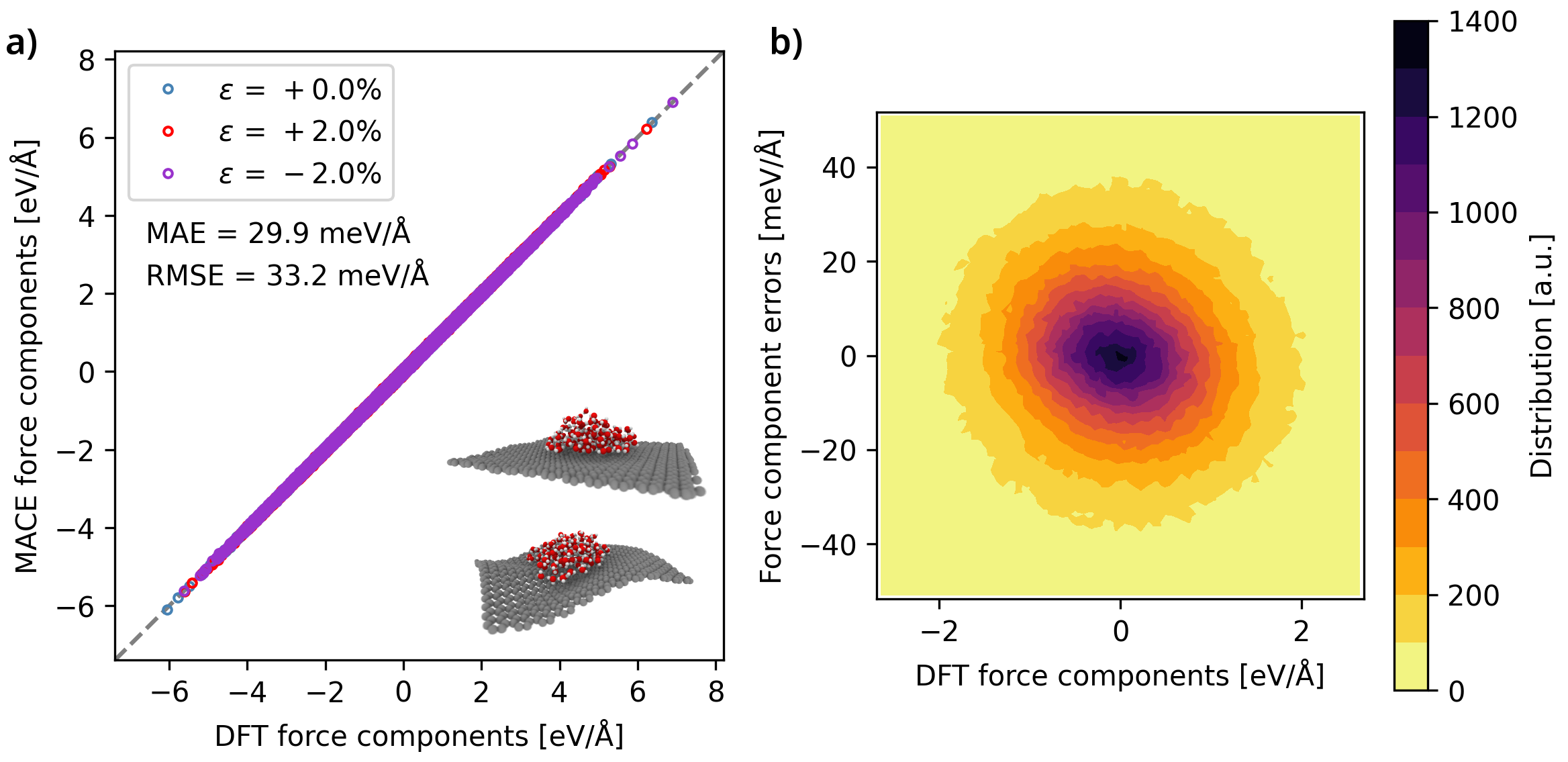}
    \caption{
    \textbf{MLP validation on miniaturized droplets.} (a) Parity plot of MLP-predicted force components ($y$-axis) against DFT-calculated reference force components ($x$-axis) for randomly-selected configurations of a small droplet on a graphene sheet under free-standing, tensile strained, and compressive strained conditions. Representative snapshots of the free-standing and compressive strained conditions are shown as insets. (b) Joint distribution of MLP-predicted force component errors against DFT-calculated force components.
    }
    \label{fig:SI-MLP-validation-droplet}
\end{figure*}

Furthermore, to verify that the MLP's description of the potential energy surface at the three-phase contact-line is accurate, additional test configurations were generated from MLP-based MD simulations of a miniaturized water droplet, containing 130 water molecules, placed on a small $\SI{51.336}{\angstrom}\times\SI{49.4}{\angstrom}$ graphene sheet under free-standing, $\varepsilon=+2\%$ tensile strain, and $\varepsilon=-2\%$ compressive strain conditions, as shown in \Cref{fig:SI-MLP-validation-droplet}. The RMSEs for energies were found to be $\SI{0.9}{\milli\electronvolt/\atom}$, and for forces to be $\SI{33.2}{\milli\electronvolt/\angstrom}$, with no significant correlations between force errors and magnitudes, indicating that the MLP still performs well despite no such droplet configurations being in the training dataset.

Overall, the validations presented in this section demonstrate that the MLP developed in this work reliably captures the physics of water-water and water-graphene interactions, and also the mechanics of graphene bending. With its energetic and force accuracy across a range of conditions, and uniform applicability to both water-water and water-carbon and carbon-carbon interactions, the MLP provides a robust and transferable framework for investigating the wetting of graphene with first-principles level accuracy.

\clearpage

\section*{Properties of liquid water predicted by the MLP model}

\subsection*{Density and radial distribution function}

\begin{figure*}[h!]
    \centering{}
    \includegraphics[width=0.9\textwidth]{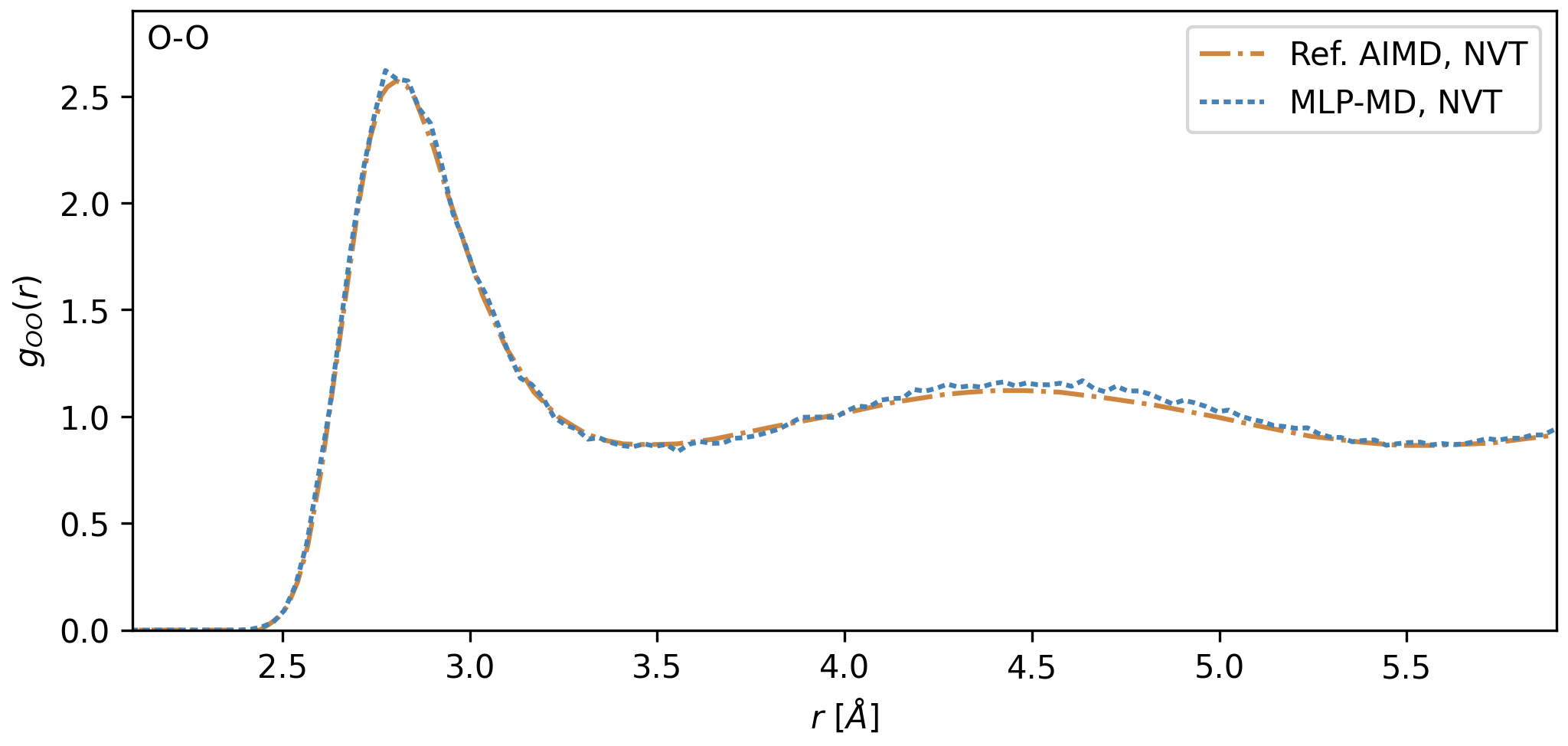}
    \caption{
    \textbf{The radial distribution function (RDF) for O-O interatomic distances in bulk liquid water} under NVT conditions, for a cubic $\SI{12.42}{\angstrom}$ box of 64 water molecules at $\SI{300}{\kelvin}$. Dash-dotted gold line: AIMD reference for $g_{\text{OO}}(r)$ using revPBE-D3, from Marsalek and Markland\cite{Marsalek2017/10.1021/acs.jpclett.7b00391}. Dotted blue line: our MLP model.
    }
    \label{fig:SI-rdf-nvt}
\end{figure*}

The O-O radial distribution function (RDF) for bulk liquid water under NVT conditions, as obtained by simulating a cubic $\SI{12.42}{\angstrom}$ box of 64 water molecules at $\SI{300}{\kelvin}$ for $\SI{1}{\ns}$ with fully periodic boundary conditions using our MLP model, is shown in \Cref{fig:SI-rdf-nvt}. We compare this directly against an \textit{ab initio} molecular dynamics (AIMD) simulation using the revPBE-D3 functional of the same configuration by Marsalek and Markland\cite{Marsalek2017/10.1021/acs.jpclett.7b00391}, and find that the structure of bulk liquid water predicted from our MLP model matches that of reference DFT calculations.

Furthermore, to obtain the equilibrium density of bulk liquid water under ambient conditions, a cubic box of 520 water molecules with fully periodic boundary conditions was simulated in the NpT ensemble at a pressure of $\SI{1}{\atm}$ and temperature of $\SI{300}{\kelvin}$; the pressure condition was maintained isotropically using a Nos{\'e}--Hoover barostat with damping time $\SI{1}{\ps}$. We find that the equilibrium density predicted by our MLP model to be $981.6\pm\SI{0.7}{\kilogram/\meter^3}$. This is within error margin of the reference value of $962\pm\SI{29}{\kilogram/\meter^3}$ found by Galib et al.\cite{Galib2017/10.1063/1.4986284}~using an AIMD simulation with revPBE-D3, and also within 2\% of the experimental value of $\SI{996.5}{\kilogram/\meter^3}$.\cite{Tanaka2001/10.1088/0026-1394/38/4/3}

\begin{figure*}[t!]
    \centering{}
    \includegraphics[width=0.9\textwidth]{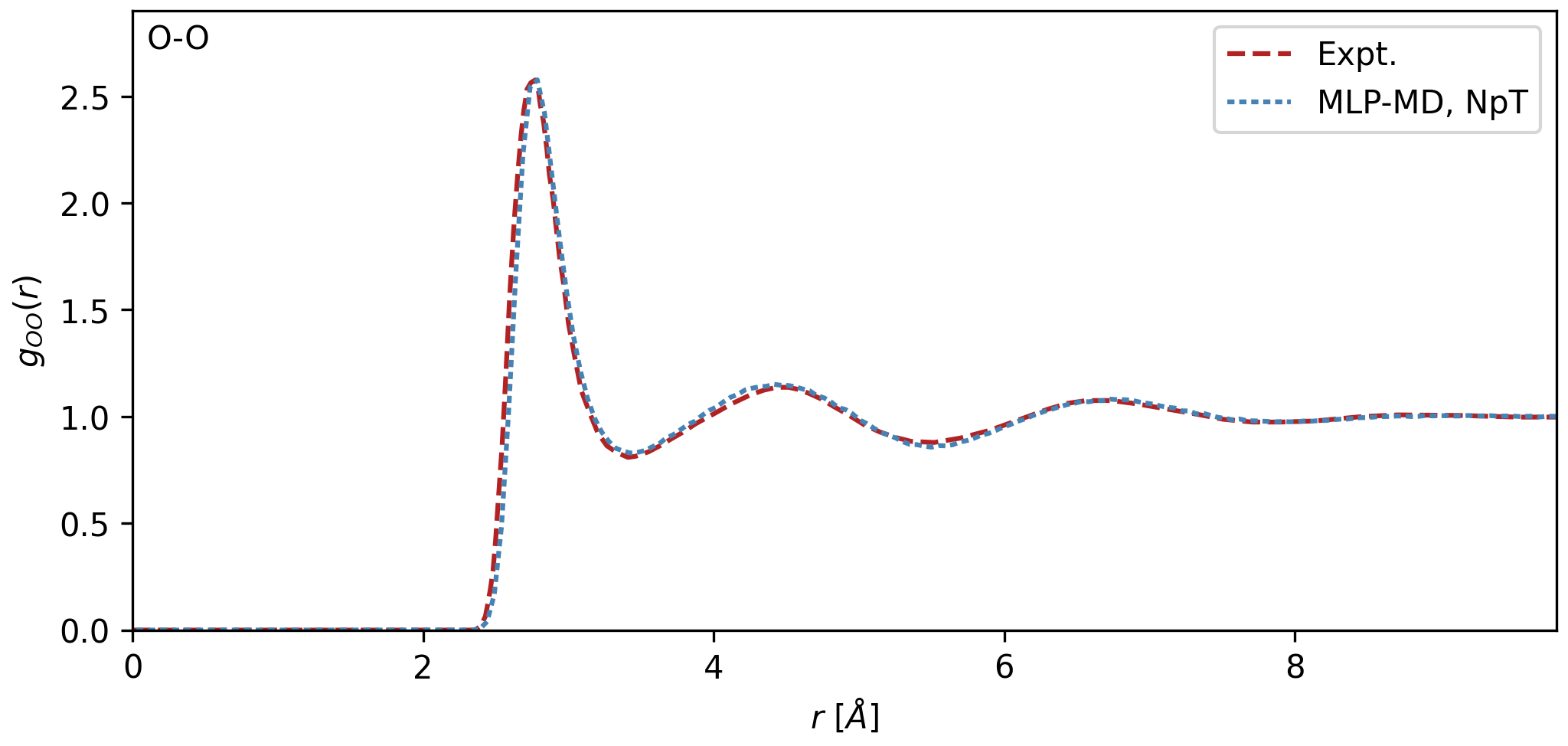}
    \caption{
    \textbf{The RDF for O-O interatomic distances in bulk liquid water under NpT conditions}. Dashed pink line: experimental references from Soper\cite{Soper2007/10.1088/0953-8984/19/33/335206}, based on joint X-ray and neutron diffraction data. Dotted blue line: predicted RDFs from our MLP model, obtained from a simulation of a cubic box of 520 water molecules in the NpT ensemble at $p=\SI{1}{\atm}$ and $T=\SI{300}{\kelvin}$.
    }
    \label{fig:SI-rdf-npt}
\end{figure*}

The RDF for the O-O interatomic distances from this NpT bulk liquid water simulation are also plotted in \Cref{fig:SI-rdf-npt}, and compared against an experimental reference measured using joint X-ray and neutron diffraction data.\cite{Soper2007/10.1088/0953-8984/19/33/335206} It is seen that the O-O RDF obtained from the MLP model almost matches the experimental reference, except for a slight shift towards larger $r$ corresponding exactly to the slightly lower predicted density.

In summary, we see that the structure of bulk liquid water as predicted by the MLP model, in terms of both the equilibrium density and the RDFs, closely match the reference DFT calculations. Furthermore, they accurately replicate realistic behaviour as compared to experimental values.

\subsection*{Surface tension}
\label{sec:SI_surface_tension}

The surface tension of the liquid-vapour interface $\gamma_{\text{lv}}$ is a key parameter in wetting behaviours and the contact angle, e.g.~via the Young equation. As such, it is also an important property pertaining to liquid water that should be validated. The surface tension was measured by simulating free-standing slabs of water under the NVT ensemble, spanning the $x$ and $y$ directions with periodic boundary conditions, and extensive vacuum in the $z$ direction; due to the broken translational symmetry along the $z$-axis, the asymmetrical part of the pressure tensor can be attributed to the surface tension of the liquid-vapour interface as follows:

\begin{equation}
    \gamma_{\text{lv}} \;\;=\;\;\frac{L_z}{2}\,\left(\,p_{zz}\,-\,\frac{p_{xx} \,+\, p_{yy}}{2}\,\right)
\end{equation}

\noindent where $L_z$ is the length of the simulation box in the $z$-axis, and $p_{\alpha\beta}$ are the components of the pressure tensor. A factor of half is inserted due to the two-sided interface of the slab.\cite{RowlinsonWidom1982,Sega2017/10.1021/acs.jpclett.7b01024}

Due to finite-size effects, the measured surface tension in these simulations are affected by the length of the simulation box $L_{\parallel}$ along the $x$ and $y$ directions, as the periodic boundary conditions constrain the spectrum of surface capillary waves allowed at the interface. This can be compensated for using a finite-size correction of the form:

\begin{equation}
    \gamma_{\text{lv}}(L_{\parallel}) \;\;\approx\;\; \gamma_{\text{lv},\infty} \,-\, k\,\frac{\ln L_z}{{L_{\parallel}}^2} \label{eq:surface_tension_finite_size}
\end{equation}

\noindent for some arbitrary fitting constant $k$, where $\gamma_{\text{lv},\infty}$ is the macroscopic surface tension.\cite{Schmitz2014/10.1103/PhysRevE.90.012128,Schmitz2014/10.1103/PhysRevLett.112.125701,Nagata2016/10.1063/1.4951710}

To perform this finite-size correction, we simulated five free-standing water slabs of between 520 to 4,680 water molecules, with width $L_{\parallel}$ ranging from $\SI{25.18}{\angstrom}$ to $\SI{75.54}{\angstrom}$. The thicknesses of the water slabs were seen to equilibrate at around ${\sim}\SI{25}{\angstrom}$ for all five slabs. We find the macroscopic surface tension to be $\gamma_{\text{lv},\infty} = 74.5\pm\SI{1.4}{\milli\newton/\meter}$, which differs by no more than 4\% from the known experimental value of $71.7\pm\SI{0.4}{\milli\newton/\meter}$.\cite{iapws-surface-tension} It is also within the error margin of the reference value of $83\pm\SI{28}{\milli\newton/\meter}$ obtained from AIMD using the same revPBE-D3 level of theory,\cite{Nagata2016/10.1063/1.4951710} although the error margin of this AIMD study is large due to high computational costs limiting the achievable simulation timescales. This gives further confidence in the MLP model's ability to simulate the liquid-vapour interface accurately.

\begin{figure*}[hb!]
    \centering{}
    \includegraphics[width=0.8\textwidth]{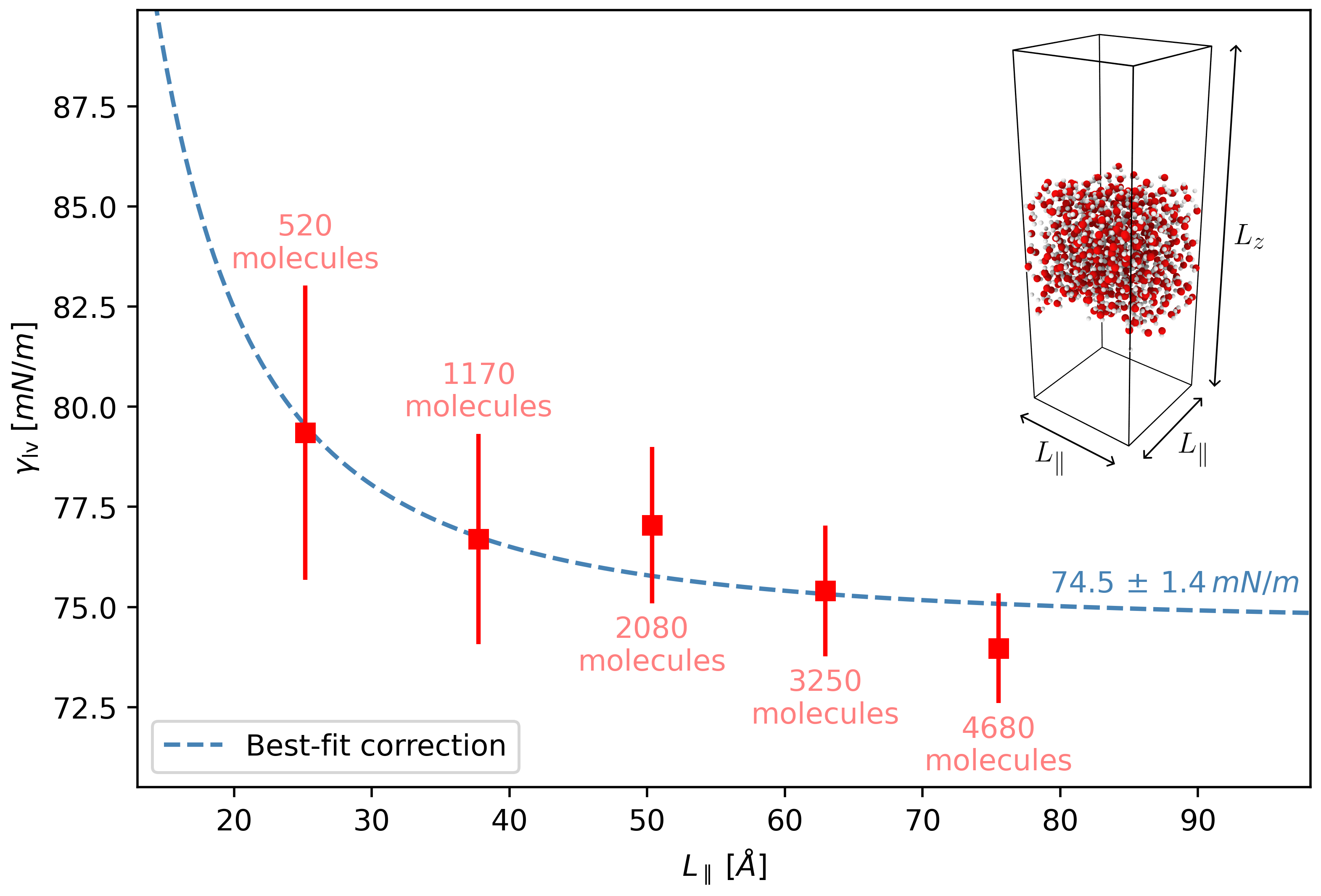}
    \caption{
    \textbf{The liquid-vapour surface tension of water} at $\SI{300}{\kelvin}$ as obtained from the MLP model, across varying sizes of water slabs; the inset shows a representative frame of the smallest slab, and the dashed blue line represents the best-fit finite-size correction following \cref{eq:surface_tension_finite_size}. Extrapolating this correction yields a macroscopic surface tension of $\gamma_{\text{lv},\infty} = 74.5\pm\SI{1.4}{\milli\newton/\meter}$.
    }
    \label{fig:SI-surface-tension}
\end{figure*}

\clearpage

\section*{Code implementation for obtaining the contact angle on non-flat surfaces}

The methodology used to obtain the contact angle throughout this work, as described in Methods, was implemented in Python for ease of development and portability. The code has been deposited in GitHub (\url{https://github.com/fast-group-cam/contact-angle}).

We use the coordinate convention, where the graphene sheet is nominally aligned to the $xy$-plane (i.e.~normal to the $z$-axis), and atomic coordinates are re-centred on every timestep such that the mean oxygen $x$- and $y$-coordinates are both zero and the mean carbon $z$-coordinate is zero at all times $t$. Note that this coordinate re-centring is only applied to the analysis of contact angles and graphene sheet dynamics from the simulation trajectories, and not to the simulations themselves.

\subsection*{Algorithm for determining the time-averaged interface}
\label{sec:SI_algorithm}

For a simulation propagated over discrete timesteps $t\in\{t_1,t_2,\ldots\}$, the time-averaging of the coarse-grained density field eq.~(2) is carried out in practice by summing over the frames:
\begin{align}
    \langle\bar{\rho}(\mathbf{r})\rangle_t \;\;:=\;\; & \frac{1}{T}\int_0^{T}\bar{\rho}(\mathbf{r},t)\dd{t} \nonumber \\
    \;\;\approx\;\; & \frac{1}{N_{t}}\sum_{\{t\}}\sum_{i=1}^{N_{\text{oxy}}} \left(2\pi\xi^2\right)^{\nicefrac{-3}{2}} \exp\!\left[\,-\,\frac{\left|\mathbf{r}-\mathbf{R}_{i}(t)\right|^2}{2\xi^2}\right] \label{eq:time_averaged_density}
\end{align}
\noindent for a fixed coarse-graining length $\xi=\SI{2.4}{\angstrom}$. The value of the time-averaged coarse-grained density field at any point $\mathbf{r}$ can be calculated efficiently in Python given the array of oxygen positions $\mathbf{R}_i(t)$, using \texttt{numpy}\cite{harris2020array} broadcasting and summation.

The time-averaged interface is then the 2-dimensional isosurface $\mathbf{r}=\mathbf{s}$ where the time-averaged coarse-grained density field reaches the cut-off value $\langle\bar{\rho}(\mathbf{s})\rangle_t=c$, which is set to $\SI{0.016}{\angstrom^{-3}}$. This isosurface can be searched for using any root-solving algorithm; a common method to do this, e.g.~as implemented by Willard and Chandler\cite{Willard2010/10.1021/jp909219k}, is to discretize space into a grid of points, calculate the density at every gridpoint, and identify points on the boundary of $\bar{\rho}>c$ and $\bar{\rho}<c$. This ``marching squares'' algorithm has a computational time complexity of $\mathcal{O}(\delta x^{-3})$ for spatial resolution $\delta x$.

In this work, we instead use a binary search algorithm to rapidly converge the position of the interface along a \textit{searching ray} emanating from a \textit{search start point}, since this is more suited to our application. The algorithm is described as follows:

\begin{center}
	\customfbox{1em}{0.7em}{\parbox{0.9\textwidth}{%
\begin{enumerate}
	\item[\textbullet] \underline{Given}: search start point $\mathbf{a}_0$ which is in the bulk liquid; search direction $\mathbf{d}$, normalized.
	\item Calculate search end point $\mathbf{b}_0 = \mathbf{a}_0 + d_{\text{max}}\mathbf{d}$ for a reasonable maximum search distance $d_{\text{max}}$, such that $\mathbf{b}_0$ is guaranteed to be outside the liquid region. This can be $d_{\text{max}} = \max_{i,t}\{\xi+\mathbf{d}\cdot(\mathbf{R}_{i}(t)-\mathbf{a}_0)\}$.
	\item At the $k$\textsuperscript{th} iteration, calculate the density $\langle\bar{\rho}(\mathbf{m}_k)\rangle_t$ at the midpoint $\mathbf{m}_k=\nicefrac{1}{2}(\mathbf{a}_k+\mathbf{b}_k)$:
	\begin{enumerate}
		\item If $\langle\bar{\rho}(\mathbf{m}_k)\rangle_t > c$, the midpoint $\mathbf{m}_k$ is inside the liquid, so set the new start point $\mathbf{a}_{k+1}$ to $\mathbf{m}_k$.
		\item If $\langle\bar{\rho}(\mathbf{m}_k)\rangle_t < c$, the midpoint $\mathbf{m}_k$ is outside the liquid, so set the new end point $\mathbf{b}_{k+1}$ to $\mathbf{m}_k$.
	\end{enumerate}
	\item Repeat step 2 until $|\mathbf{b}_k-\mathbf{a}_k|$ is converged to the precision goal $\delta x$.
\end{enumerate}
}}
\end{center}

This algorithm's time complexity scales logarithmically as $\mathcal{O}(-\log\delta x)$; throughout this work, we use a precision goal of $\delta x = \SI{0.01}{\angstrom}$. The output is thus the location of the intersection between the time-averaged interface $\mathbf{s}=\lim_{k\to\infty}\mathbf{m}_k$ and the searching ray emanating from $\mathbf{a}_0$ in direction $\mathbf{d}$.

For $N_{t}$ timesteps containing $N_{\text{oxy}}$ water molecules, the total computational cost of this interface-finding binary search algorithm --- if calculating \cref{eq:time_averaged_density} na{\"i}vely --- scales as $\mathcal{O}(-N_{t}N_{\text{oxy}}\log\delta x)$ in time complexity and $\mathcal{O}(N_{t}N_{\text{oxy}})$ in memory complexity due to \texttt{numpy} broadcasting. Memory requirements may thus exceed hardware capability if attempting to average over too many timesteps. A crucial `trick' to greatly reduce the memory footprint can be realized by excluding water molecules far away from the searching ray when calculating \cref{eq:time_averaged_density}, since the rapid fall-off of the Gaussian coarse-graining ensures that such molecules will not affect the location of the interface. In practice, this means that the algorithm only calculates $\langle\bar{\rho}(\mathbf{m}_k)\rangle_t$ using a subset of water molecules within some cut-off distance $\sigma$ of the searching ray. This `trick' therefore reduces the time complexity of the algorithm to $\mathcal{O}(-N_{t}\sigma^2 l\log\delta x)$ and the memory complexity to $\mathcal{O}(N_{t}\sigma^2 l)$, where $l$ is the expected search distance (usually $l\propto N_{\text{oxy}}^{\nicefrac{1}{3}}$), at a cost of introducing an imprecision of roughly $|\Delta\mathbf{s}|\sim\xi\int_{\nicefrac{\sigma}{\xi}}^{\infty}\zeta^2 \exp\!\left(\nicefrac{-\zeta^2}{2}\right)\dd{\zeta}$ to the interface location. We find that setting $\sigma=3.5\xi$ is sufficient to guarantee a numerical inaccuracy of less than $\SI{0.01}{\angstrom}$ in typical samples of liquid-vapour interfaces.

\subsection*{Best-fit sphere of the faraway time-averaged interface}

\begin{figure*}[b!]
    \centering{}
    \includegraphics[width=0.8\textwidth]{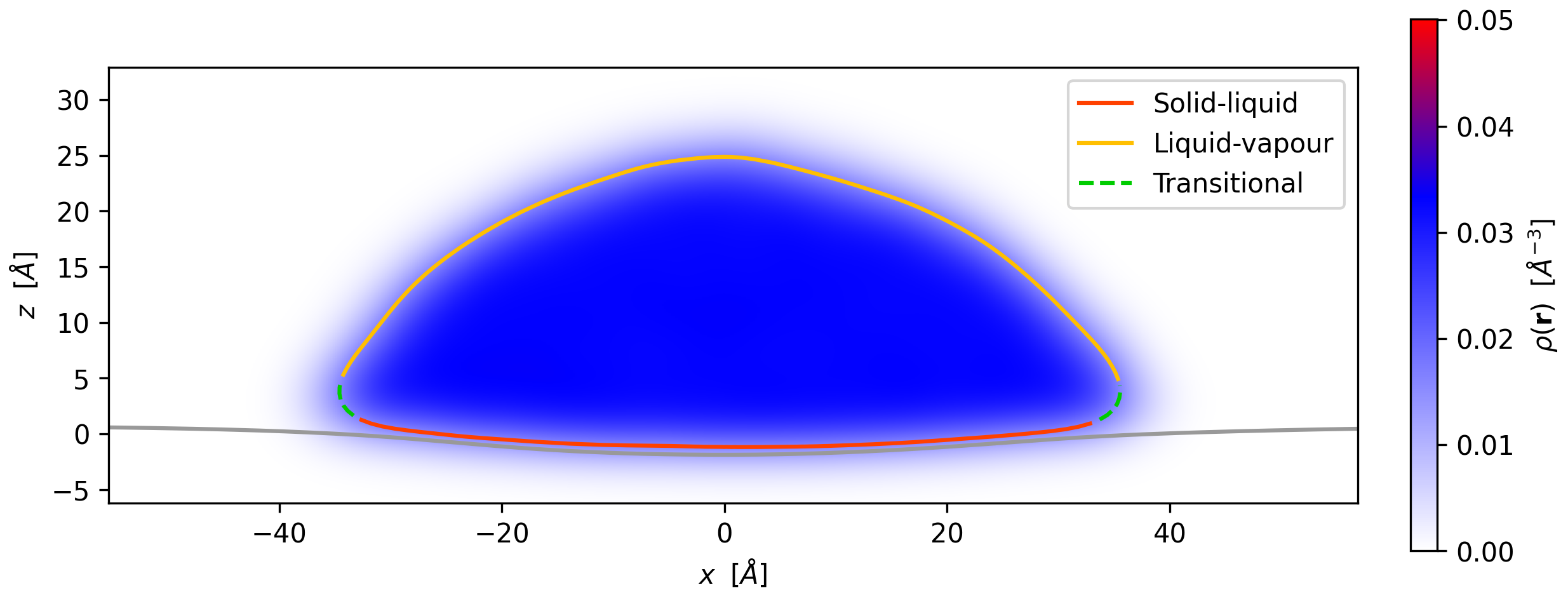}
    \caption{
    Example plot of the time-averaged coarse-grained density field $\langle\bar{\rho}(\mathbf{r})\rangle_t$, and the time-averaged interface, along the $xz$-plane for a droplet of 2,000 water molecules. The division of the time-averaged interface into `solid-liquid', `liquid-vapour', and `transitional' regions was performed by eye. In general, the analyticity of $\langle\bar{\rho}(\mathbf{r})\rangle_t$ guarantees that the time-averaged interface is a closed surface with no sharp corners, hence there will always be `solid-liquid' and `transitional' regions otherwise irrelevant to the contact angle. Instead, to match experimental definitions, the contact angle should be determined by extrapolating the liquid-vapour region into a sharp intersection with the solid surface (\Cref{fig:SI-example-best-fit-sphere}).
    }
    \label{fig:SI-example-time-averaged-interface}
\end{figure*}

The definition of the time-averaged interface, as the isosurface of a smooth analytic function, is paradoxically ``too well-behaved''; in particular, the time-averaged interface is guaranteed to be a closed surface with no sharp corners, as illustrated in \Cref{fig:SI-example-time-averaged-interface}, which means that there is no clear choice for \textit{where} the contact angle occurs on the time-averaged interface itself.

Instead, we draw inspiration from experimental techniques. Contact angles are often measured experimentally using optical goniometry, where the geometry of the droplet is imaged using a camera and the tangent lines of the liquid-vapour and solid-liquid interfaces are extrapolated to find the intersection angle\cite{yuan_contact_2013}. For the atomistic context, this therefore corresponds to extrapolating the liquid-vapour region of the time-averaged interface past the ``rounded'' transitional region, and finding the angle at the sharp intersection of the extrapolated liquid-vapour interface and the solid surface.

\begin{figure*}[b!]
    \centering{}
    \includegraphics[width=0.8\textwidth]{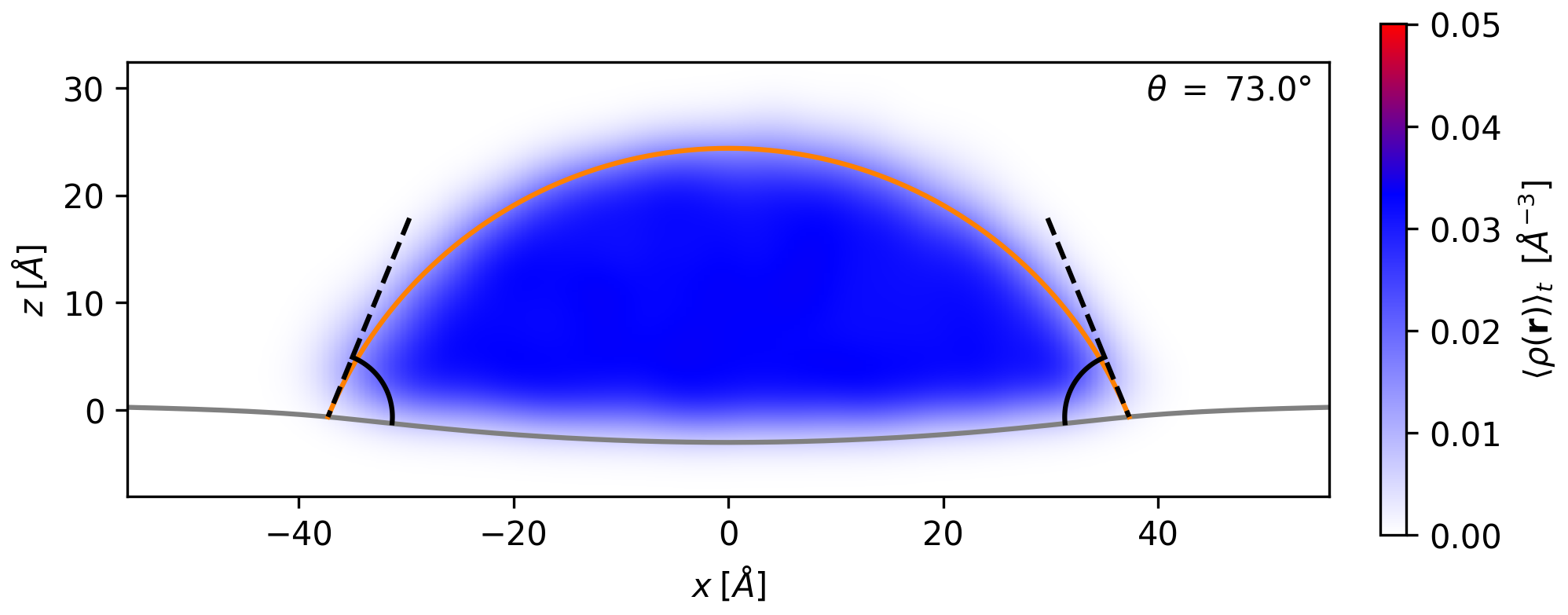}
    \caption{
    The time-averaged coarse-grained density field $\langle\bar{\rho}(\mathbf{r})\rangle_t$ along the $xz$-plane for a droplet of 2,000 water molecules (same as \Cref{fig:SI-example-time-averaged-interface}), and the best-fit sphere (orange line) fitted to randomly-selected interfacial points far from the graphene surface. The contact angle (black lines) is thus defined by the angle between this extrapolated spherical fit and the time-averaged graphene heightmap (dark gray curve).
    }
    \label{fig:SI-example-best-fit-sphere}
\end{figure*}

The liquid-vapour interface is expected to converge, in the long-time limit, to a spherical surface of uniform curvature due to the minimization of surface energy. As such, the chosen extrapolation is a sphere of the form $|\mathbf{s}-\mathbf{s}_0|=R$ for least-squares best-fit parameters $\mathbf{s}_0$ and $R$, fitted to points $\mathbf{s}$ on the time-averaged interface far away from the graphene surface (and therefore unaffected by the closure of the solid-liquid portion of the interface, or the ``rounding'' effect of the coarse-graining parameter $\xi$). These points $\mathbf{s}$ are sampled at random, by generating 150 searching rays emanating from the droplet centre-of-mass's upper hemisphere \textit{not} intersecting the time-averaged graphene heightmap. Specifically, each searching ray is generated by selecting a uniformly-distributed random azimuthal angle between $0$ and $2\pi$, and an arccosine-distributed polar angle between $0$ and $\frac{\pi}{2}$, and the ray is extended to infinity to check if it collides with the time-averaged graphene heightmap; if no such collision occurs, the interfacial point $\mathbf{s}$ found along this ray is added to the sample, otherwise the ray is re-generated.

For calculating the isotropic contact angle in axisymmetric cases, the centre of the best-fit sphere is constrained to be along the $z$-axis, so that $\mathbf{s}_0=\left(0,0,z_0\right)$. Otherwise, for anisotropic contact angles, the centre of the best-fit sphere is fully unconstrained.

The fitting residuals --- in this case the distance of the sampled interfacial points $\mathbf{s}$ from the best-fit sphere's surface --- are found to have a mean absolute value of $\SI{0.11}{\angstrom}$, and a root mean square value of $\SI{0.15}{\angstrom}$, when the spherical fit is applied isotropically to a droplet of 1,000 water molecules on free-standing graphene, and a mean absolute value of $\SI{0.16}{\angstrom}$ and root mean square value of $\SI{0.21}{\angstrom}$ when the fit is applied anisotropically to the droplet on $-2\%$ compressed graphene. For comparison, the best-fit radii for these droplets are $\SI{31.43}{\angstrom}$ and $\SI{34.56}{\angstrom}$ respectively, showing that the fitting errors are negligible compared to the scale of the droplet, and that the droplet's faraway liquid-vapour interface is correctly described as a spherical surface in both isotropic and anisotropic cases.

\subsection*{Finding the graphene heightmap}

The instantaneous graphene heightmap $h_{\text{GS}}(x,y;t)$ is obtained by constructing a Clough--Tocher interpolator using \texttt{scipy}'s built-in \texttt{CloughTocher2DInterpolator} constructor\cite{scipy_cloughtocher2dinterpolator}, feeding the $x$- and $y$-coordinates of the carbon atoms at time $t$ as the input data and the $z$-coordinates as the target, and then evaluating the interpolator over a discrete grid of $(x,y)$ points. To handle periodic boundary conditions correctly, the input data to the interpolator is padded with periodic images in the $x$ and $y$ directions. The bottleneck in this process is the construction of the triangulation mesh for the Clough--Tocher interpolator, which scales as $\mathcal{O}(N_{\text{carbons}})$ independently of the grid resolution.

The discretization of the grid is chosen to always be smaller than the carbon-carbon interatomic distance $\SI{1.426}{\angstrom}$, so that the instantaneous heightmap forms a smooth interpolation \textit{within} the hexagonal cells of the graphene sheet. Afterwards, calculating the time-averaged graphene heightmap $\langle h_{\text{GS}}(x,y)\rangle_t$ is a simple matter of averaging the instantaneous heightmaps across the simulation timesteps.

\subsection*{Calculating the anisotropic contact angle}

In the general anisotropic case, to find the contact angle along an azimuthal direction $\phi$, we first solve for the intersection between the best-fit sphere of the faraway time-averaged interfaces, and the time-averaged graphene heightmap (TAGH):
\begin{equation}
    \langle h_{\text{GS}}(x, y)\rangle_t \;=\; z_0 \,+\,\sqrt{R^2 \,-\,\left(x -x_0\right)^2 \,-\, \left(y -y_0\right)^2}
\end{equation}
\noindent where $\mathbf{s}_0=\left(x_0, y_0, z_0\right)$ and $R$ are the centre and radius of the best-fit sphere respectively; this equation can be solved 1-dimensionally over variable $r$ where $(x,y)=(r\cos\phi,r\sin\phi)$, using a numerical iterative approach. The anisotropic contact angle along direction $\phi$ is then:
\begin{equation}
    \theta(\phi) \;=\; \arctan\left(\frac{r - x_0\cos\phi - y_0\sin\phi}{\sqrt{R^2 \,-\,\left(x-x_0\right)^2 \,-\, \left(y-y_0\right)^2}}\right) \,+\, \arctan\left(\frac{\partial\langle h_{\text{GS}} \rangle}{\partial x}\cos\phi \,+\, \frac{\partial\langle h_{\text{GS}} \rangle}{\partial y}\sin\phi\right).
\end{equation}

This is the approach used to obtain the distribution of anisotropic contact angles (mapped to uniformly distributed $\phi$) for droplets under compressive strain. We sample 360 evenly-spaced values of $\phi$, finding that this is sufficient to resolve $\theta(\phi)$ continuously.

\subsection*{Calculating the isotropic contact angle}

In axisymmetric cases, it is more useful to calculate the isotropic contact angle from the azimuthally-averaged TAGH, which is a function of radial coordinate only:
\begin{equation}
    \langle h_{\text{GS}}(r)\rangle_t \;\;=\;\; \frac{1}{2\pi}\int_0^{2\pi}\langle h_{\text{GS}}(x=r\cos\phi,y=r\sin\phi)\rangle_t\dd{\phi}
\end{equation}
\noindent in which case the intersection with the best-fit sphere of the faraway time-averaged interface is simply the solution to:
\begin{equation}
    \langle h_{\text{GS}}(r)\rangle_t \;\;=\;\; z_0 \,+\,\sqrt{R^2-r^2}
\end{equation}
\noindent and the solution $r=a$ can be interpreted as the radius of the three-phase contact line. The isotropic contact angle is then:
\begin{equation}
    \theta \;=\; \arctan\left(\frac{r}{\sqrt{R^2 \,-\, r^2}}\right) \,+\, \arctan\left(\frac{\mathrm{d}\langle h_{\text{GS}} \rangle}{\mathrm{d}r}\right).
\end{equation}

This is the approach used to obtain the isotropic contact angles, and corresponding droplet radii, both for the droplets of varying size (to obtain the finite-size corrected contact angle); and also for the droplets under tensile strain.

\subsection*{Parameter sensitivity check}

Our contact angle methodology relies on two fixed parameters, the coarse-graining length $\xi$ and cut-off density $c$, which we set at $\SI{2.4}{\angstrom}$ and $\SI{0.016}{\angstrom^{-3}}$ respectively to match the original Willard--Chandler definition\cite{Willard2010/10.1021/jp909219k}. The coarse-graining length represents the scale of spatial ``blurring'' required to restore continuity of the density field $\bar{\rho}(\mathbf{r},t)$ over short sampling times, with the exact equilibrium density field being recovered under the dual limit of $\xi\to0^{+}$ and infinite sampling time. On the other hand, the cut-off density represents the threshold for deciding whether a region of space is in the liquid or vapour phase, and therefore must be bounded from below by zero (otherwise ``everywhere'' is liquid) and from above by the bulk liquid density (otherwise ``everywhere'' is vapour). The cut-off density should be set to half of the bulk liquid density, for the time-averaged interface to converge to the Gibbs equimolar dividing surface in the limit of $\xi\to0^{+}$.

\begin{figure*}[hbtp!]
    \centering{}
    \includegraphics[width=\textwidth]{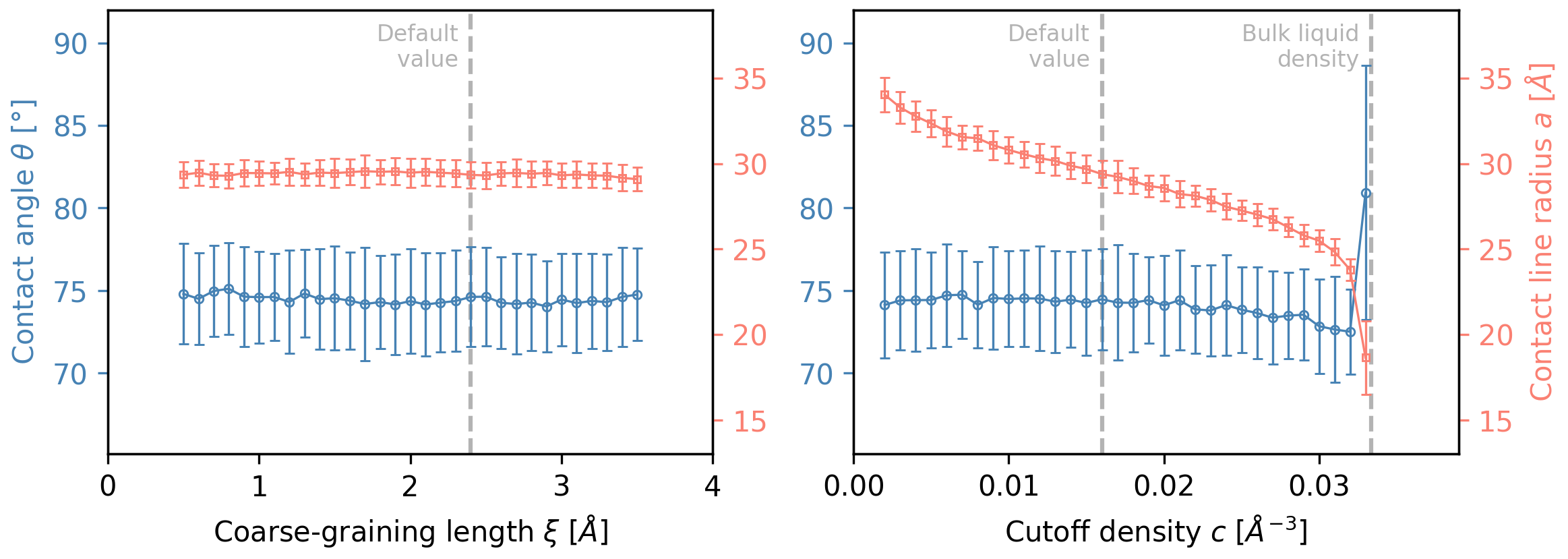}
    \caption{
    \textbf{Sensitivity check of methodology fixed parameters on contact angles and contact line radii.} Left: the microscopic contact angle (blue circle markers) and three-phase contact line radius (red square markers) for a droplet of 1,000 water molecules on free-standing graphene, analysed using varying values of the coarse-graining length $\xi$. The error bars represent the 95\% confidence interval from block averaging. Right: the same results for varying values of the cut-off density $c$.
    }
    \label{fig:SI-contact-angle-sensitivity-check}
\end{figure*}

As a quick check of the sensitivity of our reported contact angles on the methodology's fixed parameters, we performed our analysis for the droplet of 1,000 water molecules on free-standing graphene with varying values of $\xi$ and $c$, as shown in \Cref{fig:SI-contact-angle-sensitivity-check}. Both the contact angle and three-phase contact line are not sensitive to changes in the coarse-graining length, as its effect is to suppress spatial fluctuations of the droplet's density field without changing its long time average behaviour. In fact, this hints towards future room for improvement, where the coarse-graining length $\xi(T)$ could be dynamically tuned according to the time-averaging interval $T$ to smoothly connect the instantaneous Willard--Chandler parameter $\xi(T=0)=\SI{2.4}{\angstrom}$ with the long-time thermodynamic limit $\xi(T\to\infty)=0$. Nonetheless, \Cref{fig:SI-contact-angle-sensitivity-check} shows that our results are robust regardless of how this coarse-graining length is tuned. As for the cut-off density, the contact angle is similarly not sensitive to its value within the neighbourhood of the default value of half the bulk liquid density, but the contact line radius does depend weakly on $c$, following the geometric intuition that a larger value of $c$ leads to the liquid-vapour interface being identified closer in to the liquid region, thus shrinking the droplet's fitted shape. Indeed, the reported results diverge when the cut-off density is too close to the bulk liquid density, as the liquid region becomes undefined. We therefore recommend that the cut-off density should be kept at $c=\frac{1}{2}\rho_{\text{bulk}}$, to maintain connection with the equimolar dividing surface, noting that the reported contact angles do not depend on this choice anyways.

\clearpage

\section*{Comparison of contact angle methodologies on spatially-fixed flat graphene and dynamical graphene (unstrained)}

\begin{figure*}[bp!]
    \centering{}
    \includegraphics[width=0.8\textwidth]{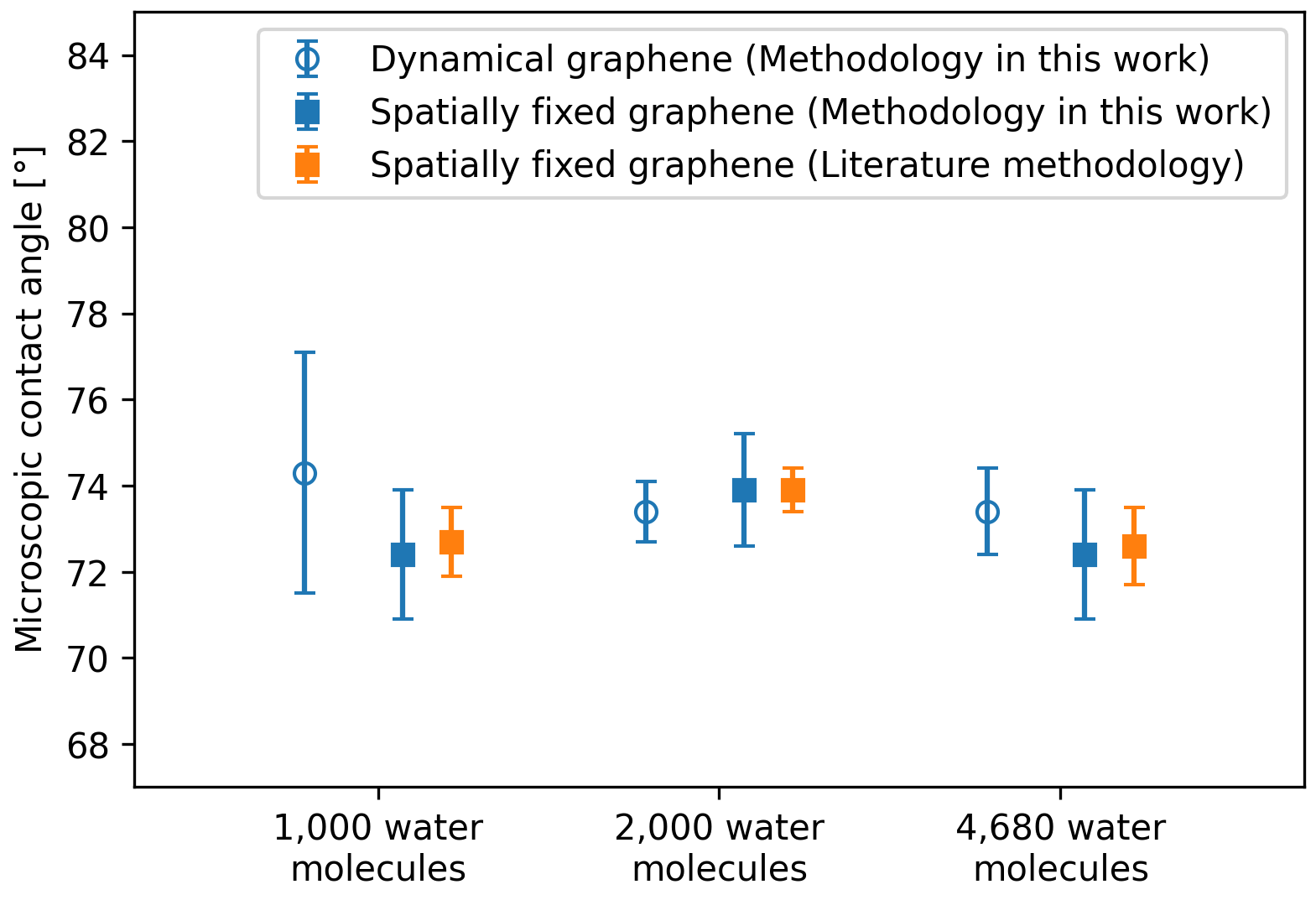}
    \caption{
    \textbf{Contact angles of water droplets of various sizes, as obtained by the MLP, on either dynamical free-standing graphene or spatially fixed flat graphene} with no strain applied. For the fixed sheet, contact angles were measured using both the methodology introduced in this work and also the methodology of Werder et al.~(Ref.~\citenum{Werder2003/10.1021/jp0268112}), which gives consistent results. In all three cases, the contact angle on free-standing graphene is within the error margin of the contact angle on spatially-fixed graphene, indicating that the contact angle is not significantly affected by whether or not the graphene is capable of deforming flexibly, as long as the sheet is unstrained.
    }
    \label{fig:SI-methodology-comparison}
\end{figure*}

In order to compare our novel methodology for measuring contact angles against the literature-established methodology, e.g.~as detailed by Werder et al.~(Ref.~\citenum{Werder2003/10.1021/jp0268112}), we simulated three droplets of different sizes on spatially fixed, flat graphene sheets and measured their contact angles using both methodologies. These separate simulations are necessary, as the Werder et al.~methodology is limited to flat solid surfaces by nature of the $z$-coordinate based histogram. We find that both methodologies give consistent results for the contact angles on these frozen sheets, as shown in \Cref{fig:SI-methodology-comparison}, verifying that our methodology as introduced in this work is consistent with previous interpretations of the contact angle.

More interestingly, we compare the contact angles obtained for the droplets of the same sizes on dynamical free-standing graphene (also \Cref{fig:SI-methodology-comparison}), which can only be measured using our methodology. We find that the contact angles on free-standing graphene sheets are within error margin of those on spatially frozen graphene sheets. These findings are in line with the observations reported by both Werder et al.\cite{Werder2003/10.1021/jp0268112}~and Liao et al.\cite{Liao2022/10.1016/j.apsusc.2022.154477}~concerning graphene-water contact angles modelled using empirical force fields. This shows that the contact angle is not significantly affected by whether or not the graphene is capable of deforming flexibly, when the sheet is unstrained.

This presents a contrast against the contact angle of the 1,000 water molecules droplet on graphene under tensile strain, which increases with strain more dramatically for dynamical graphene than for spatially fixed graphene. Such an anomalous effect would not have been discovered using only spatially fixed simulations, highlighting the importance of the machine learning potential based simulations to handle both water-carbon and carbon-carbon interactions at the same level of accuracy.

\clearpage

\section*{Interaction energy of water with graphene under tensile strain}

The increase of contact angle when tensile strain is applied on the graphene sheet, from $74.3\degree$ in the free-standing case up to $84.8\degree$ under +2.0\% tensile strain, corresponds to a very large decrease of 67\% in the magnitude of the interfacial energy difference $\Delta\gamma=\gamma_{\text{sv}}-\gamma_{\text{sl}}$. This decrease is too large to be explained purely by the increased interatomic spacing of the graphene sheet, which would na{\"i}vely account for only a ${\sim}4\%$ reduction in the area density of carbon-water interactions. Instead, this reduction of $\Delta\gamma$ must be driven either by a significant change in the solid-vapour surface tension $\gamma_{\text{sv}}$, which corresponds to the free energy associated with surface deformations of the graphene sheet, or by a significant change in the solid-liquid surface tension $\gamma_{\text{sl}}$, which corresponds to the interaction energy between the liquid and solid phases.

To clarify which mechanism dominates the drastic decrease in hydrophilicity under tensile stress, we calculate the single-molecule static binding energy between a water molecule and a graphene sheet, as a function of the distance $d$. The binding energy is defined as:

\begin{equation}
    E_b(d) \;\;=\;\; E_{\text{W}+\text{G}}(d) \,-\, E_{\text{W}} \,-\, E_{\text{G}}
\end{equation}

\noindent where $E_{\text{W}+\text{G}}(d)$ is the total energy of the system where the water molecule is placed at distance $d$ from the graphene sheet, $E_{\text{W}}$ is the energy of the isolated water molecule, and $E_{\text{G}}$ is the energy of the isolated graphene sheet. Note that this involves only the static electronic energy, without thermal or nuclear quantum effects.

\begin{figure*}[bp!]
    \centering{}
    \includegraphics[width=\textwidth]{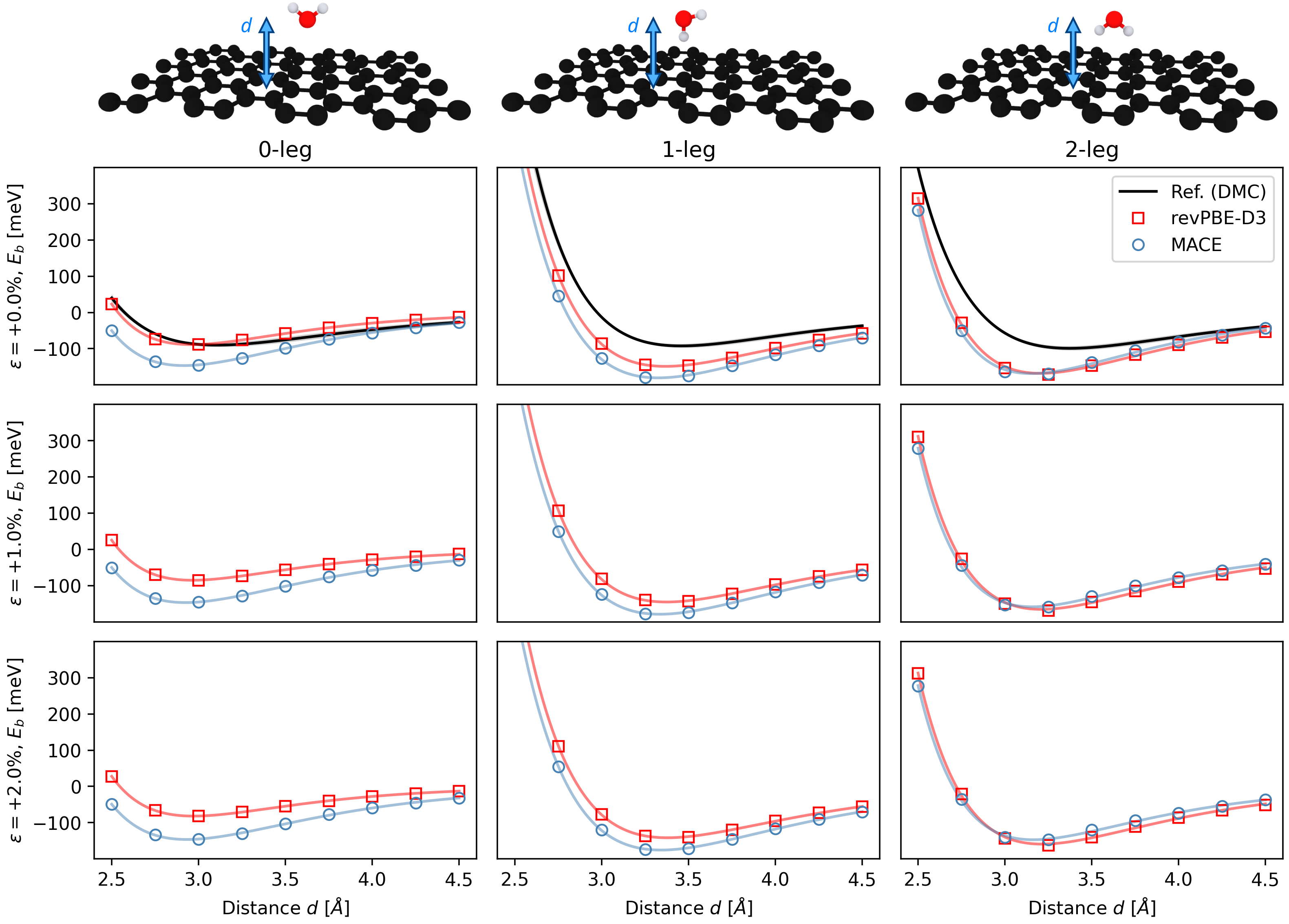}
    \caption{
    \textbf{Single-molecule binding energy curves of water on graphene} in the (left) 0-leg, (middle) 1-leg, and (right) 2-leg motifs, under varying tensile strains $\varepsilon$ applied on the graphene sheet from (top) 0\% to (bottom) 2\%. The red square markers correspond to DFT calculations made using the revPBE-D3 functional in CP2K/Quickstep, while the blue circular markers correspond to the MLP model; the continuous curves are best-fit Morse potentials for each set of calculations. The black curves for the top row for unstrained graphene indicate the reference diffusion Monte Carlo calculations, taken from Ref.~\citenum{Brandenburg2019/10.1021/acs.jpclett.8b03679}.
    }
    \label{fig:SI-interaction-energy-scan}
\end{figure*}

We calculate this binding energy curve $E_b(d)$ for varying water molecule orientations (0-leg, 1-leg, and 2-leg) and tensile strains applied to the graphene sheets, using both the MLP and using DFT calculations at the revPBE-D3 level from CP2K/Quickstep under the same settings as those used to train the MLP. These binding energy curves are plotted in \Cref{fig:SI-interaction-energy-scan}. In all cases, the MLP model is able to replicate the DFT calculations closely, regardless of the orientation of the water molecule or the strain on the graphene sheet. It is also seen that the application of strain does not significantly affect the binding energy curves, for all water orientations.

The effects of strain on the interaction energy between liquid water and the graphene sheet can be quantified through the single-molecule adsorption energy, which is the magnitude of the minimum of the binding energy curve. We find that the DFT calculated adsorption energy, for the 0-leg motif, decreases from $88.7\pm\SI{1.0}{\milli\electronvolt}$ in the unstrained case to $81.8\pm\SI{1.0}{\milli\electronvolt}$ under +2.0\% tensile strain; and for the 1-leg motif, from $148.7\pm\SI{7.0}{\milli\electronvolt}$ when unstrained to $141.6\pm\SI{6.1}{\milli\electronvolt}$ under +2.0\% tensile strain; and for the 2-leg motif, from $168\pm\SI{18}{\milli\electronvolt}$ when unstrained to $159\pm\SI{19}{\milli\electronvolt}$ under +2.0\% tensile strain. Averaged across all three orientations, the mean adsorption energy thus effectively decreases by only 5.6\% when under strain.

\begin{figure*}[bp!]
    \centering{}
    \includegraphics[width=0.6\textwidth]{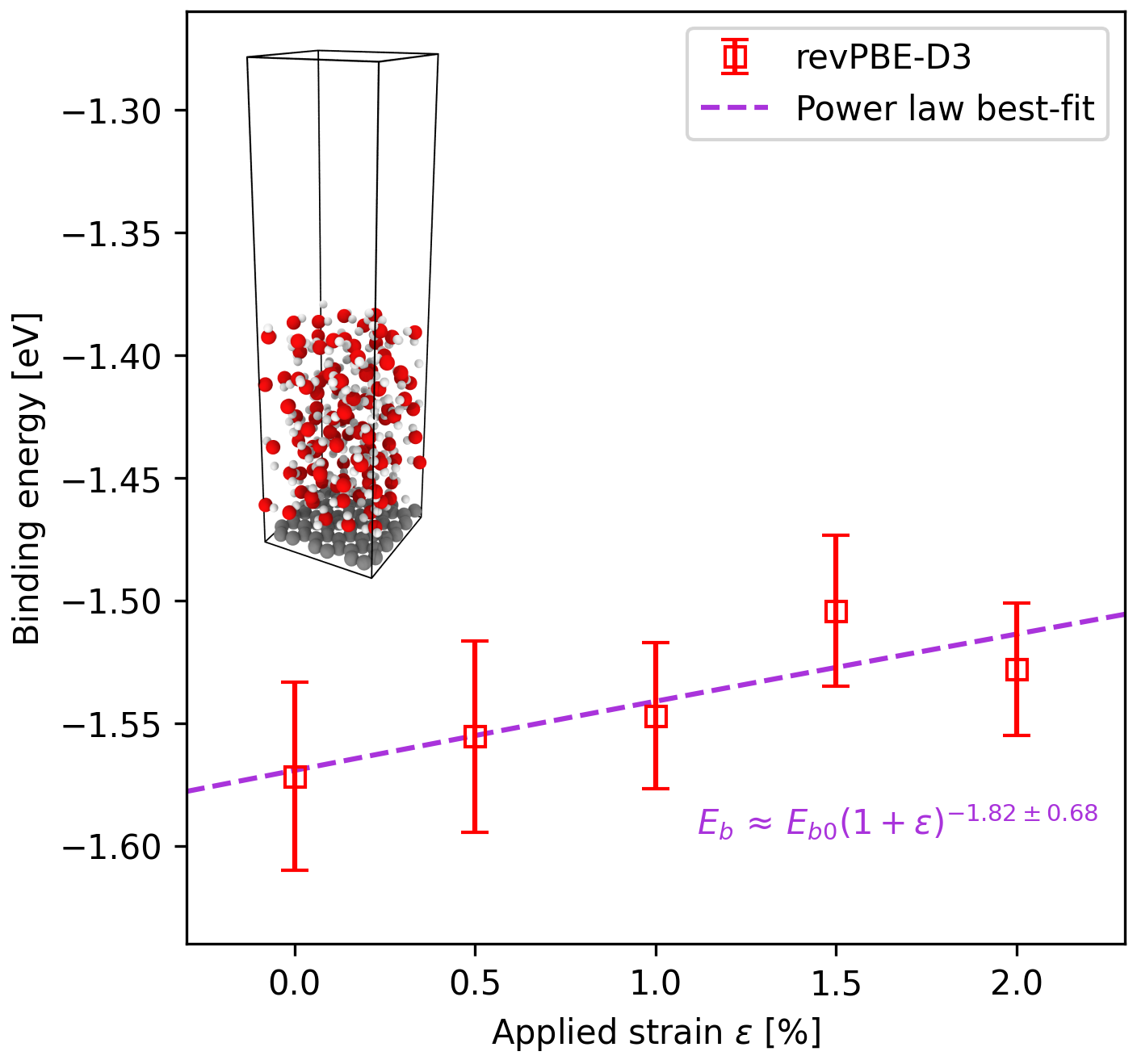}
    \caption{
    \textbf{Binding energy of water slabs on graphene under varying tensile strain conditions,} representing the static DFT interaction energy between water and fully-wetted graphene. The binding energies for each strain condition were averaged over 50 samples randomly selected from a thermally-equilibrated MD simulation, with error bars indicating the 95\% confidence interval.
    }
    \label{fig:SI-interaction-energy-slab}
\end{figure*}

As a further test of the effects of strain on the static water-graphene interaction energy, we extracted random configurations of a fully-wetted $\SI{12.834}{\angstrom}\times\SI{12.35}{\angstrom}$ graphene sheet, with a water slab of ${\sim}\SI{20}{\angstrom}$ thickness atop, from a thermally-equilibrated MD simulation under varying tensile strain conditions, and computed the binding energy of the entire water slab at its equilibrium distance. The resulting binding energies, as shown in \Cref{fig:SI-interaction-energy-slab}, are well-explained by the na{\"i}ve area-scaling law of $E_b\propto\left(1+\varepsilon\right)^{-2}$, which therefore gives only a ${\sim}4\%$ reduction in the water slab's binding energy at 2\% tensile strain.

The small decrease in both the single-molecule adsorption energy, on the order of ${\sim}6\%$, and the water slab binding energy, on the order of ${\sim}4\%$, indicates that the static contribution of the solid-liquid interaction energy likely varies by a similar amount in the original ``droplet on graphene'' system. Noting that the difference of the solid-vapour and solid-liquid surface tensions $\Delta\gamma$ can be interpreted as a change in the interfacial free energy, which is generally driven by both static interaction energies and entropic contributions from dynamical degrees of freedom, the fact that the solid-liquid interaction energy decreases by a much smaller amount than the 67\% decrease of $\Delta\gamma$ indicates that dynamical contributions must drive the strain-modulated weakening of graphene's hydrophilicity. These dynamical contributions may be associated with any part of the system, but the most likely candidate is the free energy associated with surface deformations of the graphene sheet, or in other words by changes in the morphological dynamics of graphene surface rippling, which is illustrated in the localized perturbation of the long-time angle-angle autocorrelation $\mathcal{C}_{\text{GS}}(\tau\to\infty)$ at the droplet edge.

\clearpage

\section*{Contact angle on spatially-fixed graphene versus dynamical graphene under +2.0\% tensile strain}

The evidence of the DFT-calculated interaction energy for a single water molecule upon spatially fixed graphene, which only decreases by 8\% when under +2.0\% tensile strain, demonstrates that the change in static interaction energies is insufficient to explain the drastic weakening of hydrophilicity of free-standing graphene under tensile strain. To further compare whether or not the observed increase of contact angle under tension is related to the dynamical rippling motion of the graphene membrane, we also simulated a droplet of 1,000 water molecules on a spatially fixed, flat graphene sheet under +2.0\% tensile strain.

\begin{figure*}[b!]
    \centering{}
    \includegraphics[width=0.75\textwidth]{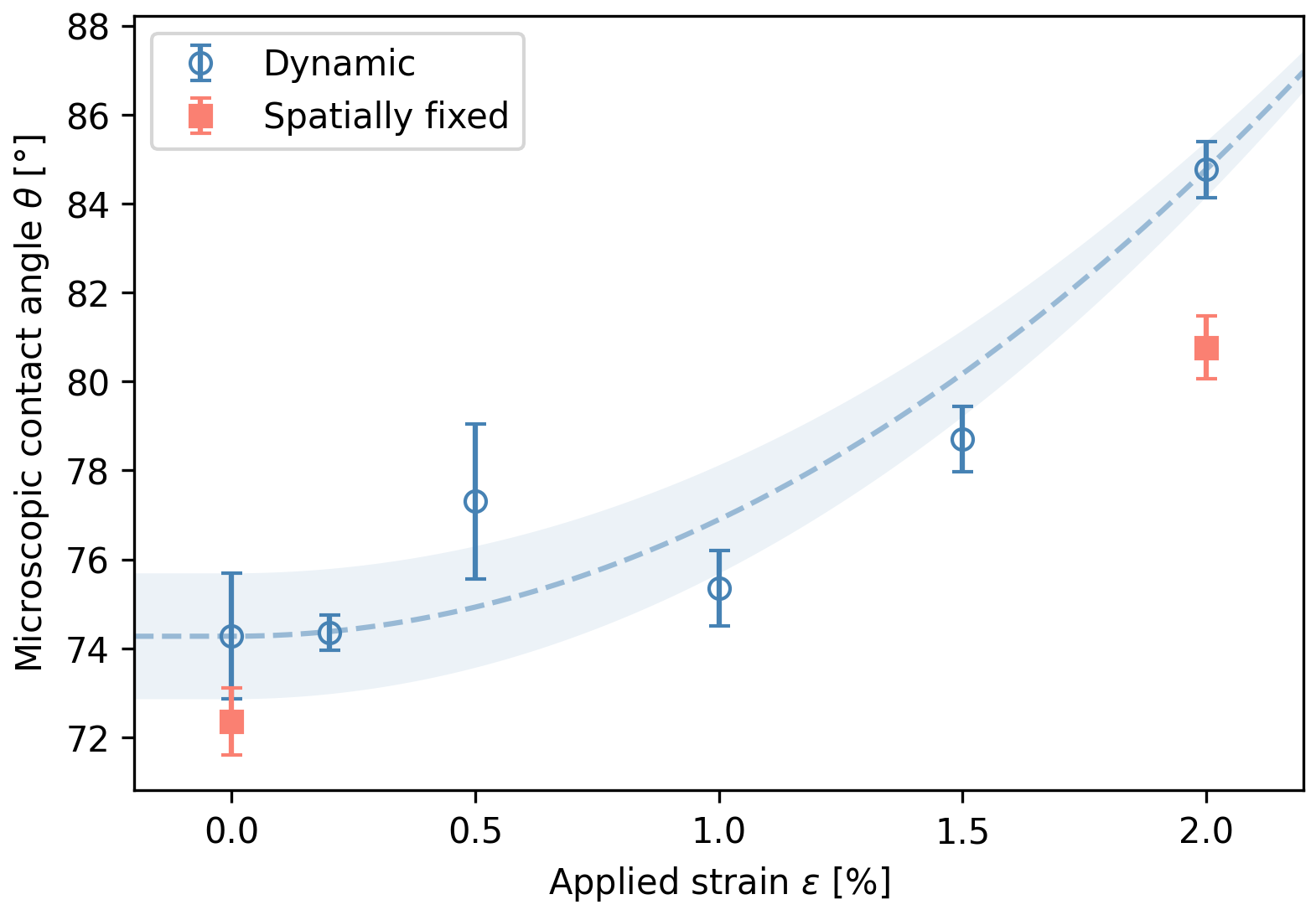}
    \caption{
    \textbf{Contact angles for the droplet of 1,000 water molecules under varying tensile strain, on dynamical versus spatially fixed flat graphene sheets.} The dashed blue line is a guide to the eye, and does not represent any specific functional form.
    }
    \label{fig:SI-fixed-contact-angles}
\end{figure*}

The resulting contact angle on this spatially-fixed strained sheet is $80.8 \pm 1.4 \degree$. This represents an increase in the contact angle (i.e.~a weakening of hydrophilicity) compared to the free-standing value of $74.3 \pm 2.8 \degree$, but still significantly less than the contact angle of $84.8 \pm 1.2 \degree$ seen on dynamical graphene under the same tensile strain. In particular, this implies that the surface energy difference $\Delta\gamma=\gamma_{\mathrm{lv}}\cos\theta$ between the solid-vapour and solid-liquid interfaces is decreased by 41\% on spatially-fixed graphene under only a 2\% strain, which is surprisingly large but nonetheless insufficient to account for the nearly 67\% decrease of $\Delta\gamma$ on dynamical graphene.

These results point towards the idea that the strain-driven modulation of wetting is partially explained by weakening static intermolecular effects, e.g.~as supported by the decrease in DFT-calculated single-molecule interaction energy, whose effect can be captured by spatially fixed simulations with first-principles level accuracy; however, the remaining anomalous increase of contact angle is driven from the unique dynamical properties of graphene's surface ripples and their coupling to the three-phase contact line, which is then modified by strain.

\clearpage

\section*{``Surfing'' motion of droplet on long-ranged coherent ripple waves}

\begin{figure*}[b!]
    \centering{}
    \includegraphics[width=\textwidth]{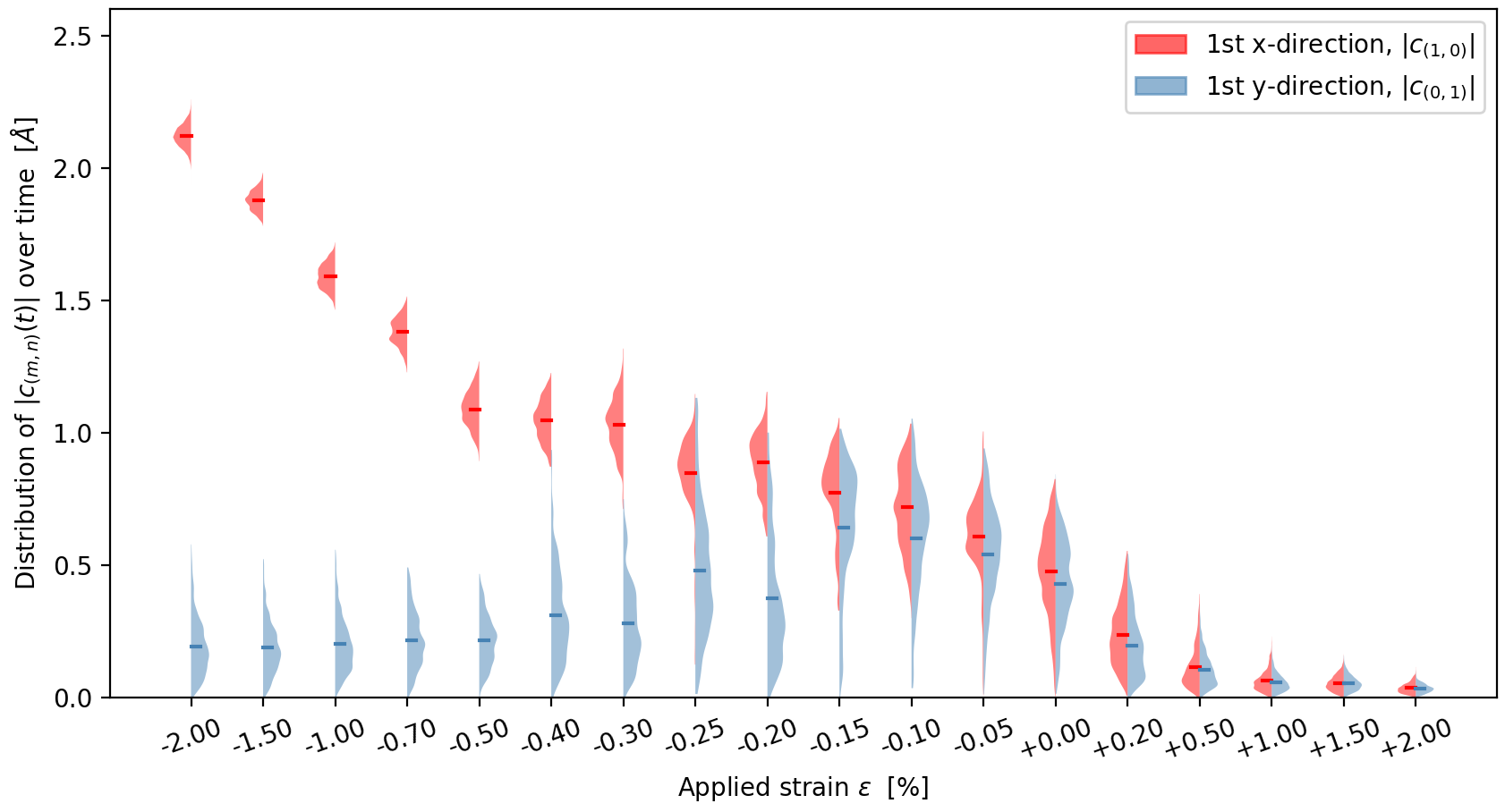}
    \caption{
    \textbf{The distributions of the magnitudes of the first Fourier coefficients} for the graphene heightmap $h_{\text{GS}}(x,y;t)$ in the $x$-direction (red) and $y$-direction (blue) over time, for each applied strain condition. These coefficients were calculated from the simulation trajectories of the 1,000 water molecules droplet present on the graphene sheet. For compressive strains greater than or equal to $-0.20\%$, a bifurcation forms between the dominant $(1,0)$ mode, which gains a non-zero time-average,  and the thermally distributed $(0,1)$ mode which is centred around zero. On the other hand, for all other strain conditions, both modes are thermally distributed around zero. This indicates that the critical strain for the random-to-coherent rippling phase transition occurs between $-0.15\%$ and $-0.20\%$ compressive strain. Note that, for the purposes of clarity in plotting this graph, the $x$- and $y$-directions were swapped for some of the simulations such that the direction of spontaneous symmetry breaking for the long-ranged coherent wave is in the $x$-direction; in the original simulation coordinates, roughly half of the simulations with a long-ranged coherent wave had the wave occur in the $y$-direction (i.e.~dominant $(0,1)$ mode) instead of the $x$-direction.
    }
    \label{fig:SI-fourier-coefficients}
\end{figure*}

The application of compressive strain on a graphene sheet results in a phase transition, where the sheet spontaneously ``buckles'', and transitions from thermally random small-amplitude surface ripples to a large-amplitude long-ranged coherent ripple wave. This has been reported in both the experimental and theoretical literature before, e.g.~in Refs.~\citenum{Ma2016/10.1038/nmat4449,Meng2013/10.1063/1.4857115,Thiemann2020/10.1021/acs.jpcc.0c05831,Wang2017/10.1088/2053-1591/aa7324}. For our simulations, we establish the critical point of this phase transition by studying the $(m,n)$\textsuperscript{th} Fourier coefficients for the graphene sheet heightmap $h_{\text{GS}}(x,y;t)$, defined as follows:
\begin{equation}
    c_{(m,n)}(t) \;=\; \frac{1}{L_{x}L_{y}} \int_{0}^{L_x}\int_{0}^{L_y} \, h_{\text{GS}}(x,y;t) \,\exp\left(i \frac{2\pi m}{L_x} x \right)\,\exp\left(i \frac{2\pi n}{L_y} y\right) \dd{x}\dd{y}
\end{equation}
\noindent where $L_x$ and $L_y$ are the lengths of the simulation box in the $x$ and $y$ directions respectively. Since the longest-ranged coherent ripple corresponds to the lowest $(m,n)$, the magnitudes of the first Fourier coefficients for each direction $|c_{(1,0)}|$ and $|c_{(0,1)}|$ serve as a useful indicator for the random-to-coherent rippling phase transition. The distribution of these magnitudes is plotted in \Cref{fig:SI-fourier-coefficients}.

The phase transition is marked by a bifurcation of the two modes, which occurs for compressive strains greater than or equal to $-0.20\%$; the dominant $(1,0)$ mode gains a non-zero time-average, whereas the $(0,1)$ mode is thermally distributed around zero. Visually, this corresponds to the appearance of a long-ranged coherent ripple wave on the graphene sheet. On the other hand, for all tensile strains, as well as compressive strains less than or equal to $-0.15\%$, both modes are thermally distributed around zero, and correspondingly there is no visual appearance of a long-ranged wave. The critical compressive strain for this random-to-coherent rippling phase transition therefore lies between $-0.15\%$ to $-0.20\%$, although it should be noted that this transition may be affected by the presence of the droplet and the wetting-induced curvature.

\begin{figure*}[b!]
    \centering{}
    \includegraphics[width=\textwidth]{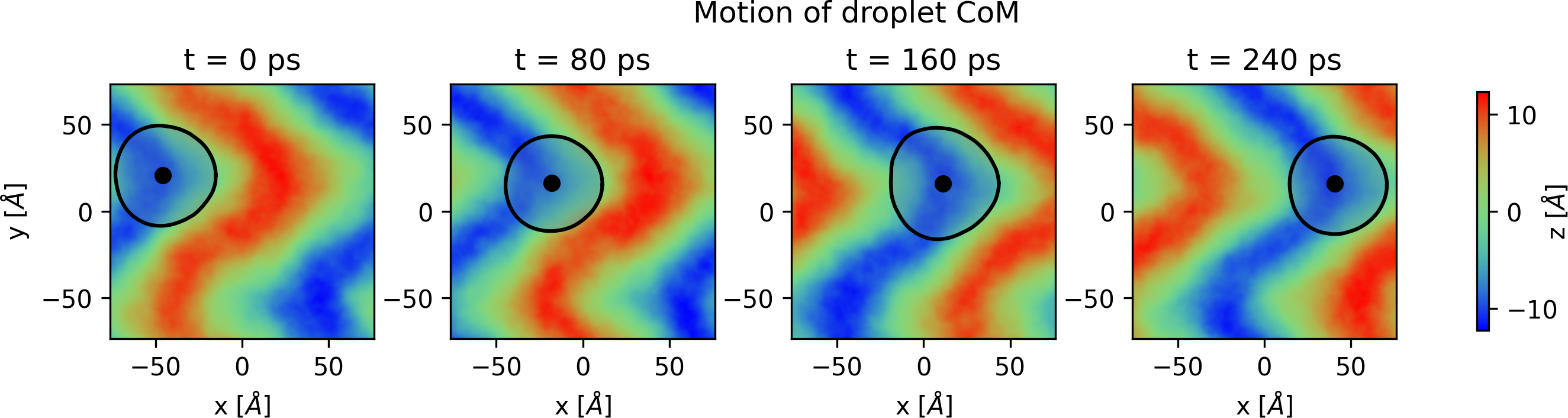}
    \caption{
    \textbf{The ``surfing'' motion of the droplet} on the long-ranged coherent rippling of the graphene sheet under compressive strain, as depicted in these series of snapshots of the $xy$-plane in the ``raw'' simulation coordinates (rather than droplet-centred coordinates). Here a droplet of 1,000 water molecules is placed upon a graphene sheet with $-1.0\%$ compressive strain. The graphene sheet is coloured according to the instantaneous $z$-coordinate; while the translucent blue shape indicates the location of the droplet, with the black outline showing the instantaneous three-phase contact line, and the black dot showing the droplet centre-of-mass. Across these snapshots, the long-ranged coherent wave moves from left to right, with the droplet ``surfing'' together with the wave in order to reside in the valley.
    }
    \label{fig:SI-surfing-motion}
\end{figure*}

In the long-ranged coherent rippling phase, it is seen that the water droplet resides in the valley of the wave. In particular, the long-ranged wave on the graphene sheet appears to be stationary in droplet-centred coordinates, and is reflected in the time-averaged graphene heightmap $\langle h_{\text{GS}}(x,y)\rangle_t$ over the droplet-centred $x$ and $y$ coordinates. This means that, in absolute coordinates, the motions of the long-ranged wave and the droplet are coupled tightly together, with the droplet ``surfing'' together with the wave in order to maintain its position in the wave valley. An example of this motion, for the sheet under $-1.0\%$ compressive strain, is shown in \Cref{fig:SI-surfing-motion}.

This consistent motion of the droplet together with the long-ranged coherent wave means that, relative to the droplet, the contact surface is not rotationally symmetric. Instead, one side of the droplet interface is continuously receding while the opposite side is advancing, being driven by the ``surfing'' motion. This results in the anisotropy of the three-phase contact line, with a range of anisotropic contact angles representing the possible dynamical contact angles between the minimum value of the receding contact angle and the maximum value of the advancing contact angle.

\clearpage

\section*{Droplet footprint on compressed graphene}

The presence of the droplet on free-standing and tensile-strained graphene induces a persistent distortion on the graphene sheet, causing a ``footprint'' on the time-averaged graphene heightmap $\langle z(x,y) \rangle$ as well as a localized perturbation of the long-time inclination correlation $\mathcal{C}_{\text{GS}}(\tau\to\infty)$ at the three-phase contact line. The droplet's footprint breaks the graphene sheet's translational symmetry, giving rise to both $\langle z(x,y) \rangle$ and $\mathcal{C}_{\text{GS}}(\tau\to\infty)$ being perturbed from a uniform background value in a manner centred around the droplet, as shown in Figure 3 of the main text.

For compressive strains greater than or equal to $-0.20\%$, however, this footprint no longer acts on a uniform background, as the translational symmetry of the graphene sheet is already broken by the long-ranged coherent ripple wave. A large static value of the time-averaged graphene heightmap $\langle z(x,y) \rangle$ in droplet-centred coordinates is to be expected, following the shape of the ripple wavefront, as the droplet is co-moving with the ripple wave under its ``surfing'' motion. As for the long-time inclination correlation $\mathcal{C}_{\text{GS}}(\tau\to\infty)$, corrugation-induced stiffening suppresses thermal ripples depending on the local curvature, thus causing $\mathcal{C}_{\text{GS}}(\tau\to\infty)$ to fall at the ripple wave's peaks and troughs, and to rise in between. The droplet's footprint must therefore act `on top of' the ripple wave's effects on $\langle z(x,y) \rangle$ and $\mathcal{C}_{\text{GS}}(\tau\to\infty)$.

\begin{figure*}[b!]
    \centering{}
    \includegraphics[width=\textwidth]{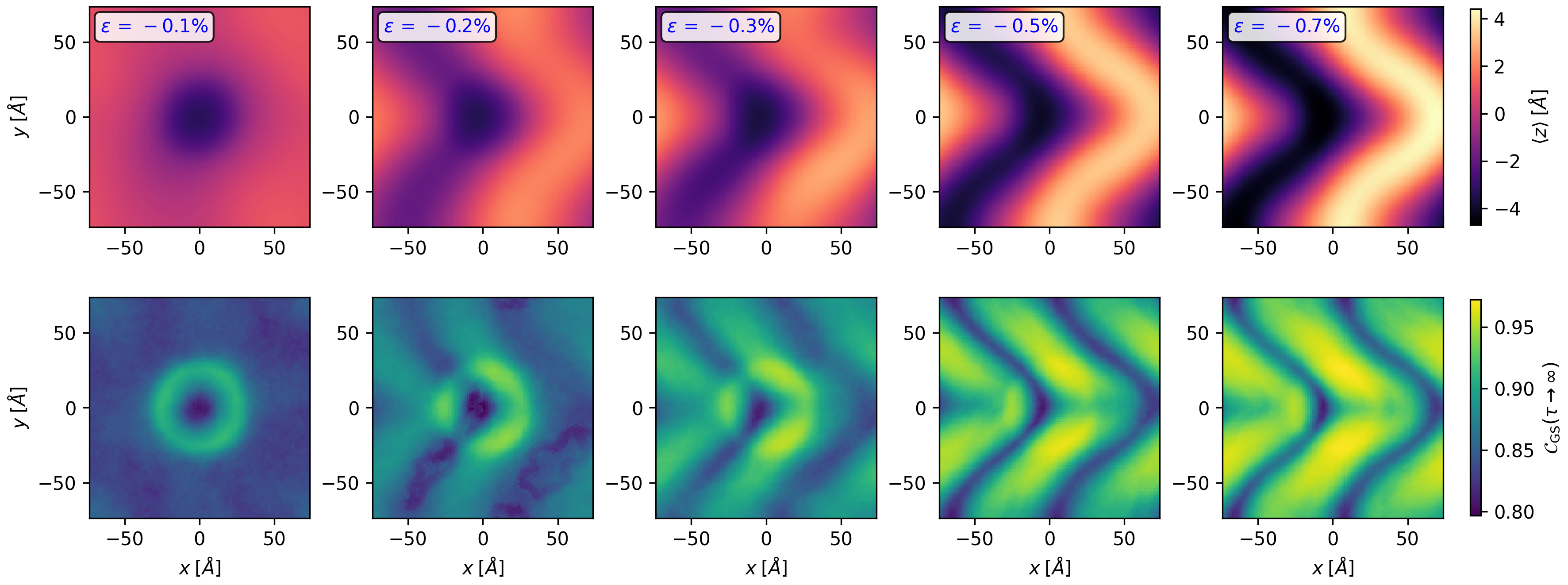}
    \caption{
    \textbf{The droplet's footprint on compressed graphene sheets} in droplet-centred coordinates, for compressive strains ranging from $-0.10\%$ to $-0.70\%$. Note that the random-to-coherent rippling phase transition occurs between $-0.15\%$ to $-0.20\%$, i.e.\ between the leftmost and second-leftmost columns of these plots. The top row depicts the time-averaged graphene heightmap $\langle z \rangle$, and the bottom row depicts the long-time inclination correlation $\mathcal{C}_{\text{GS}}(\tau\to\infty)$. As the compressive strain is increased, the coherent ripple wave becomes stronger in amplitude, as seen in the increasingly intense ``background'' induced in both $\langle z \rangle$ and $\mathcal{C}_{\text{GS}}(\tau\to\infty)$; consequently, while the droplet's footprint is still present, it becomes increasingly insignificant compared to the ripple wave's background.
    }
    \label{fig:SI-compressed-footprint}
\end{figure*}

Examples of the droplet's footprint on a few representative compressive strains, not much larger than the random-to-coherent rippling phase transition's critical strain, are shown in \Cref{fig:SI-compressed-footprint}. It is seen that, as the compressive strain is increased, the long-ranged coherent ripple wave becomes stronger in amplitude, resulting in an increasingly intense background variation of both $\langle z(x,y) \rangle$ and $\mathcal{C}_{\text{GS}}(\tau\to\infty)$. Consequently, the droplet's localized perturbation for both observables becomes increasingly insignificant compared to the background variation, ``disappearing'' into the ripple wave pattern.

The continuous ``disappearance'' of the droplet's footprint under increasing compressive strain, in terms of its effect on $\mathcal{C}_{\text{GS}}(\tau\to\infty)$ at the droplet perimeter, is reminiscent of the similar suppression of the $\mathcal{C}_{\text{GS}}(\tau\to\infty)$ `bump' under increasing tensile strain (main text, Figure 3f). This hints at a qualitative similarity between the maximum anisotropic contact angle under compressive strain, and how it increases with applied strain, versus the isotropic contact angle under tensile strain: namely, that corrugation-induced stiffening due to the long-ranged coherent ripple wave could also suppress the thermal ripple-wetting coupling, driving graphene to become less hydrophilic. That said, a final conclusion should be taken with the caveat that the dynamic angles for the droplet ``surfing'' on compressed graphene do not represent equilibrium contact angles, and may not be directly comparable to the angles under tension; nonetheless, this may explain the non-monotonic trend of graphene becoming less hydrophilic under either type of applied strain, as seen in Figure 2a of the main text.

\clearpage

\input{si.bbl}

%% file: main.bbl
\providecommand{\latin}[1]{#1}
\makeatletter
\providecommand{\doi}
  {\begingroup\let\do\@makeother\dospecials
  \catcode`\{=1 \catcode`\}=2 \doi@aux}
\providecommand{\doi@aux}[1]{\endgroup\texttt{#1}}
\makeatother
\providecommand*\mcitethebibliography{\thebibliography}
\csname @ifundefined\endcsname{endmcitethebibliography}
  {\let\endmcitethebibliography\endthebibliography}{}

%% file: si.bbl
\providecommand{\latin}[1]{#1}
\makeatletter
\providecommand{\doi}
  {\begingroup\let\do\@makeother\dospecials
  \catcode`\{=1 \catcode`\}=2 \doi@aux}
\providecommand{\doi@aux}[1]{\endgroup\texttt{#1}}
\makeatother
\providecommand*\mcitethebibliography{\thebibliography}
\csname @ifundefined\endcsname{endmcitethebibliography}
  {\let\endmcitethebibliography\endthebibliography}{}